\documentclass[english]{article}
\usepackage[T1]{fontenc}
\usepackage[latin9]{inputenc}
\usepackage{geometry}
\usepackage{color}
\usepackage{array}
\usepackage{float}
\usepackage{booktabs}
\usepackage{url}
\usepackage{multirow}
\usepackage{amsbsy}
\usepackage{amstext}
\usepackage{amssymb}
\usepackage{graphicx}
\usepackage{setspace}
\usepackage[authoryear]{natbib}
\makeatletter

\providecommand{\tabularnewline}{\\}
\floatstyle{ruled}
\newfloat{algorithm}{tbp}{loa}
\providecommand{\algorithmname}{Algorithm}
\floatname{algorithm}{\protect\algorithmname}

\usepackage{algpseudocode}
\usepackage{pdflscape}
\usepackage{pdfpages}
\date{}
\usepackage{colortbl}
\usepackage{xcolor}

\makeatother

\usepackage{babel}
\begin{document}
\title{A Scalable Recommendation Engine for New Users and Items}
\author{Boya Xu, Yiting Deng, and Carl F. Mela\thanks{Boya Xu is a PhD student at the Fuqua School of Business, Duke University
(email: \protect\url{boya.xu@duke.edu}). Yiting Deng is an Assistant
Professor at the UCL School of Management, University College London
(email: \protect\url{yiting.deng@ucl.ac.uk}). Carl F. Mela is the
T. Austin Finch Foundation Professor of Business Administration at
the Fuqua School of Business, Duke University (email: \protect\url{mela@duke.edu}).
A portion of this work was completed during the first author's internship
at JD.com. The authors would like to thank Allison Chaney, Rex Du,
Alina Ferecatu, Dokyun Lee, Bora Keskin, Siddharth Prusty, Julian
Runge, Wen-Yun Yang, Levin Zhu, participants at the 2021 Marketing
Science Conference, the 1st London Quant Marketing Conference, the
2022 MSI Data Analytics Conference, and 2022 Harvard Conference on
AI/ML/BA, as well as seminar participants at Seoul National University,
Dartmouth College, Northeastern University, Syracuse University and
Assetario for useful comments. All errors are our own.}}

\maketitle
\thispagestyle{empty}
\begin{abstract}
In many digital contexts such as online news and e-tailing with many
new users and items, recommendation systems face several challenges:
i) how to make initial recommendations to users with little or no
response history (i.e., cold-start problem), ii) how to learn user
preferences on items (test and learn), and iii) how to scale across
many users and items with myriad demographics and attributes. While
many recommendation systems accommodate aspects of these challenges,
few if any address all. This paper introduces a Collaborative Filtering
(CF) Multi-armed Bandit (B) with Attributes (A) recommendation system
(CFB-A) to jointly accommodate all of these considerations. Empirical
applications including an offline test on MovieLens data, synthetic
data simulations, and an online grocery experiment indicate the CFB-A
leads to substantial improvement on cumulative average rewards (e.g.,
total money or time spent, clicks, purchased quantities, average ratings,
etc.) relative to the most powerful extant baseline methods.

\textbf{Keywords}: recommendation, data reduction, multi-armed bandit,
cold start

\newpage{}
\end{abstract}

\section{Introduction\label{sec:Introduction}}

Recommender systems are ubiquitous in businesses (e.g., \citealp{Ying2006,Ansari2018,Bernstein2019,kokkodis2020good,kumar2020scalable,song2019and}),
having been generating personalized product lists on e-commerce sites
such as Amazon and Alibaba, as well as content sites such as Netflix
and YouTube. For instance, Netflix is believed to have benefited \$1
billion from its personalized recommendations,\footnote{\url{https://www.businessinsider.com/netflix-recommendation-engine-worth-1-billion-per-year-2016-6}}
and 64\% of YouTube's video recommendations bring more than 1 million
views.\footnote{\url{https://www.theatlantic.com/technology/archive/2018/11/how-youtubes-algorithm-really-works/575212/}}

Among the most prevalent algorithms used for online recommendation
is collaborative filtering (CF), which leverages information from
multiple data sources to find similarities among users in the items
they consume in order to recommend items based on those similarities
(\citealp{Sarwar2001}). In the context of movies, this might imply
that two individuals with overlaps in the genres they consume would
share the same preferences for genre. However, when new items or users
exist, CF lacks sufficient usage history to compute user similarities,
a problem known as the ``cold start'' issue (\citealp{Ahn2008}).
In many online contexts, such as news, music, and movies, new items
and users are common and the consideration is exacerbated by the scale
of users, items, and the attributes to describe them.

\begin{figure}[h]
\noindent \begin{centering}
\caption{Apple News Recommendations}
\par\end{centering}
\noindent \centering{}\includegraphics[scale=0.18]{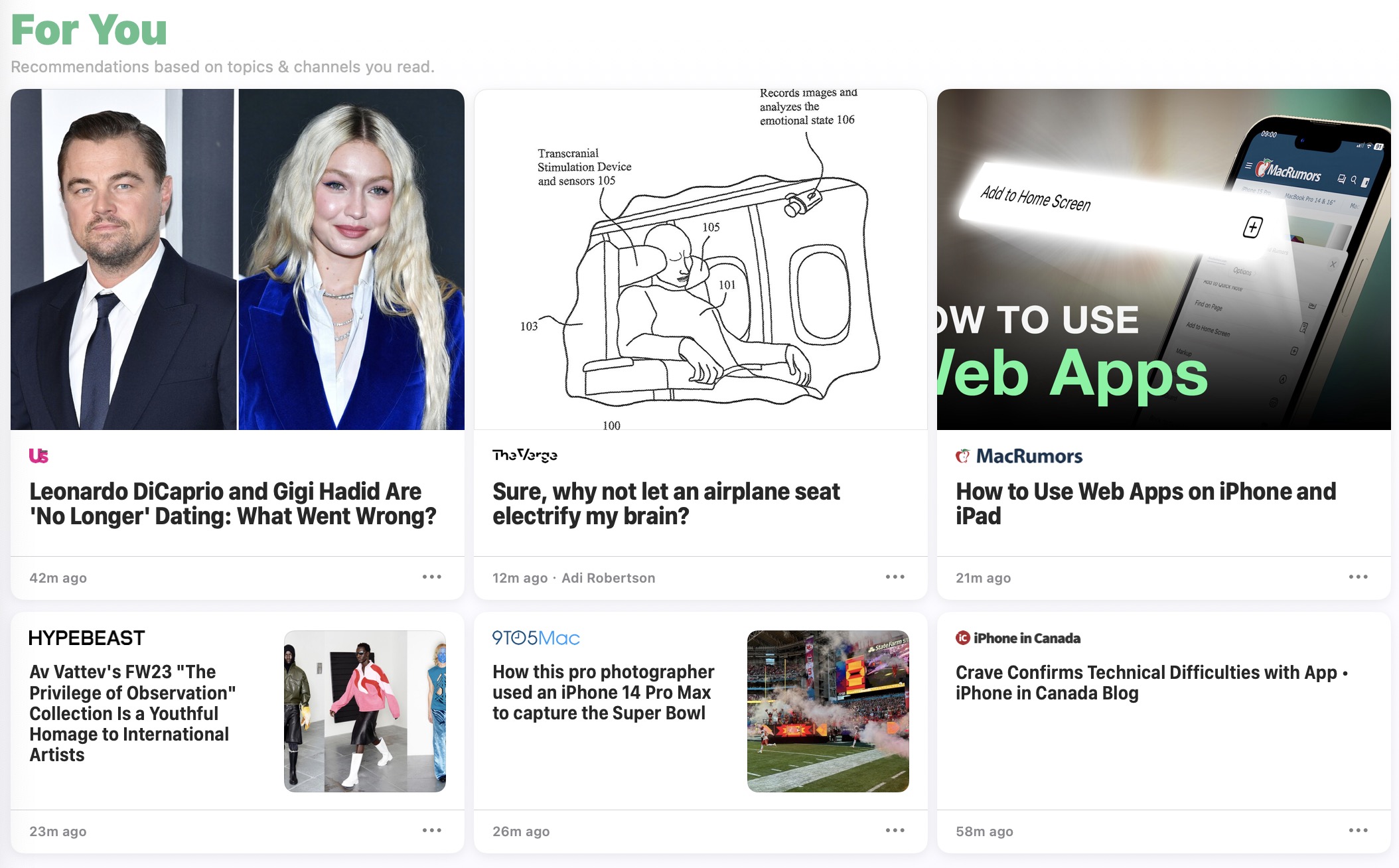}\label{Apple News}
\end{figure}

To illustrate these points, consider the Apple News recommendations
shown in Figure \ref{Apple News}. Each day, new items appear. Each
day, new readers appear, defined by a large set of demographic variables
and usage information. Each story they read is associated with potentially
hundreds of tags to reflect content, such as the topics, publisher,
location, length, time, and words in the document. Moreover, there
are more than 125 million active readers of Apple News and more than
300 publishers. Thus, there are many observations and many attributes
in the data, making recommendations in these contexts a ``big data''
problem. Collectively, these aspects raise the question of what sets
of new stories to recommend to users, and how to quickly learn the
preferences of new users in a large-scale data environment.

In sum, three key recommendation challenges are presented by many
online two-sided platforms: i) how to make initial recommendations
with little or no response history (cold-start); ii) how to efficiently
improve recommendations as user responses to recommendations are observed
(test and learn); and iii) how to cope with the large number of users
and items, each respectively described by a large number of features
(data reduction). The goal of this paper is to addresses all three
challenges.

We do so in two complementary and interwoven components. The first
component combines the CF with user demographic information\footnote{Demographic information refers to user-specific attributes, including
demographic variables (such as age and gender) and contextual variables
(such as browsing histories).} and item attribute information to initialize recommendations (e.g.,
a new user seeing a new article). The intuition is that similar users
will prefer similar items, where user similarity is reflected in similar
demographics and item similarity is reflected in similar attributes.
Hence, when a new user arrives, we can generate recommendations to
them based on what those with similar demographics liked. We call
this step the collaborative filtering with attributes (CFA), and it
solves problems i) and iii).

In the second component, given the preferences inferred from CF, bandit
algorithms are used to balance recommendations predicated upon what
is known about user preferences in the current period with novel recommendations
intended to explore their preferences to improve recommendations in
future periods (\citealp{elena2021survey,Schwartz2017,silva2022multi}).
This bandit learning step solves problem ii). In particular, the bandit
is combined with the CF to expedite learning user preferences for
items in high-dimensional environments with a large number of users
and items. The intuition is that it is more computationally efficient
to learn preferences that groups of users have for groups of items
rather than learning each user's preferences for each item. We call
this approach a collaborative filtering bandit (CFB), and it makes
recommendations at each stage of the user browsing process based on
what is known about user preferences entering that stage. As a result,
CFB is a scalable method that quickly learns how recommendations can
be further improved, and solves problems ii) and (iii). To address
i) - iii) simultaneously, we propose a method called CFB-A, a combination
of CFA and CFB, which can make scalable recommendations with good
priors in a cold-start context. While each component (the CFA and
CFB) has been deployed separately in prior research, to our knowledge
our combination of them is unique (see Section \ref{sec:Literature-Review}
for more detail).

Of note, challenges i) and ii) are complementary. To the extent that
available demographic and attribute information is (is not) highly
predictive of user-item match, there is little (much) need to test
and learn about user preferences. Hence the proposed recommendation
system is robust to both contexts. Oftentimes demographic data are
sourced from third-party vendors such as Experian Marketing Services
while user responses to product or content recommendations are warehoused
in a first-party user data platform. In the face of increasing interest
in consumer privacy (\citealp{Gordon_et_al_2021}), third-party data
become less available to help with cold-start considerations, and
the utility of bandits in learning preferences with first-party data
becomes more important.

Our model is tested on static and historical panel data from MovieLens,
synthetic simulations, and a live experiment on a simulated e-tailing
site that allows us to alter recommendations conditioned on past behaviors.
In the MovieLens data, the CFB-A method yields a modest 1\% improvement
in user ratings relative to the second-best extant benchmark model
which incorporates only data reduction and cold start. The finding
suggests that test and learn (problem ii) is not as critical a concern
as cold start (problem i) and data reduction (problems iii) in that
data. We conjecture this small improvement in the CFB-A over the second-best
model in the MovieLens data is due in part to the limited 1-5 categorical
movie rating scale; the top-performing approaches all generate high
ratings and face a ceiling effect that limits their variation in performance.
The simulated data use an unrestricted continuous rating scale and
evidence a 19\% improvement in the CFB-A over the second-best extant
model (the CFB which combines CF with test and learn). Because CFB
and CFB-A both include test and learn and data reduction components,
this result suggests that, for the parameters of the simulated data,
test and learn (problem ii) and data reduction (problem iii) are most
relevant. In the live experiment, the CFB-A increases cumulative homepage
purchase rates by 8\% compared to the best baseline method involving
either CF (data reduction), A (cold start), or B (test and learn)
components. Moreover, we find that data reduction, cold start, and
test and learn respectively enhance performance by 69\%, 19\%, and
8\%. Overall, we find the CFB-A approach to be adaptable across contexts.
In categories where attributes do not explain choices well, test and
learn becomes relatively important (and vice-versa). This finding
speaks to the importance of integrating both components into a recommendation
engine. The simulated data indicate our approach appears more robust
to different data generating processes than a number of other prominent
recommendation engines.

The rest of the paper is arranged as follows. Section \ref{sec:Literature-Review}
provides a literature review and illustrates our contributions. Section
\ref{sec:Model-Descriptions} details the CF, CFA, and CFB-A models.
Section \ref{sec:Simulations} presents simulations on MovieLens data
and synthetic data to evaluate the CFB-A method, with rationales behind
method advantages. Section \ref{sec:Live-Experiment} tests the CFB-A
method via an online grocery shopping experiment, and examines the
conditions under which either A or B component makes a dominant contribution.
Finally, Section \ref{sec:Discussions} concludes with a discussion
on future research directions.

\section{\label{sec:Literature-Review}Literature Review}

In addressing the aforementioned challenges in recommender systems
(cold start, test and learn, and data reduction including both large
numbers of users and items and the tags used to describe each of them),
this work builds upon several literature streams. Table \ref{tab:Literature-Summary}
outlines how the prior literature maps to the three challenges. Of
note, while the CF, CFA, CFB, and BA (bandits with attributes, or
contextual bandits) literature streams each addresses specific subsets
of challenges, none to our knowledge addresses them all. This is because
each of these models contains one or two components of the CFB-A (CF,
CFA, CFB, and BA), but not all three: CF, B, and A. Moreover, in later
sections, we show via simulation, benchmark data sets, and a live
experiment that addressing all three components (cold-start, data-reduction,
and test and learn) makes material differences in recommendation system
performance.

\begin{table}[h]
\caption{\label{tab:Literature-Summary}Literature Summary}

\,
\noindent \begin{centering}
\begin{tabular}{>{\centering}p{2cm}>{\centering}p{4cm}c>{\centering}p{2.2cm}>{\centering}p{2.5cm}>{\centering}p{1.5cm}}
\hline 
\centering{}Technology & \centering{}Representative Research & Cold Start & \multicolumn{2}{c}{Data Reduction} & \centering{}Test and Learn\tabularnewline
\cline{3-6} 
\centering{} & \centering{} &  & \centering{}Many Users and Items & \centering{}Many Demographics and Attributes & \centering{}\tabularnewline
\hline 
\centering{}CF & \centering{}\citet{Resnick1994,sardianos2017scaling,Wang2011,geng2022representation,koren2009matrix} &  & \centering{}$\surd$ & \centering{} & \centering{}\tabularnewline\arrayrulecolor{gray!30}\tabularnewline
\hline 
\centering{}CFA & \centering{}\citealp{Kumar2020,Wei2017,zhang2020collaborative,hu2019hybrid,strub2016hybrid,cortes2018cold,abernethy2009new,farias2019learning} & $\surd$ & \centering{}$\surd$ & \centering{}$\surd$ & \centering{}\tabularnewline
\hline 
\centering{}CFB & \centering{}\citet{jain2022online,kveton2017stochastic,Zhao2013,li2016collaborative,waisman2019online,Wang_Wu_Wang_2017,kawale2015efficient,lu2021low,trinh2020solving,Xing2014} &  & \centering{}$\surd$ & \centering{} & \centering{}$\surd$\tabularnewline
\hline 
\centering{}BA & \centering{}\citet{chu2011contextual,Li2010a,Zhou2015,wang2016learning} &  & \centering{} & \centering{}$\surd$$^{*}$ & \centering{}$\surd$\tabularnewline
\hline 
\centering{}\textcolor{black}{This paper (CFB-A)} & \centering{}-- & \textbf{\textcolor{black}{$\mathbf{\surd}$}} & \centering{}\textbf{\textcolor{black}{$\mathbf{\surd}$}} & \centering{}\textbf{\textcolor{black}{$\mathbf{\surd}$}} & \centering{}\textbf{\textcolor{black}{$\mathbf{\surd}$}}\tabularnewline\arrayrulecolor{black}\tabularnewline
\hline 
\end{tabular}
\par\end{centering}
\noindent \raggedright{}{\footnotesize{}$^{*}$Some classes of contextual
bandits (BA) do not reduce the demographic or attribute space, though
the ones most relevant to our analysis do.}{\footnotesize\par}
\end{table}

\subsection{Data Reduction (CF)\label{subsec:Collaborative-Filtering}}

The general logic of collaborative filtering (CF) recommendation systems
involves collaboration among multiple agents and data sources for
filtering information and data patterns, thereby reducing the dimensionality
of the user-by-item matrix of choices, ratings, or other measures
of user-item match. There are three main CF approaches (\citealp{shi2014collaborative}):
memory-based, model-based, and graph-based. Memory-based CF (\citealp{Resnick1994})
directly clusters users and items according to the similarity of their
observed features and predicts user-item interactions based on this
clustering. This neighborhood-based approach is infeasible for big
data due to its high computational complexity, and suffers from the
overfitting problem due to the lack of data regularization. Graph-based
CF (\citealp{He2020}) is another effective approach, which depicts
the relationships between users and items by a bipartite graph network
with a weighted link in each user-item pair. While scalability remains
a concern for graph-based CF (\citealp{sardianos2017scaling}), model-based
CF (\citealp{Wang2011,koren2009matrix}) overcomes these shortcomings
by modeling user-item interactions with low-dimensional latent factors
(often extracted via matrix factorization) to make predictions and
recommendations.

However, model-based CF relies on rich historical records, a condition
which is not satisfied in the contexts of new users and new items.
Hence, it is necessary in these contexts to augment the model-based
CF with user and item attributes to address the cold-start problem,
similar to recent advances incorporating deep learning into CF (e.g.,
\citealp{dong2020hybrid,geng2022representation}). We discuss this
literature next.

\subsection{Collaborative Filtering with Cold Start (CFA)}

Cold start is a common challenge in the digital environment. Proposed
approaches to address the cold-start problem in marketing outside
the context of collaborative filtering span contexts such as ad bidding
algorithms for new ads (\citealp{ye2020cold}), customer relationship
management for new users (\citealp{padilla2021overcoming}), personalized
website design for new visitors (\citealp{liberali2022morphing}).
\citet{gardete2020no} leverage consumer browsing data to mitigate
the cold-start problem, and \citet{hu2022zero} utilize characteristics
of salespeople (such as demographics) to overcome the cold-start problem
in deep learning-based recommendations matching salespeople with customers.

In the context of collaborative filtering, the existing literature
addresses the cold-start problem by incorporating user or item attributes
(\citealp{Kumar2020,Wei2017,zhang2020collaborative,hu2019hybrid,strub2016hybrid,cortes2018cold,abernethy2009new,farias2019learning}),
network information (\citealp{GonzalezCamacho2018}), or similarity
measures that link new users (items) with existing users (items) (\citep{Bobadilla2012,zheng2018spectral}).
This linkage enables the use of attribute information to address the
cold-start problem, in essence assuming those new users (or items)
will have similar preferences to other users (items) with the same
demographics (attributes). We denote approaches that integrate attribute
and demographic information (A) with the collaborative filtering approach
(CF) as the collaborative filtering with attributes (CFA).

Conceptually, one can think of the CFA as generating a prior belief
about user-item match when user-level purchase history is not available
and demographics are informative of preferences. Yet in and of itself,
the CFA cannot efficiently learn users' preferences from their initial
choices. This is especially problematic when attributes and demographics
explain a limited or negligible proportion of variation in user ratings
or choices. For this reason, bandits have been sometimes integrated
with CF into a CFB (albeit in the absence of A) as we discuss next.
Conceptually, such bandit methods can be thought of as an efficient
means by which to update prior beliefs about user-item match obtained
via the CF or CFA.

\subsection{\label{subsec:Multi-armed-Bandits}Collaborative Filtering with Test
and Learn (CFB)}

Our method also builds on literature applying online experimentation
to learn about uncertainties in marketing (\citealp{Schwartz2017,Misra2019,liberali2022morphing,ye2020cold,gardete2020no,aramayo2022multi}),
operations management (\citealp{Bertsimas2007,Bernstein2019,johari2021matching,keskin2022data}),
and computer science (\citealp{Li2010a,Gomez-Uribe2015,Wu2018InteractiveOL,Zhou2015,silva2022multi,elena2021survey,chu2011contextual,McInerney2018,guo2020deep}).
Specifically, we focus on Multi-armed Bandits (MAB), a classic reinforcement
learning problem (\citealp{Katehakis1987}). MABs seek experimental
designs that can maximize cumulative expected gains (or equivalently,
minimize cumulative expected regrets) when allocating arms (items
in our context) to an agent (in our case the user receiving the recommendation)
over time. By setting this long-term objective instead of focusing
on optimizing rewards (e.g., clicks, views, purchases) only in a single
time period (exploitation), the bandit approach balances exploitation
and learning about uncertainties of user preferences (exploration).
Though learning might reduce rewards in the current period, it enables
better recommendations and thus higher rewards in future periods.
The bandit approach ensures the learning costs are not too excessive
by limiting the divergence between the allocated arm and true optimality.\footnote{Conceptually, bandits relate to serendipity and diversity in recommendation
systems (\citealp{kotkov2016survey}). Serendipity is the ability
of a recommendation system to suggest an alternative that is both
unexpected, in the sense that it has not been recommended in the past,
and relevant in the sense that there is a match with consumer preferences.
Diversity is characterized by the breadth of items recommended. As
bandits recommend items with high match uncertainty (exploration)
and often deviate from recommending past choices, bandit-based recommendation
systems tend to be more serendipitous and diverse than their counterparts
without the exploration component.} Collectively, this research demonstrates the utility of the bandit
approach for learning. For example, \citet{Misra2019} use MABs coupled
with economic choice theory to capture an unknown demand curve with
ten alternative prices in order to maximize revenue. \citet{Schwartz2017}
use MABs to optimize ad impression allocation across twelve different
websites.

However, in contexts such as online retailing and media recommendation
systems, where users and items are both of large scale and with high-dimensional
characteristics, it can be challenging for current bandit approaches
to scale efficiently, making them difficult to implement. Hence, bandits
(B) have been combined with collaborative filtering to create CFBs
\citep{Zhao2013,li2016collaborative,Wang_Wu_Wang_2017,kawale2015efficient,Xing2014},
sometimes called low rank bandits \citep{jain2022online,kveton2017stochastic,lu2021low,trinh2020solving,bayati2022speed}.\footnote{Please see \citet{silva2022multi} and \citet{elena2021survey} for
surveys of applications of multi-armed bandits in recommendation systems.} The problem of large-scale users and items has been addressed with
user or item segmentation approaches, including both fixed clustering
(\citealp{Bertsimas2007,Agarwal2009a}) and dynamically updated clustering
(\citealp{Zhao2013,li2016collaborative,Christakopoulou2016TowardsCR,Bernstein2019,kawale2015efficient}).

Yet these CFB approaches do not incorporate the high-dimensional characteristics
of users and items that enable one to address the cold start problem.
Conceptually, CFBs help to update priors about user-item match but
do not incorporate cold-start to generate the better priors for new
users and items. This omission can be a limitation when attributes
and demographics are predictive of consumer feedback such as ratings
or choices.

\subsection{\label{subsec:Bandits-with-Attributes}Bandits with Attributes (BA)}

There exists a literature combining MABs and attribute data. We denote
this a bandit with attributes approach (BA), though it is oftentimes
described as ``contextual bandits'' (\citealp{chu2011contextual,Li2010a,Zhou2015,tang2014ensemble,nabi2022bayesian,li2013generalized,korkut2021disposable}).
Contextual bandits consider user demographics or item attributes as
contextual information to guide bandit experimentation. Principal
Component Analysis (PCA) is often applied to reduce data if the raw
demographics or attributes are high-dimensional (\citealp{Li2012,Zhao2013,wang2016learning}).
In practice, one can use either raw attribute data or the reduced
attribute data in a contextual bandits setting. However, the contextual
information in these BA models does not factor in user-item match
outcomes (e.g., purchases, ratings, etc.) so cannot be used to address
the cold-start issue, a notable limitation when considering new users
or items. For example, when contextual bandits use PCA to reduce data,
the principal components are extracted only from user-by-demographic
or item-by-attribute matrices; by construction they are therefore
less predictive of user-item matches than CFA-solved latent factors,
which specifically decompose the user-item match matrix. The CFB-A
which we characterize next accounts for not only the user-by-demographic
or item-by-attribute matrices, but also the user-by-item match matrix.

\subsection{Collaborative Filtering with Cold Start and MABs (CFB-A)}

In light of the foregoing discussion, this paper combines all three
components, CF, B and A into a CFB-A to address data reduction, cold-start,
and test and learn. By adding the bandit exploration (B) to the CFA,
the CFB-A affords more efficient learning on user preferences, especially
in contexts where attributes are not as informative about preferences
as past behaviors. In the same vein, by adding characteristics of
users and items (A), the CFB is extended into the CFB-A, which addresses
the cold-start problem by providing better initial guesses about user
preferences for items, thereby improving the bandit exploration.

Notably, combining a CFB and a CFA into a CFB-A generates two key
synergies. First, it enhances the efficiency of bandit learning, as
learning about the attribute (demographics) of one item (user) is
informative about the same attribute (demographics) of another item
(user). In other words, with informative priors, the bandit searches
over a dramatically reduced space of parameters, making estimation
more computationally efficient and scalable. Second, the combination
of CFB and CFA is more robust. The less informative attribute (demographics)
becomes about preferences across items (users) to address cold start
(as in the CFA), the more important the bandit becomes in enhancing
recommendations for new items/users.

To be clear, the CFB and CFA approaches in and of themselves are not
novel; rather it is their integration that is novel. In the ensuing
sections, using simulated data, a benchmark data set commonly applied
to evaluate recommendation systems, and a live experiment, we show
that this innovation yields material gains in recommendation performance.
In the next section, we describe how the CF, B and A components are
all combined.

\section{\label{sec:Model-Descriptions}Model Description}

\subsection{Model Overview}

This section describes the joint integration of the bandit and attribute
components into the collaborative filtering model. To simplify exposition,
we first layer the A component into the CF model top obtain the CFA
as a waypoint in our final model development. Next we layer on the
B component to the CFA into the CFB-A, detailing our key contribution
of jointly considering the CFA and CFB components.

As the first step, Section \ref{subsec:CF} outlines the CFA component
that models user preferences for items. CFA incorporates two types
of matrix factorizations, where i) the user-item preferences matrix
is reduced into latent user and item factors (CF), and ii) user-demographics
and item-attribute matrices (A) are projected respectively onto these
latent user and item factors. Intuitively, the double (user-item and
attribute) matrix factorizations alleviate the cold-start problem
by enabling users with similar demographics to have similar preferences
on items with similar attributes. Thus, as long as some demographics
are observed for new users or some attributes are observed for new
items, the initial recommendation can be facilitated.

While demographics and attributes are informative of preferences,
they cannot entirely capture users' preferences owing to unobservable
factors that influence these user preferences. Therefore, in the second
step, a learning stage (B) is implemented to improve the initial recommendations
from the CFA, which is described in Section \ref{subsec:Bandit-Learning-Stage}.
The bandit stage takes place at each recommendation occasion and seeks
to learn user preferences in order to improve future recommendations.
It experiments with current-period recommendations to learn user preferences
while, at the same time, minimizing the cost of experimentation in
terms of potentially worse current-period outcomes. As such, it trades
off expected current-period gains from recommending items with high
preference likelihood against future gains by recommending items with
high variance in preference likelihoods.

When there are too many items to learn about, the efficiency and accuracy
of bandit learning methods decline. Thus by implementing the B component
on the reduced dimension achieved by CFA, one can considerably improve
the learning efficiency of the bandit step. Section \ref{subsec:CFBA}
discusses the integration of the CFA and B components, where observed
user responses to recommendations are fed into the model for the preference
learning and future recommendations.

The main notation used to characterize our model is detailed in Appendix
\ref{sec:Notations}.

\subsection{CF with Matrix Factorizations (MF) \label{subsec:CF}}

This section begins by detailing the case where preferences are not
modeled as functions of user demographics and item attributes (CF),
and then proceeds to the case where they are (CFA).

\subsubsection{\label{subsec:No-Attributes-(CF)}No Attributes (CF)}

Denote $\boldsymbol{\text{\ensuremath{\mu}}}$ as an $I\times J$
preference matrix, where $I$ and $J$ are respectively the total
number of users and items, and each element $\mu_{ij}$ is user $i$'s
preferences for item $j$. This preference matrix can be decomposed
into latent spaces with dimensions $I\times K$ and $K\times J$ respectively,
thereby drastically reducing the preference matrix to $I\times K+K\times J$
parameters as one usually adopts a small value of $K$. $K$ is a
hyper-parameter. This decomposition technique is matrix factorization
(MF), a model-based collaborative filtering method (\citealt{koren2009matrix}).
Specifically, following \citet{Zhao2013} and \citet{Wang_Wu_Wang_2017},
we model the mean preferences (measured by feedback such as clicks,
ratings, purchased quantities, time spent, etc.) as a function of
$U\in\mathbb{R}^{I\times K}$ and $V\in\mathbb{R}^{J\times K}$:

\begin{equation}
\text{\ensuremath{\boldsymbol{\mu}}}=UV^{T}+\boldsymbol{\eta}\label{eq:mu_uv}
\end{equation}
where $\boldsymbol{\eta}\sim\mathcal{N}\left(\boldsymbol{0},\sigma^{2}\mathbf{I}_{I\times J}\right)$
are the observation errors. Similarly, define the predicted mean preference
as $\widehat{\boldsymbol{\mu}}=\text{\ensuremath{\widehat{U}\widehat{V}^{T}}}$.
The latent representations $\widehat{U}$ and $\widehat{V}$ are estimated
by minimizing the regularized squared error with respect to $U=\left(\boldsymbol{u}_{i}\right)_{i=1}^{I}$
and $V=\left(\boldsymbol{v}_{j}\right)_{j=1}^{J}$ as below: 
\begin{equation}
\min_{U,V}\sum_{i}\sum_{j}\left(\mu_{ij}-\boldsymbol{u}_{i}^{T}\boldsymbol{v}_{j}\right)^{2}+\lambda_{u}\left\Vert \boldsymbol{u}_{i}\right\Vert ^{2}+\lambda_{v}\left\Vert \boldsymbol{v}_{j}\right\Vert ^{2},\label{object_cfb}
\end{equation}
where $\lambda_{u}$ and $\lambda_{v}$ are regularization parameters
to avoid overfitting.

$\widehat{U}$ and $\widehat{V}$ can be imputed in a Bayesian fashion
(\citealp{Wang2011,Zhao2013}). Assume the following prior distributions
for $K$-dimensional vectors $\boldsymbol{u}_{i}$ and $\boldsymbol{v}_{j}$:
$\boldsymbol{u}_{i}\sim\mathcal{N}\left(0,\lambda_{u}^{-1}\mathbf{I}_{K}\right),$
$\boldsymbol{v}_{j}\sim\mathcal{N}\left(0,\lambda_{v}^{-1}\mathbf{I}_{K}\right)$.
The prior distribution for $\mu_{ij}$ is specified as $\mu_{ij}\sim\mathcal{N}\left(\boldsymbol{u}_{i}^{T}\boldsymbol{v}_{j},\sigma^{2}\right)$.
$\left\{ \lambda_{u},\lambda_{v},\sigma^{2}\right\} $ are hyper-parameters.

Therefore, the distribution of preference matrix $\boldsymbol{\mu}$
given $U$, $V$ and hyper-parameters can be expressed as the joint
probability,
\begin{equation}
P\left(\boldsymbol{\mu}\mid U,V,\lambda_{u},\lambda_{v},\sigma^{2}\right)=\Pi_{i=1}^{I}\Pi_{j=1}^{J}\left[\mathcal{N}\left(\mu_{ij}\mid\boldsymbol{u}_{i}^{T}\boldsymbol{v}_{j},\lambda_{u},\lambda_{v},\sigma^{2}\right)\right]^{y_{ij}}\label{mu_distribution_cf}
\end{equation}
where $y_{ij}=1$ if feedback of $i$ to $j$ is observed and $y_{ij}=0$,
otherwise.

Consequently, with the observed $\boldsymbol{\mu}$, we implement
MCMC to solve latent spaces $\{U,V\}$ based on their Bayesian updated
posterior distributions:
\begin{equation}
U\sim\mathcal{N}\left[\mathbf{F}_{U}\left(U^{prior},V,{\color{black}\boldsymbol{\mu}},\lambda_{u},\lambda_{v},\sigma^{2}\right)\right]\label{eq:cf_post_u}
\end{equation}
\begin{equation}
V\sim\mathcal{N}\left[\mathbf{F}_{V}\left(V^{prior},U,{\color{black}\boldsymbol{\mu}},\lambda_{u},\lambda_{v},\sigma^{2}\right)\right]\label{cf_post_v}
\end{equation}

Using Equation (\ref{eq:cf_post_u}) as an example, Web Appendix \ref{sec:Derivations-of-Posteriors_CF}
shows that for user $i$, $u_{i}\sim N\left(\overline{\mathbf{u}}_{i},\Sigma_{\boldsymbol{u}_{i}}\right)$,
where

\begin{equation}
\overline{\boldsymbol{u}}_{i}=\left(\frac{1}{\sigma^{2}}\sum_{j}y_{ij}\cdot\mu_{ij}\boldsymbol{v}_{j}\right)\Sigma_{\boldsymbol{u}_{i}},\label{u_posterior_mean_cf-1}
\end{equation}

\begin{equation}
\Sigma_{\boldsymbol{u}_{i}}=\left(\lambda_{u}\mathbf{I}_{K}+\sum_{j}y_{ij}\cdot\frac{1}{\sigma^{2}}\boldsymbol{v}_{j}\boldsymbol{v}_{j}^{T}\right)^{-1}.\label{u_posterior_covariance_cf-1}
\end{equation}

Details on Equation (\ref{cf_post_v}) are in Web Appendix \ref{sec:Derivations-of-Posteriors_CF}.

\subsubsection{Attributes (CFA)\label{subsec:Attributes-(CF-A)}}

The basic CF can be extended to CFA by incorporating the observable
characteristics (A) of users and items (\citealp{cortes2018cold}).
This step has two benefits. First, it enables the recommendation system
to ``borrow'' information across users and items to the extent that
those with similar demographics have preferences for similar attributes.
Second, it helps resolve the cold-start problem, because a user's
demographics can be used to form an initial guess about item preferences.

We augment the CF model with user demographics and item attributes
by creating two additional latent spaces to supplement the user-item
latent spaces. The first additional space links demographics with
the latent space $U$, allowing for matches between demographics and
user locations in the latent preference space. The second additional
space links attributes with the latent space $V$, allowing for matches
between attributes and item locations in the latent preference space.
The user demographics and item attribute matrices are denoted as $D\in\mathbb{R}^{I\times P}$
and $A\in\mathbb{R}^{J\times Q}$ respectively, where $P$ and $Q$
represent respectively the dimensions of the user-demographic vector
and the item-attribute vector. For each user, we specify $d_{ip}=\boldsymbol{u}_{i}^{T}\boldsymbol{w}_{p}+\eta_{ip}^{d}$
for $p=1,...,P$, and $a_{jq}=\boldsymbol{v}_{j}^{T}\boldsymbol{\psi}_{q}+\eta_{jq}^{a}$
for $q=1,...,Q$, where $\boldsymbol{w}_{p}\in\mathbb{R}^{K\times1}$
is the vector that maps user locations in the latent preference onto
user demographics, $\boldsymbol{\psi}_{q}\in\mathbb{R}^{K\times1}$
is the vector that projects item locations in the latent preference
onto item attributes, and $\eta_{ip}^{d}\stackrel{iid}{\sim}\mathcal{N}\left(0,\sigma_{d}^{2}\right)$
and $\eta_{jq}^{a}\stackrel{iid}{\sim}\mathcal{N}\left(0,\sigma_{a}^{2}\right)$
are observation errors. That is, $D=UW^{T}+\boldsymbol{\eta}_{d}$,
$A=V\Psi^{T}+\boldsymbol{\eta}_{a}$, where $W\in\mathbb{R}^{P\times K}$,
$\Psi\in\mathbb{R}^{Q\times K}$, $\boldsymbol{\eta}_{d}\sim\mathcal{N}\left(\boldsymbol{0},\sigma_{d}^{2}\mathbf{I}_{I\times J}\right)$,
and $\boldsymbol{\eta}_{a}\sim\mathcal{N}\left(\boldsymbol{0},\sigma_{a}^{2}\mathbf{I}_{I\times J}\right)$.
Therefore, Equation (\ref{object_cfb}) is extended to be

\begin{eqnarray}
 & \min_{U,V,W,\Psi} & \sum_{i,j}\left(\mu_{ij}-\boldsymbol{u}_{i}^{T}\boldsymbol{v}_{j}\right)^{2}+\sum_{i,p}\left[\left(d_{ip}-\boldsymbol{u}_{i}^{T}\boldsymbol{w}_{p}\right)^{2}+2\lambda_{u}\left\Vert \boldsymbol{u}_{i}\right\Vert ^{2}+\lambda_{w}\left\Vert \boldsymbol{w}_{p}\right\Vert ^{2}\right]\label{object_cfa}\\
 &  & +\sum_{q,j}\left[\left(a_{qj}-\boldsymbol{\psi}_{q}^{T}\boldsymbol{v}_{j}\right)^{2}+2\lambda_{v}\left\Vert \boldsymbol{v}_{j}\right\Vert ^{2}+\lambda_{\psi}\left\Vert \boldsymbol{\psi}_{q}\right\Vert ^{2}\right],\nonumber 
\end{eqnarray}
where $\lambda_{w}$ and $\lambda_{\psi}$ are two additional regularization
parameters.

Similar to Section \ref{subsec:No-Attributes-(CF)}, we assume the
following priors to estimate Equation (\ref{object_cfa}) in a Bayesian
fashion: i) $w_{p}\sim N\left(\textbf{0},\lambda_{w}^{-1}\mathbf{I}_{K}\right)$,
ii) $\psi_{q}\sim N\left(\textbf{0},\lambda_{\psi}^{-1}\mathbf{I}_{K}\right)$,
iii) $d_{ip}\sim N\left(\boldsymbol{u}_{i}^{T}\boldsymbol{w}_{p},\sigma_{d}^{2}\right)$,
and iv) $a_{qj}\sim N\left(\boldsymbol{v}_{j}^{T}\boldsymbol{\psi}_{q},\sigma_{a}^{2}\right)$.
$\left\{ \lambda_{w},\lambda_{\psi},\sigma_{d}^{2},\sigma_{a}^{2}\right\} $
are hyper-parameters, adding to $\left\{ \lambda_{u},\lambda_{v},\sigma^{2}\right\} $.
The priors for $\boldsymbol{u}_{i}$, $\boldsymbol{v}_{j}$, and $\mu_{ij}$
are the same as in Section \ref{subsec:No-Attributes-(CF)}.

The joint probability of observed $\boldsymbol{\mu}$, $D$ and $A$
conditional on the latent matrices $U,V,W,\Psi$, and the hyper-parameters
$\Sigma=\left\{ \sigma^{2},\sigma_{d}^{2},\sigma_{a}^{2},\lambda_{u},\lambda_{v},\lambda_{w},\lambda_{\psi}\right\} $
is as follows: 
\begin{eqnarray}
 &  & P\left(\boldsymbol{\mu},D,A\mid U,V,W,\Psi,\Sigma\right)\label{mu_distribution_cfa}\\
 & = & \Pi_{i=1}^{I}\Pi_{j=1}^{J}\left[\mathcal{N}\left(\mu_{ij}\mid\textbf{d}_{i},\textbf{a}_{i},U,V,W,\Psi,\Sigma\right)\right]^{y_{ij}}\Pi_{p=1}^{P}\mathcal{N}\left(d_{ip}\mid U,W,\Sigma\right)\Pi_{q=1}^{Q}\mathcal{N}\left(a_{jq}\mid V,\Psi,\Sigma\right)\nonumber 
\end{eqnarray}
Consequently, with the observed $\boldsymbol{\mu}$, $D$ and $A$,
the full conditional posteriors for the latent spaces $\{U,V,W,\Psi\}$
are given by: 
\begin{eqnarray}
U & \sim & \mathcal{N}\left[\mathbf{F}_{U}\left(U^{prior},V,W,{\color{blue}{\color{black}D,\boldsymbol{\mu}}},\Sigma\right)\right]\label{U_posterior}\\
V & \sim & \mathcal{N}\left[\mathbf{F}_{V}\left(V^{prior},U,\Psi,{\color{blue}{\color{black}A,\boldsymbol{\mu}}},\Sigma\right)\right]\label{eq:V_posterior}\\
W & \sim & \mathcal{N}\left[\mathbf{F}_{W}\left(W^{prior},U,{\color{black}D,}\Sigma\right)\right]\label{W_posterior}\\
\Psi & \sim & \mathcal{N}\left[\mathbf{F}_{\Psi}\left(\Psi^{prior},V,{\color{blue}{\color{black}A}},\Sigma\right)\right]\label{eq:Psi_posterior}
\end{eqnarray}
Using Equation (\ref{U_posterior}) as an example, Web Appendix \ref{sec:Derivations-of-Posteriors}
shows that for user $i$, $u_{i}\sim N\left(\overline{\mathbf{u}}_{i},\Sigma_{\boldsymbol{u}_{i}}\right)$,
where

\begin{eqnarray}
\overline{\mathbf{u}}_{i} & = & \left(\frac{1}{\sigma^{2}}\sum_{j=1}^{J}y_{ij}\cdot\mu_{ij}\boldsymbol{v}_{j}+\frac{1}{\sigma_{d}^{2}}\sum_{p=1}^{P}d_{ip}\boldsymbol{w}_{p}\right)\Sigma_{\boldsymbol{u}_{i}},\label{u_posterior_mean}\\
\Sigma_{\boldsymbol{u}_{i}} & = & \left[\lambda_{u}\mathbf{I}_{K}+\sum_{j=1}^{J}y_{ij}\cdot\frac{1}{\sigma^{2}}\boldsymbol{v}_{j}\boldsymbol{v}_{j}^{T}+\frac{1}{\sigma_{d}^{2}}\sum_{p=1}^{P}\boldsymbol{w}_{p}\boldsymbol{w}_{p}^{T}\right]^{-1}.\label{eq:u_posterior_covariance}
\end{eqnarray}

Equations (\ref{u_posterior_mean}) and (\ref{eq:u_posterior_covariance})
address the cold-start problem by extending Equations (\ref{u_posterior_mean_cf-1})
and (\ref{u_posterior_covariance_cf-1}) by adding a term with $\boldsymbol{w}_{p}$,
the vector that maps user locations in the latent preference onto
user demographics. Equation (\ref{u_posterior_mean}) demonstrates
the tradeoff between cold start and test and learn. If $y_{ij}=0$
for all $j$, that is no feedback from user $i$ on any items is observed,
then the posterior relies on user $i$'s demographics $\boldsymbol{d}_{i}$
and therefore user $i$'s location in the preference matrix is inferred
from the demographics, $\sum_{p=1}^{P}d_{ip}\boldsymbol{w}_{p}$,
where $W$ is inferred from existing users via Equation (\ref{W_posterior}).\footnote{Note that the demographics $D$ shift the prior means for $U$ if
that user's $D$ is informative about their preferences (that is,
$W$ is non-zero so that demographics are informative about preferences).} Thus information about demographics alleviates the cold-start problem
for new users.

When feedback of user $i$ to item $j$ is observed ($y_{ij}=1$),
the posterior user preference is a weighted sum of the contribution
from user feedback, $\mu_{ij}\boldsymbol{v}_{j}$, and the contribution
from user demographics, $\sum_{p=1}^{P}d_{ip}\boldsymbol{w}_{p}$.
The weights are functions of the precision in the respective component
(feedback or demographics). The greater the precision (or the smaller
the variance) of a component, the higher its weight is. Hence, as
more user feedback is collected, a greater weight is given to this
feedback (test and learn), and demographics (cold start) play a less
important role. In contrast, in the cold-start period, no feedback
is observed, so $\frac{1}{\sigma^{2}}\sum_{j}y_{ij}\cdot\mu_{ij}\boldsymbol{v}_{j}=\boldsymbol{0}$
and all the weight is placed on the demographic component. Additional
details on Equations (\ref{eq:V_posterior})-(\ref{eq:Psi_posterior})
are in Web Appendix \ref{sec:Derivations-of-Posteriors}.

In sum, there are three matrix factorizations in the CFA: i) a reduction
of the user-item preferences matrix, ii) a reduction of the user-demographic
matrix, and iii) a reduction of the item-attribute matrix. Collectively,
these reductions lead to a drastic decrease in the number of parameters
necessary to forecast user preferences, both enhancing forecast reliability
and expediting the bandit learning of preferences as discussed next.

\subsection{Bandit Learning Stage\label{subsec:Bandit-Learning-Stage}}

Following the general bandit problem, the objective function for optimization
is defined as: 
\begin{equation}
Max_{\left\{ j\left(\tau\right)\right\} }\,CAR=\sum_{i}E\left[\sum_{\tau}^{t-1}\mu_{ij\left(\tau\right)\tau}\right],\label{object_bandit}
\end{equation}
where $CAR$ is the cumulative average response (where the response
$\boldsymbol{\mu}$ can be measured with clicks, purchase quantities,
money or time spent, ratings, etc.) and the subscript $ij\left(\tau\right)\tau$
indicates the item recommended by an algorithm to user $i$ at a given
decision occasion $\tau$ (e.g., a movie viewing or article reading
occasion), where indexing item $j$ by $\tau$ (i.e., $j(\tau)$)
indicates that the recommended item is a function of available information
at decision occasion $\tau$. Because both current and future customer
feedback is considered in the objective function, it induces a tradeoff
between current and future period rewards.

Thus, the bandit problem solves for the sequence of recommended items
$\left\{ j\left(\tau\right)\right\} $ that maximizes $CAR$. Intuitively,
this goal implies finding the stream of recommendations that leads
to the highest possible average outcomes by trading off exploring
unknown preferences with exploiting them as learned.

Following \citet{Zhao2013}, we adopt the Upper Confidence Bound (UCB)
policy for contextual bandits (\citealp{chu2011contextual}) to solve
Equation (\ref{object_bandit}).\footnote{Two commonly applied approaches to solve the bandit problem are the
Upper Confidence Bound (UCB) approach (\citealp{auer2002using}) and
the Thompson Sampling (TS) approach (\citealp{Samples1933}). See
Algorithm \ref{alg:CFB-A_ts} in Appendix \ref{sec:CFB-A-algorithm-flow}
for more details on TS.} The UCB algorithm requires that either the user latent space ($U$)
or item ($V$) latent space is fixed while the other one is regarded
as coefficients to estimate in the linear payoff function specified
in Equation (\ref{eq:mu_uv}) (\citealp{chu2011contextual,Zhao2013}).\footnote{In this setting, the use of linear UCB policy (\citealp{chu2011contextual,Zhao2013})
is predicated upon the conditional independence of arm rewards specified
in Equation (\ref{eq:mu_uv}), where elements in the reward matrix
$\boldsymbol{\mu}$ are independent of each other because the observation
errors $\boldsymbol{\eta}\sim\mathcal{N}\left(\boldsymbol{0},\sigma^{2}\mathbf{I}_{I\times J}\right)$
are independent conditional on $U$ or $V$.} This condition is congruent with treating users (items) as new and
items (users) as existing if $U$($V$) is to be estimated and $V$
($U$) is fixed. In the case where users and items are both new, one
would iterate the UCB bandit recommendation step over new users for
existing items and new items for existing users.

In the case of new users (new items), the UCB recommends item $j^{*}$
at time $t$ that maximizes $\widehat{\mu}_{ij,t}+\alpha\cdot var\left(\widehat{\mu}_{ij,t}\right)$,
where $\widehat{\mu}_{ij,t}=\overline{\boldsymbol{u}}_{i,t}\widehat{\boldsymbol{v}}_{j}^{T}$
($\widehat{\mu}_{ij,t}=\widehat{\boldsymbol{u}}_{i}\overline{\boldsymbol{v}}_{j,t}^{T}$)
is the CFA-solved prediction of user $i$'s feedback to item $j$
and $var\left(\widehat{\mu}_{ij,t}\right)=\widehat{\boldsymbol{v}}_{j}\Sigma_{\boldsymbol{u}_{i,t}}\widehat{\boldsymbol{v}}_{j}^{T}$
($var\left(\widehat{\mu}_{ij,t}\right)=\widehat{\boldsymbol{u}}_{i}\Sigma_{\boldsymbol{v}_{j,t}}\widehat{\boldsymbol{u}}_{i}^{T}$)
is the prediction uncertainty from Section \ref{subsec:CF}. With
new users, $\overline{\boldsymbol{u}}_{i,t}$ and $\Sigma_{\boldsymbol{u}_{i,t}}$
are the posterior mean and covariance in Equations (\ref{u_posterior_mean})
and (\ref{eq:u_posterior_covariance}), respectively.\footnote{The posterior mean $\overline{\boldsymbol{v}}_{j,t}^{T}$ and covariance
$\Sigma_{\boldsymbol{v}_{j,t}}$ of new items are detailed in Equations
(\ref{eq:v_post_mean}) and (\ref{eq:v_post_cov}) in Web Appendix.} Following the probabilistic formulation of CFA (Equation \ref{mu_distribution_cfa}),
we adopt the maximum a posterior (MAP) for $\widehat{\boldsymbol{v}}_{j}$
($\widehat{\boldsymbol{u}}_{i}$) for existing items (users).\footnote{\citet{cortes2018cold} presents a non-Bayesian solution for Equation
(\ref{object_cfa}), which can be applied to the historical interaction
data ($\boldsymbol{\mu}$) to obtain $\widehat{\boldsymbol{v}}_{j}$
($\widehat{\boldsymbol{u}}_{i}$) for existing items (users).} Hyper-parameter $\alpha$ is the UCB scale parameter to balance exploration
and exploitation, such that a larger $\alpha$ places a greater emphasis
on exploration. Intuitively, a larger $\alpha$ favors an alternative
with high statistical uncertainty in $\widehat{\boldsymbol{\mu}}$,
as there is little to learn by recommending an item if a user's response
to it has been well known.

\subsection{The CFB-A\label{subsec:CFBA}}

There are three steps to the CFB-A: i) estimate the CFA in section
\ref{subsec:Attributes-(CF-A)} to obtain user locations in the factor
space and impute their implied preferences; ii) solve the UCB bandit
(B) in section \ref{subsec:Bandit-Learning-Stage} to recommend items
and observe choices, and iii) repeat the steps for the next purchase.
We detail each step below:

\paragraph{Step i):}

The input to this step is the demographic data, $D$, the attribute
data, $A$, and the observed preference matrix, $\boldsymbol{\mu}_{t}$.
The key outputs are the factorizations of the matrices, $\left\{ U_{t},V_{t},W_{t},\Psi_{t}\right\} $,
which imply user-item matches. The process is initialized by $\left\{ U_{0},V_{0},W_{0},\Psi_{0}\right\} $
as a new user or item arrives; because choices are not observed for
new users, these estimates are formed using the user's demographics,
$\boldsymbol{d}_{i}$ and the item's attributes $\boldsymbol{a}_{j}$.
This initialization builds on prior CFB methods (\citealp{jain2022online,kveton2017stochastic,Zhao2013,li2016collaborative,waisman2019online,Wang_Wu_Wang_2017,kawale2015efficient,lu2021low,trinh2020solving})
that initiate bandit explorations with naive values such as zero vectors.
When the user is not new, then their past choices, $y_{ij}$ and $\mu_{ij}$,
are also used to infer their preferences. Of note, two things update
(are learned) after observing a user rates or chooses an item: the
location of that user in the factor space and the factor space itself.
With just one user, the former effect is far more substantial. With
many users updated each period (such as at the end of the day), both
effects can become sizable.

\paragraph{Step ii):}

Next, the bandit determines the item(s) recommended to a user $i$
(see Section \ref{subsec:Bandit-Learning-Stage}). The inputs to this
step are the user's preferences, and the output is a recommended item.

\paragraph{Step iii)}

Finally, the user's response to the bandit suggested item is observed.
The input to this step is the recommendation and user response to
recommended item, $y_{ij}$ and $\mu_{ij}$, and the output to this
step is the updated preference matrix, $\boldsymbol{\mu}_{t+1}$.
The process then repeats to Step i).

In sum, this model solves the aforementioned three key challenges
as follows: First, by borrowing information across users and items
(CFA), step i) solves the cold-start problem and makes initial recommendations
based on $W$ and $\Psi$ for users or items with little or no response
history. Second, applying a bandit algorithm on preferences inferred
from CFA to choose the optimal item to recommend to each user in each
period, step ii) improves the recommendation efficiency. Third, by
reducing the full preference matrix $\mathbf{\boldsymbol{\mu}}$ to
$U$, $V$, $W$ and $\Psi$ with matrix factorizations in CFA in
step i), we alleviate both the large scale of users and items, and
the large scale of features on users and items. The algorithm is detailed
in Appendix \ref{sec:CFB-A-algorithm-flow}.

\section{\label{sec:Simulations}Simulations}

This section outlines tests of the CFB-A against various benchmarks
via simulations. These benchmarks are chosen to delineate the relative
contributions of our model outlined in Table \ref{tab:Literature-Summary}.
We do so using MovieLens and synthetic data. The MovieLens data (\citealt{Harper2015TheMD},
\url{https://grouplens.org/datasets/movielens/100k/}), a standard
data repository used for evaluating recommender systems, have the
advantage of capturing actual consumer behaviors. However, as archival
data, one cannot condition recommendations based on past choices.
To address this concern, we draw on precedent (\citealt{Zhao2013,chen2019generative,Christakopoulou2018,kille2015news,tang2012dynamic,wang2018online,elena2021survey,wang2022})
and apply the replay method (\citealp{Li2011}) on MovieLens data
to evaluate CFB-A method against benchmarks. The replay method only
retains observations where the algorithm's recommendations are the
same as the ones observed in the historical dataset, as if the recommendations
had actually been made.

To enable dynamic recommendations, we create data and recommendations
dynamically where simulated agents interact with recommendations.
However, the synthetic data are not actual user behaviors. In Section
\ref{sec:Live-Experiment}, therefore, a live experiment accommodates
both dynamic recommendations and actual behaviors.

This section first outline the various benchmarks for algorithm comparisons
(Section \ref{subsec:Benchmark-Approaches}). Next, Section \ref{subsec:MovieLens-Simulation}
presents the static MovieLens simulation (i.e., where choices of movies
have already been made in the absence of a recommendation system)
where we describe data details, introduce experimental design and
evaluation metric, and report results. Afterwards, Section \ref{subsec:Synthetic-Data-Simulation}
describes the dynamic synthetic data simulation with different data
generation processes.

\subsection{\label{subsec:Benchmark-Approaches}Benchmark Approaches}

Table \ref{tab:Benchmark-Models} outlines the various benchmark models
selected from the literature and their features, which allows us to
infer the key contrasts with the CFB-A approach. For example, relative
to UCB-pca, the CFB-A provides cold-start recommendations.

\begin{table}[h]
\noindent \begin{centering}
\caption{\label{tab:Benchmark-Models}Benchmark Models}
\par\end{centering}
\vspace{0.2cm}

\noindent \centering{}%
\begin{tabular}{>{\raggedright}m{2.7cm}>{\raggedright}m{2.7cm}c>{\centering}p{2cm}>{\centering}p{2.2cm}c}
\hline 
\centering{}Technology & \centering{}Method & Cold Start & \multicolumn{2}{>{\centering}p{4cm}}{Data Reduction} & Test and Learn\tabularnewline
\cline{3-6} 
\centering{} & \centering{} &  & Many Users and Items & Many Demographics and Attributes & \tabularnewline
\hline 
\multirow{2}{2.7cm}{\centering{}Null} & \centering{}Random &  &  &  & \tabularnewline\arrayrulecolor{white!30}\tabularnewline
 & \centering{}Popularity &  &  &  & \tabularnewline\arrayrulecolor{gray!30}\tabularnewline
\hline 
\centering{}CFA & \centering{}Active Learning & $\surd$ & $\surd$ & $\surd$ & \tabularnewline\arrayrulecolor{gray!30}\tabularnewline
\hline 
\multirow{4}{2.7cm}{\centering{}BA} & \centering{}TS &  &  &  & $\surd$\tabularnewline\arrayrulecolor{white!30}\tabularnewline
 & \centering{}UCB &  &  &  & $\surd$\tabularnewline\arrayrulecolor{white!30}\tabularnewline
 & \centering{}TS\_pca &  &  & $\surd$ & $\surd$\tabularnewline\arrayrulecolor{white!30}\tabularnewline
 & \centering{}UCB\_pca &  &  & $\surd$ & $\surd$\tabularnewline\arrayrulecolor{gray!30}\tabularnewline
\hline 
\centering{}CFB & \centering{}CFB &  & $\surd$ &  & $\surd$\tabularnewline\arrayrulecolor{gray!30}\tabularnewline
\hline 
\centering{}CFB-A & \centering{}CFB-A & $\surd$ & $\surd$ & $\surd$ & $\surd$\tabularnewline\arrayrulecolor{black}\tabularnewline
\hline 
\end{tabular}
\end{table}
\noindent The descriptions of the various benchmark models are as
follows:

\paragraph{Random:}

The recommendation system randomly recommends items among all available
ones to a new user.

\paragraph{Popularity-based (POP):}

The recommendation system allocates the most popular items (i.e.,
items with the highest average rating) to a new user. The popularity
score of an item is calculated by historical records (or referred
to as the training set).

\paragraph{Active Learning (AL):}

The recommendation system recommends items with the highest new user
preference uncertainties (i.e., the standard deviation of $\widehat{\boldsymbol{\mu}}$)
, based on the idea of minimizing these uncertainties (\citealp{harpale2008personalized,rubens2015active}).
For comparability, in this paper the AL adopts CFA-solved latent factors
$\left\{ U,V,W,\Psi\right\} $ to compute uncertainties of user feedback.
However, there is no bandit experimentation used in the AL algorithm.

\paragraph{Thompson Sampling (TS) and Upper Confidence Bound (UCB):}

TS and UCB are both predominant heuristics used to solve the multi-armed
bandit (\citealp{elena2021survey}). Both algorithms dynamically update
the probability for an item to be chosen. TS selects items based on
the sampling probability that an item is optimal. UCB, which was described
in Section \ref{subsec:Bandit-Learning-Stage}, explores items with
higher uncertainty and a strong potential to be the optimal choice.
Random recommendations can be made to each user at the initial period
to initialize the estimated preferences, and then allocate items over
the remaining periods via TS or UCB to update preference estimates
over time.

\paragraph{TS Principal Components (TS\_pca) and UCB Principal Components (UCB\_pca):}

In the case of high-dimensional attributes, the dimension of item
attributes are reduced by using Principal Component Analysis (PCA).
Then, we apply the retained components to recommend items following
TS and UCB rules, respectively. A notable difference between the CFB-A
and the two contextual bandit methods is that the latter reduces the
attribute space based only on the attribute correlations, without
considering which of the attributes are informative about preferences.
This might hamper prediction because the reduced space is not created
with prediction in mind.

\paragraph{Collaborative Filtering Bandit (CFB):}

The CFB method is a combination of CF (Section \ref{subsec:No-Attributes-(CF)})
and bandit (Section \ref{subsec:Bandit-Learning-Stage}). The difference
between CFB and CFB-A is that CFB-A also incorporates the observed
characteristics of users and items via two additional matrix factorizations
(the user-by-demographic matrix and the item-by-attribute matrix).

The details of hyper-parameter tuning for CFB-A and benchmark models
are provided in Web Appendix \ref{sec:Hyper-parameter-Tuning}.

\subsection{\label{subsec:MovieLens-Simulation}MovieLens Simulation}

\subsubsection{MovieLens Dataset}

The MovieLens 100k dataset\footnote{https://grouplens.org/datasets/movielens/100k/}
was collected by the GroupLens Research Project at the University
of Minnesota, and it is often used as a stable benchmark dataset in
computer science literature for recommendation algorithm comparisons.
This dataset consists of 100,000 ratings (1-5) from 943 users on 1682
movies. The data contain basic information on viewers (e.g., age,
gender, occupation, zip) and movies (e.g., title, release date, genre,
etc.). The number of movies rated by each user ranges from 20 to 737
with a mean of 106 and a standard deviation of 101. The number of
ratings received by each movie ranges from 1 to 583 with a mean of
59 and a standard deviation of 80.

Given that 16 occupations (e.g., technician, writer, etc.), age and
gender are observed, demographics are represented by an 18-dimensional
vector, with 16 dummy variables corresponding to the 16 occupations
and two variables capturing age and gender (i.e., 1 for female, and
0 otherwise). Similarly, movie attributes are represented by an 18-dimensional
attribute vector corresponding to 18 movie genres.

\subsubsection{Simulation Design and Evaluation}

We consider different numbers of periods $T$: 40, 60, 80, 100, and
120. To ensure that sufficient data are collected for the replay method,
we randomly select 200 users from those with more than $T$ periods
of records to form the testing set, and the remaining 743 users form
the training set.

Recall, the replay method only uses alternatives that a user has seen
in the archival data. Each user is presented with a slate of items
in each period instead of a single recommendation to ensure that a
sufficient number of recommended items are collected by the replay
method for algorithm evaluations. If the slate were too small, then
there would often be no feedback collected because the user would
not have seen the recommended movie in the slate, making it hard to
use the replay method to compare alternative recommendation systems.
Users in the training set are used to tune the hyper-parameters for
each method, including the exploration rate $\alpha$ for methods
with a bandit stage, $K$ and $\Sigma$ for CFB-A and Active Learning,
$K$ and $\sigma^{2}$ for CFB, and item popularity for the Popularity
policy.

The measure used to evaluate algorithm performance is the aggregate
cumulative average rewards, defined as follows:
\begin{equation}
CAR=\frac{1}{\widetilde{N}}\sum_{i=1}^{\widetilde{N}}\frac{1}{T_{i}}\sum_{t=1}^{T_{i}}r_{it}\label{eq: CAR}
\end{equation}

\noindent where $\widetilde{N}\leq200$ is the total number of new
users who contribute at least one observed rating, $r_{it}$, on a
scale of 1-5 in response to an algorithm's recommendations. $T_{i}\in\left[1,T\right]$
is the total number of periods during which user $i$ has at least
one observed feedback to an algorithm's recommendations.

\subsubsection{\label{subsec:Results}Results}

Table \ref{tab:movielens} summarizes the aggregate cumulative average
rewards under different methods and number of periods, with a slate
size of 10.\footnote{Because each duration T is associated with a different sample, comparisons
should be made across methods for the same T, instead of across Ts
for the same method. Section \ref{subsec:Synthetic-Data-Simulation}
reports that, for the synthetic data not subject to this sampling
issue, the performance of methods that incorporate test and learn
improves over the duration of time.}

\begin{figure}[h]
\caption{\label{fig:stacked_barplot-of-Rating}Distribution of Received Ratings
by Method (T=120)}

\begin{centering}
\includegraphics[scale=0.6]{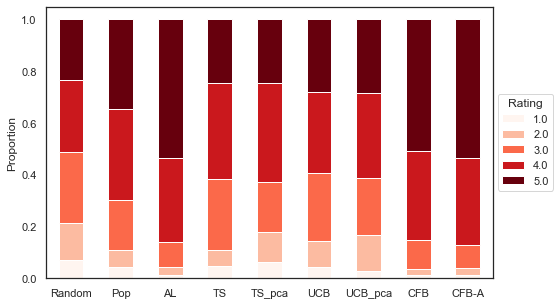}
\par\end{centering}
{\footnotesize{}Notes: Each color represents the proportion of each
categorical rating (1-5) received by recommended movies over T=120.
A larger area of a color within a bar implies a higher proportion
of its corresponding rating category.}{\footnotesize\par}
\end{figure}

Results in Table \ref{tab:movielens} indicate that the CFB-A method
outperforms other benchmarks under different settings of period length.
Compared to the most powerful extant baseline model (Active Learning),
the maximum improvement by the CFB-A method is about 1\% ($(4.38-4.33)/4.33$
for $T=60$). Thus, for the MovieLens data, data reduction and cold
start are sufficient features for making recommendations as evidenced
by the strong performance of Active Learning and CFB-A, both of which
share these components. Yet because ratings are capped at 5 and use
a highly limited categorical range, the MovieLens data are not ideal
for discriminating between these highest performing methods. Figure
\ref{fig:stacked_barplot-of-Rating} depicts this ceiling effect,
showing that more than half of all movie ratings across consumers
and occasions for the AL, CFB, and CFB-A are capped at 5 out 5. Loosely
speaking, these approaches are nearly ``maxed out'' on the MovieLens
rating scale.

\begin{table}[h]
\caption{\label{tab:movielens}$CAR$ for the MovieLens Dataset}

\vspace{0.2cm}

\noindent \begin{centering}
{\scriptsize{}}%
\begin{tabular}{>{\raggedright}m{1.5cm}>{\raggedright}m{1.2cm}>{\centering}m{1.2cm}>{\centering}m{1.2cm}>{\centering}m{1.5cm}>{\centering}m{1.7cm}ccccc}
\hline 
\centering{}{\scriptsize{}Technology} & \centering{}{\scriptsize{}Method} & {\scriptsize{}Cold Start} & \multicolumn{2}{c}{{\scriptsize{}Data Reduction}} & {\scriptsize{}Test and Learn} & \multicolumn{5}{c}{{\scriptsize{}Number of Periods}}\tabularnewline
\cline{3-11} 
\centering{} & \centering{} &  & {\scriptsize{}Many Users and Items} & {\scriptsize{}Many Demographics and Attributes} &  & {\scriptsize{}T=40} & {\scriptsize{}T=60} & {\scriptsize{}T=80} & {\scriptsize{}T=100} & {\scriptsize{}T=120}\tabularnewline
\hline 
\multirow{2}{1.5cm}{\centering{}{\scriptsize{}Null}} & \centering{}{\scriptsize{}Random} &  &  &  &  & {\scriptsize{}3.50} & {\scriptsize{}3.36} & {\scriptsize{}3.42} & {\scriptsize{}3.43} & {\scriptsize{}3.47}\tabularnewline\arrayrulecolor{white!30}\tabularnewline
 & \centering{}{\scriptsize{}Popularity} &  &  &  &  & {\scriptsize{}3.93} & {\scriptsize{}3.94} & {\scriptsize{}3.90} & {\scriptsize{}3.92} & {\scriptsize{}3.91}\tabularnewline\arrayrulecolor{gray!30}\tabularnewline
\hline 
\centering{}{\scriptsize{}CFA} & \centering{}{\scriptsize{}Active Learning} & {\scriptsize{}$\surd$} & {\scriptsize{}$\surd$} & {\scriptsize{}$\surd$} &  & \textbf{\scriptsize{}4.38} & {\scriptsize{}4.33} & {\scriptsize{}4.36} & {\scriptsize{}4.40} & {\scriptsize{}4.38}\tabularnewline\arrayrulecolor{gray!30}\tabularnewline
\hline 
\multirow{4}{1.5cm}{\centering{}{\scriptsize{}BA}} & \centering{}{\scriptsize{}TS} &  &  &  & {\scriptsize{}$\surd$} & {\scriptsize{}3.69} & {\scriptsize{}3.77} & {\scriptsize{}3.78} & {\scriptsize{}3.76} & {\scriptsize{}3.71}\tabularnewline\arrayrulecolor{white!30}\tabularnewline
 & \centering{}{\scriptsize{}UCB} &  &  &  & {\scriptsize{}$\surd$} & {\scriptsize{}3.50} & {\scriptsize{}3.54} & {\scriptsize{}3.67} & {\scriptsize{}3.68} & {\scriptsize{}3.71}\tabularnewline\arrayrulecolor{white!30}\tabularnewline
 & \centering{}{\scriptsize{}TS\_pca} &  &  & {\scriptsize{}$\surd$} & {\scriptsize{}$\surd$} & {\scriptsize{}3.56} & {\scriptsize{}3.58} & {\scriptsize{}3.59} & {\scriptsize{}3.63} & {\scriptsize{}3.63}\tabularnewline\arrayrulecolor{white!30}\tabularnewline
 & \centering{}{\scriptsize{}UCB\_pca} &  &  & {\scriptsize{}$\surd$} & {\scriptsize{}$\surd$} & {\scriptsize{}3.89} & {\scriptsize{}3.75} & {\scriptsize{}3.74} & {\scriptsize{}3.76} & {\scriptsize{}3.73}\tabularnewline\arrayrulecolor{white!30}\tabularnewline
\hline 
\centering{}{\scriptsize{}CFB} & \centering{}{\scriptsize{}CFB} &  & {\scriptsize{}$\surd$} &  & {\scriptsize{}$\surd$} & {\scriptsize{}4.37} & {\scriptsize{}4.32} & {\scriptsize{}4.32} & {\scriptsize{}4.33} & {\scriptsize{}4.32}\tabularnewline\arrayrulecolor{gray!30}\tabularnewline
\hline 
\centering{}{\scriptsize{}CFB-A} & \centering{}{\scriptsize{}CFB-A} & {\scriptsize{}$\surd$} & {\scriptsize{}$\surd$} & {\scriptsize{}$\surd$} & {\scriptsize{}$\surd$} & \textbf{\scriptsize{}4.38} & \textbf{\scriptsize{}4.38} & \textbf{\scriptsize{}4.39} & \textbf{\scriptsize{}4.41} & \textbf{\scriptsize{}4.39}\tabularnewline\arrayrulecolor{black}\tabularnewline
\hline 
\end{tabular}{\scriptsize\par}
\par\end{centering}
{\footnotesize{}Notes: Each cell depicts the average CAR (in terms
of movie ratings) across users by method and period length. Higher
numbers imply better ratings received by recommended movies.}{\footnotesize\par}
\end{table}
To provide a better basis of comparison across approaches, unrestricted
feedback ranges are used in the synthetic data simulation in Section
\ref{subsec:Synthetic-Data-Simulation} and more categorical levels
(purchase rates on a 0-100 categorical scale) are used in the experiment
in Section \ref{sec:Live-Experiment}. As the additional data sets
will indicate, i) the sufficiency of the data reduction and cold-start
features for recommendation engines is not generally the case, and
ii) the advantages of the CFB-A become more apparent with a broader,
more sensitive scale.

\begin{figure}[h]
\caption{\label{fig:The-Boxplot-of} Boxplots of Number of Items Searched (by
User and Phase, T=120)}

\textcolor{white}{\textbackslash\textbackslash\textbackslash}
\noindent \begin{centering}
\includegraphics[scale=0.5]{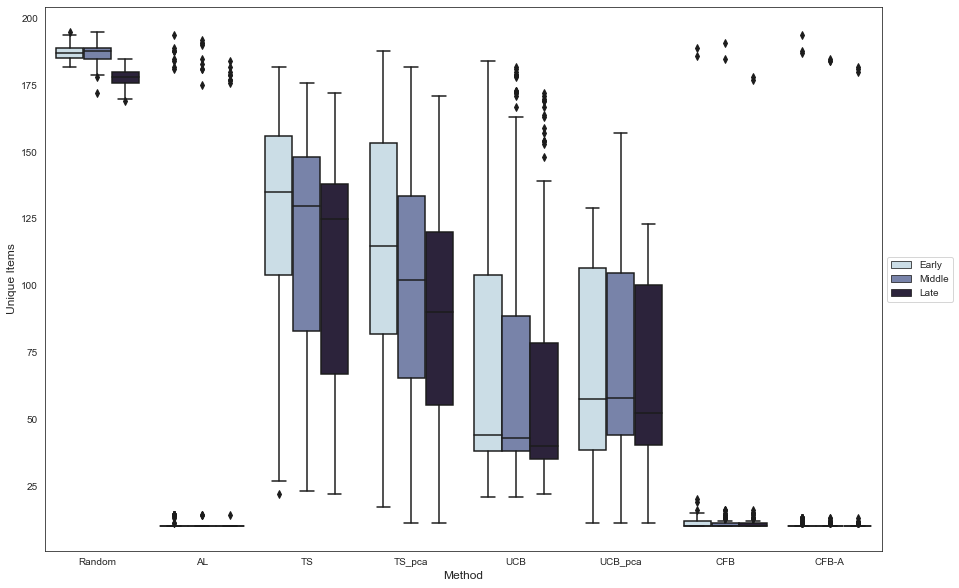}
\par\end{centering}
{\footnotesize{}Notes: Early phase includes periods 61-80; Middle
Phase includes periods 81-100; Late Phase includes periods 101-120.
The number of unique items for each phase represents the number of
non-redundant alternatives recommended to each user in the phase averaged
over users. A smaller number implies that less experimentation is
needed to infer user preferences and thus less inefficiency in making
higher rated recommendations.}{\footnotesize\par}
\end{figure}

Next we consider the search area dynamics on the bandit stage within
the MovieLens data, which investigates the search area of optimal
recommendations by each algorithm over time. An algorithm is more
efficient if the optimality search is more concentrated over the item
space. One advantage of CFB-A is the improved search efficiency, as
the bandit stage only needs to search user preferences in a reduced
space. Figure \ref{fig:The-Boxplot-of} shows the boxplots of number
of unique items recommended to all 200 new users over three phases
when $T=120$. This number decreases as search becomes more efficient
(fewer unique items are suggested within the slates of items presented
to them).

The CFB-A and Active Learning, two methods using CFA-solved latent
factors, have the highest item concentrations (and therefore the greatest
search efficiency) even in early periods when information about consumer
preferences is most limited. This implies that CFA-solved latent factors
used for data reduction are highly informative for narrowing the search
area of bandit learning even at the early stage. Unsurprisingly, the
random policy has the most items to explore and therefore the lowest
item concentration. Moreover, all traditional bandit approaches have
an increasing trend of item concentrations (and of search efficiency)
over time, implying a shrinking area of search for the best consumer-item
match as they learn. Given TS methods incorporate explorations via
sampling which may involve options with higher variances in practice,
the TS and TS\_pca have lower item concentrations compared to UCB
and UCB\_pca, respectively. An interesting insight, comparing across
methods, is that the cold-start solution generates more search efficiency
than the bandit solution, suggesting that the attribute information
is relatively informative of user preferences. 

\subsection{\label{subsec:Synthetic-Data-Simulation}Synthetic Data Simulation}

This section first outlines the approach to generate the synthetic
data, and then reports the results of our benchmarking analysis.

\subsubsection{\label{subsec:Data-Generation}Data Generation}

Recall, the goal of our simulation is to compare how well the various
alternative models fare in making recommendations for new users (items)
facing a large set of alternatives (users) in contexts with a large
number of attributes and demographics. To accomplish this aim, 1000
users are simulated in a market with 1000 items, each characterized
by a set of attributes.

Once users and items are created, the next step is to generate the
user-item match utility over items. We create two different match
utility datasets that vary in how the utility is computed. First,
a \emph{linear} model generates utilities as linear functions of user
demographics and item attributes. Second, \emph{nonlinear} match utilities
are constructed where demographics, attributes and utilities are determined
using a factor model. The linear model is a standard approach used
to construct match utilities in marketing and economics, and the factor
model more closely comports to the data reduction approach for specifying
match utilities in the CFB-A. Having two different data generating
strategies enables an assessment of the robustness of the CFB-A in
making recommendations across them.

The\emph{ linear setting} specifies the matrix $Utility\in\mathbb{R}^{I\times J}$
for user $i\in\left\{ 1,...,I\right\} $ and item $j\in\left\{ 1,...J\right\} $
as: 
\begin{eqnarray}
Utility & = & \boldsymbol{\beta}\widetilde{A}^{T}+\boldsymbol{\epsilon},\label{eq:linear_utility}\\
\boldsymbol{\beta} & = & D\Gamma+\boldsymbol{e},\label{eq:linear_preference}
\end{eqnarray}
where $D\in\mathbb{R}^{I\times P}$ is the demographic matrix, $\widetilde{A}=\left[\boldsymbol{1}\quad A\right]\in\mathbb{R}^{J\times(Q+1)}$
is the attribute matrix $A$ (appended with a vector of ones to allow
for an intercept), and $\boldsymbol{\beta}\in\mathbb{R}^{I\times(Q+1)}$
is the matrix of user preferences for item attributes. Note that these
preferences, $\boldsymbol{\beta}$, are a function of user demographics
where $\Gamma\in\mathbb{R}^{P\times(Q+1)}$ maps these preferences
to demographics, and each element $\Gamma_{pq}$ is i.i.d. $\mathcal{N}(0,1)$.
The i.i.d. shocks $\boldsymbol{\epsilon}$ and $\boldsymbol{e}$ are
both drawn from a standard normal distribution.\footnote{Note that the error variance assumption is consequential. As the variance
becomes large, demographics begin to only weakly explain initial preferences,
mitigating the efficacy of using them to solve the cold-start problem.
A separate set of simulations finds that the improvement of CFB-A
over CFB increases as ratio $\frac{|e|}{|e|+|D\Gamma|}$ becomes smaller.} The demographic matrix, $D\in\mathbb{R}^{I\times P}$ is constructed
by creating a total number of $P=50$ user demographic variables.
These demographic variables correspond to indicators for gender, income
level, and location. Gender is specified to be a binary variable drawn
from a Bernoulli $(p=0.5)$ distribution. In addition, income level
for a user is randomly drawn from across one of five categorical levels
with equal probability, while location is randomly drawn from across
one of 44 categorical locales with equal probability. The attribute
matrix, $\widetilde{A}=\left[\boldsymbol{1}\quad A\right]\in\mathbb{R}^{J\times(Q+1)}$
is constructed by creating $Q=300$ categories and randomly assigning
items with equal probability to one of those categories or some alternative
as reflected by an intercept vector.\footnote{Note that this linear simulation creates categorical attribute and
demographic variables. As a robustness check, we alternatively create
continuous attribute and demographic variables. All simulation findings
remain substantively similar, implying our results are robust to this
categorical variable construction.}

In the \emph{non-linear setting}, the latent spaces and utilities
are specified as

\begin{equation}
D=UW^{T},\quad A=V\Psi^{T},\label{eq:  D and A}
\end{equation}
\begin{equation}
Utility=UV^{T}+\boldsymbol{\boldsymbol{\boldsymbol{\varepsilon}}},
\end{equation}
 where $\left\{ U\in\mathbb{R}^{I\times K},V\in\mathbb{R}^{J\times K},W\in\mathbb{R}^{P\times K},\Psi\in\mathbb{R}^{Q\times K}\right\} $
are latent spaces that determine demographics, attributes, and utilities.
The number of latent dimensions in the factor space is set to $K=5$.
To be consistent with our CFB-A model, all elements in $\left\{ U,V,W,\Psi\right\} $
and the random shock $\boldsymbol{\boldsymbol{\varepsilon}}$ are
independently drawn from standard normal distributions. As in the
linear model, the number of demographic variables is set to $P=50$
and the number of attributes is set to $Q=300$.

After generating the two synthetic datasets, 200 of the 1000 users
from each dataset are randomly selected as ``new users'' to compare
the CFB-A against the benchmark models outlined in Section \ref{subsec:Benchmark-Approaches},
and the remaining 800 users from each dataset are regarded as the
training set (i.e., existing users). Similar to the MovieLens application,
the ``existing users'' are used to tune the hyper-parameters for
each method, including the exploration rate $\alpha$ for methods
with a bandit stage, $K$ and $\Sigma$ for CFB-A and Active Learning,
$K$ and $\sigma^{2}$ for CFB, and item popularity for the Popularity
policy. The total number of periods (rating occasions) over which
to generate recommendations is set to $T=15$. Model performance is
determined by computing the average cumulative utilities (CAR defined
in Equation (\ref{eq: CAR})) across the 200 new users. A higher value
of this metric indicates a better recommendation.

\subsubsection{Results}

\begin{figure}[h]
\caption{\label{fig:Dynamics-of-Model}Average Cumulative Utilities of Recommended
Items by Model and Period}

\begin{centering}
\begin{tabular}{cc}
\includegraphics[scale=0.5]{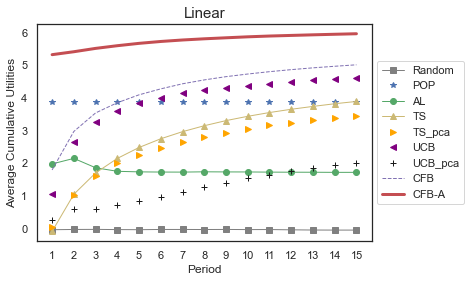} & \includegraphics[scale=0.5]{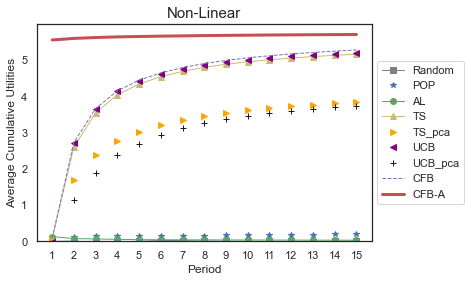}\tabularnewline
(a) & (b)\tabularnewline
\end{tabular}
\par\end{centering}
{\footnotesize{}Notes: A period represents a choice occasion, and
average cumulative utilities represent the average utilities of recommended
alternatives over periods and users. Panel (a) corresponds to the
linear setting and Panel (b) corresponds to the non-linear setting.}{\footnotesize\par}
\end{figure}

Table \ref{tab:Representative-Model-Performance} reports model performance
in three representative periods (i.e., the initial, middle, and final
periods in the simulated data) and Figure \ref{fig:Dynamics-of-Model}
shows the performance by period and simulated data set.
\begin{table}[h]
\caption{\label{tab:Representative-Model-Performance}Average Cumulative Utilities
of Recommended Items by Model and Setting}

\vspace{0.2cm}

\noindent \begin{centering}
{\scriptsize{}}%
\begin{tabular}{>{\raggedright}m{1.3cm}>{\raggedright}m{1.2cm}>{\centering}m{1.2cm}>{\centering}m{1.2cm}>{\centering}m{1.3cm}>{\centering}m{1.7cm}>{\centering}p{0.7cm}>{\centering}p{0.7cm}>{\centering}p{0.7cm}>{\centering}p{0.7cm}>{\centering}p{0.7cm}>{\centering}p{0.7cm}}
\hline 
\centering{}{\scriptsize{}Technology} & \centering{}{\scriptsize{}Method} & {\scriptsize{}Cold Start} & \multicolumn{2}{c}{{\scriptsize{}Data Reduction}} & {\scriptsize{}Test and Learn} & \multicolumn{6}{c}{{\scriptsize{}Utility Type}}\tabularnewline
\cline{3-12} 
\centering{} & \centering{} &  & {\scriptsize{}Many Users and Items} & {\scriptsize{}Many Demographics and Attributes} &  & \multicolumn{3}{>{\centering}p{1.4cm}}{{\scriptsize{}Linear}} & \multicolumn{3}{>{\centering}p{1.4cm}}{{\scriptsize{}Non-Linear}}\tabularnewline
\hline 
\centering{} & \centering{} &  &  &  &  & {\scriptsize{}T=1} & {\scriptsize{}T=8} & {\scriptsize{}T=15} & {\scriptsize{}T=1} & {\scriptsize{}T=8} & {\scriptsize{}T=15}\tabularnewline
\multirow{2}{1.3cm}{\centering{}{\scriptsize{}Null}} & \centering{}{\scriptsize{}Random} &  &  &  &  & {\scriptsize{}-0.04} & {\scriptsize{}-0.04} & {\scriptsize{}-0.06} & {\scriptsize{}0} & {\scriptsize{}0.01} & {\scriptsize{}0}\tabularnewline\arrayrulecolor{white!30}\tabularnewline
 & \centering{}{\scriptsize{}Popularity} &  &  &  &  & {\scriptsize{}3.87} & {\scriptsize{}3.87} & {\scriptsize{}3.87} & {\scriptsize{}0.10} & {\scriptsize{}0.16} & {\scriptsize{}0.20}\tabularnewline\arrayrulecolor{gray!30}\tabularnewline
\hline 
\centering{}{\scriptsize{}CFA} & \centering{}{\scriptsize{}Active Learning} & {\scriptsize{}$\surd$} & {\scriptsize{}$\surd$} & {\scriptsize{}$\surd$} &  & {\scriptsize{}1.97} & {\scriptsize{}1.73} & {\scriptsize{}1.71} & {\scriptsize{}0.09} & {\scriptsize{}0.05} & {\scriptsize{}0.05}\tabularnewline\arrayrulecolor{gray!30}\tabularnewline
\hline 
\multirow{4}{1.3cm}{\centering{}{\scriptsize{}BA}} & \centering{}{\scriptsize{}TS} &  &  &  & {\scriptsize{}$\surd$} & {\scriptsize{}-0.10} & {\scriptsize{}3.14} & {\scriptsize{}3.88} & {\scriptsize{}0.06} & {\scriptsize{}4.80} & {\scriptsize{}5.16}\tabularnewline\arrayrulecolor{white!30}\tabularnewline
 & \centering{}{\scriptsize{}UCB} &  &  &  & {\scriptsize{}$\surd$} & {\scriptsize{}1.05} & {\scriptsize{}4.22} & {\scriptsize{}4.61} & {\scriptsize{}0} & {\scriptsize{}4.85} & {\scriptsize{}5.20}\tabularnewline\arrayrulecolor{white!30}\tabularnewline
 & \centering{}{\scriptsize{}TS\_pca} &  &  & {\scriptsize{}$\surd$} & {\scriptsize{}$\surd$} & {\scriptsize{}0.02} & {\scriptsize{}2.80} & {\scriptsize{}3.44} & {\scriptsize{}0.04} & {\scriptsize{}3.45} & {\scriptsize{}3.83}\tabularnewline\arrayrulecolor{white!30}\tabularnewline
 & \centering{}{\scriptsize{}UCB\_pca} &  &  & {\scriptsize{}$\surd$} & {\scriptsize{}$\surd$} & {\scriptsize{}0.24} & {\scriptsize{}1.26} & {\scriptsize{}2.01} & {\scriptsize{}-0.16} & {\scriptsize{}3.26} & {\scriptsize{}3.74}\tabularnewline\arrayrulecolor{white!30}\tabularnewline
\hline 
\centering{}{\scriptsize{}CFB} & \centering{}{\scriptsize{}CFB} &  & {\scriptsize{}$\surd$} &  & {\scriptsize{}$\surd$} & {\scriptsize{}1.79} & {\scriptsize{}4.54} & {\scriptsize{}5.00} & {\scriptsize{}0.07} & {\scriptsize{}4.90} & {\scriptsize{}5.28}\tabularnewline\arrayrulecolor{gray!30}\tabularnewline
\hline 
\centering{}{\scriptsize{}CFB-A} & \centering{}{\scriptsize{}CFB-A} & {\scriptsize{}$\surd$} & {\scriptsize{}$\surd$} & {\scriptsize{}$\surd$} & {\scriptsize{}$\surd$} & \textbf{\scriptsize{}5.32} & \textbf{\scriptsize{}5.80} & \textbf{\scriptsize{}5.96} & \textbf{\scriptsize{}5.55} & \textbf{\scriptsize{}5.67} & \textbf{\scriptsize{}5.70}\tabularnewline\arrayrulecolor{black}\tabularnewline
\hline 
\end{tabular}{\scriptsize\par}
\par\end{centering}
\noindent \begin{centering}
{\scriptsize{}}{\scriptsize\par}
\par\end{centering}
{\footnotesize{}Notes: This table reports the average cumulative utilities
of items recommended to users by period, utility type, and method.
A higher number reflects better recommendations. For example, Thompson
sampling generates an average cumulative utility of 3.14 over 8 periods
when utilities are generated using the linear method. Utilities generated
by the linear simulation range between $\left[-13.64,15.45\right]$
with mean -0.06 and standard deviation 2.10; utilities generated by
the non-linear setting range between $\left[-16.71,14.64\right]$
with mean 0 and standard deviation 2.42; a more detailed distributions
of utilities in the linear and nonlinear settings can be found in
Web Appendix \ref{sec:Synthetic-Data-Distribution}.}{\footnotesize\par}
\end{table}
 
Four key findings emerge. First, the CFB-A yields the most favorable
customer feedback for the recommended items in both the linear and
non-linear cases. This suggests the performance enhancement of the
CFB-A is robust across various data generating mechanisms. Second,
algorithms with the bandit component evidence improved performance
over time as a result of learning from feedback. In contrast, approaches
without a bandit component (random, popularity, and active learning)
do not improve customer responses to their recommendations over time
as much. Third, in the linear case, the CFB-A outperforms the second-best
approach (POP) in the first period by 37\%. Likewise, in the non-linear
case, the CFB-A outperforms the second-best approaches (POP) by 5450\%.
These findings suggest that the CFB-A is effective at addressing cold
start. The effect is stronger in the nonlinear case because the underlying
data structure aligns with the latent factor model applied by CFB-A,
but the finding that performance is enhanced in both cases suggests
the robustness of the approach across different data. Fourth, when
the data generating process (i.e., linear) differs from the CFB-A
model specifications, the initial priors about consumer match are
less informative, and therefore the learning stage of CFB-A becomes
more important as evidenced by the rapid improvement of CFB-A in initial
periods.\footnote{The linear simulation employs user preferences on item attributes,
$\boldsymbol{\beta}$, that are highly informed by user demographics.
As such, even models without test and learn (POP and Active Learning)
perform well because users' initial preferences are captured well
by their demographics. To assess how less informative attributes alter
the simulation findings, $\boldsymbol{\beta}$ is specified to be
independent of demographics. In this case, algorithms with test and
learn perform well (CFB-A performs the best), while POP and Active
Learning evidence substantially degraded performance, similar to Random.}

\section{\label{sec:Live-Experiment}Live Experiment}

\noindent Using the MovieLens and simulated data, Section \ref{sec:Simulations}
demonstrates that the CFB-A outperforms competing models. While the
simulated data are live (i.e., recommendations are updated dynamically)
and the MovieLens data are real (i.e., a result of consumers' rating
behaviors), it is desirable to test the CFB-A against competing models
in a context that is both live and real. In addition, the CFB-A embeds
three key innovations: data reduction, cold-start, and test and learn.
Decomposing the relative contribution of each component in terms of
improving recommendations would give a sense of their relative import.
An experiment accomplishes these twin aims.

\subsection{Experimental Context}

The Open Science Online Grocery (OSOG) platform,\footnote{\url{https://openscience-onlinegrocery.com/}}
a free research tool established by \citet{howeopen}, is used to
conduct an experiment to test the relative performance of the CFB-A
in a live setting.\footnote{Here ``live'' means that recommendations are updated every session,
factoring in user response in previous periods, instead of recommendations
being updated in real-time within a session.} The OSOG context affords the engineering infrastructure to implement
a recommendation system and allows for many new items and users with
many demographics and attributes, making it an appropriate model testing
context.

\begin{figure}[h]
\caption{\label{fig:OSOG-participant-interface}OSOG Product Listing Page}

\noindent \centering{}\includegraphics[scale=0.25]{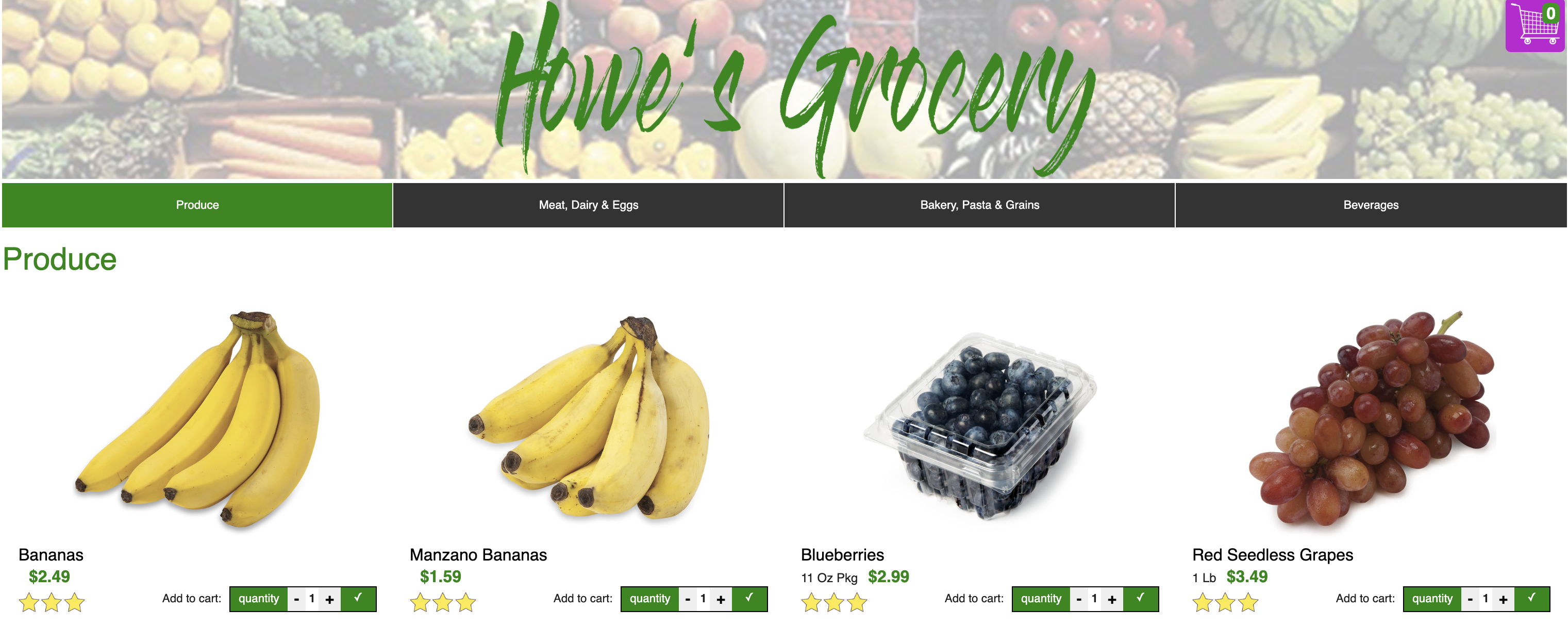}
\end{figure}

The participant interface mimics e-commerce websites (e.g., Amazon
Fresh, Instacart, Walmart), and products on the platform are actual
products purveyed at a major grocer in 2018. Upon entering the experimental
site via a directed link, participants are presented the category
\textit{home product listing page} (called category \textit{home page}
for short) for the produce category, where they can browse the recommended
products in that page, visit another category's home page, or scroll
forward to the second (and beyond) product listing page for the produce
category. Figure \ref{fig:OSOG-participant-interface} portrays a
subset of products on the produce category's home page at the online
store, as well as the banner menu that can be used to navigate to
an alternative category's home page. Each \textit{product listing
page}, including the category home page and all subsequent product
listing pages, includes up to 100 products. The challenge for a recommender
system in this context is to determine the order of products presented
to a participant on the product listing pages for each category.

\noindent 
\begin{figure}[H]
\caption{\label{fig:OSOG-Product-Detail}OSOG Product Detail Page}

\noindent \centering{}\includegraphics[scale=0.4]{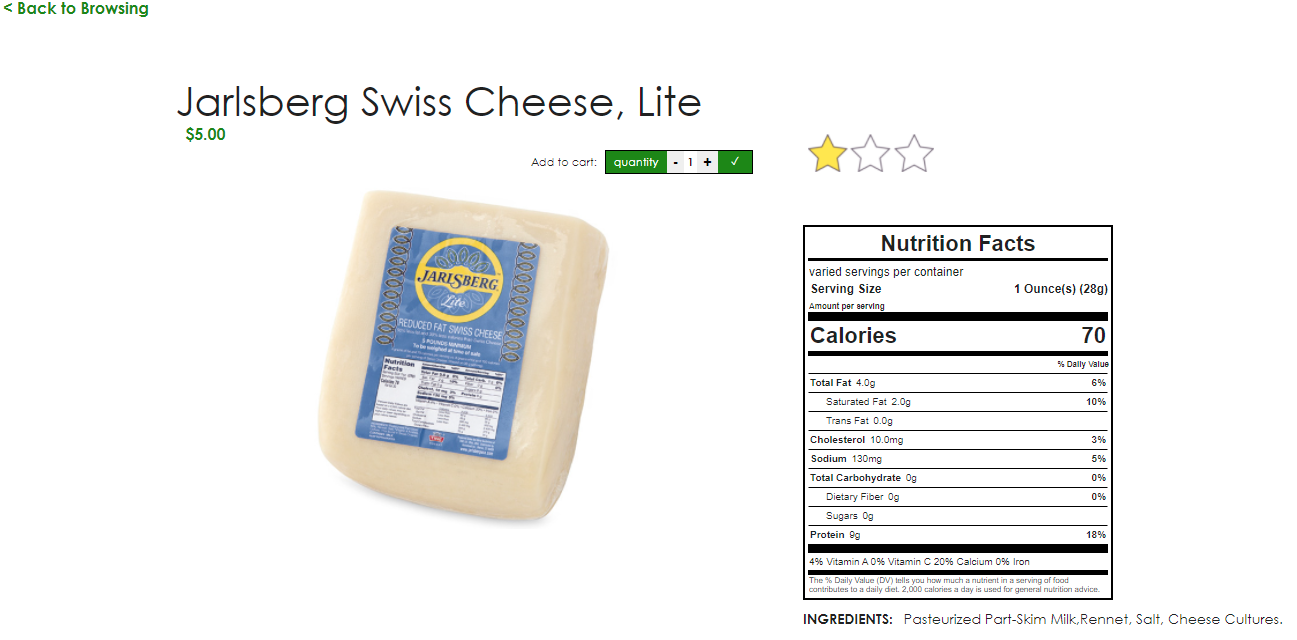}
\end{figure}

Upon clicking an item on the product listing page, a participant is
directed to the \textit{product detail page}, which displays detailed
attribute-level information about the product. An example of the product
detail page is shown in Figure \ref{fig:OSOG-Product-Detail} for
Jarlsberg Swiss Cheese, Lite. 

\begin{figure}[H]
\caption{\label{fig:OSOG-Checkout-Page}OSOG Checkout Page}

\noindent \centering{}\includegraphics[scale=0.4]{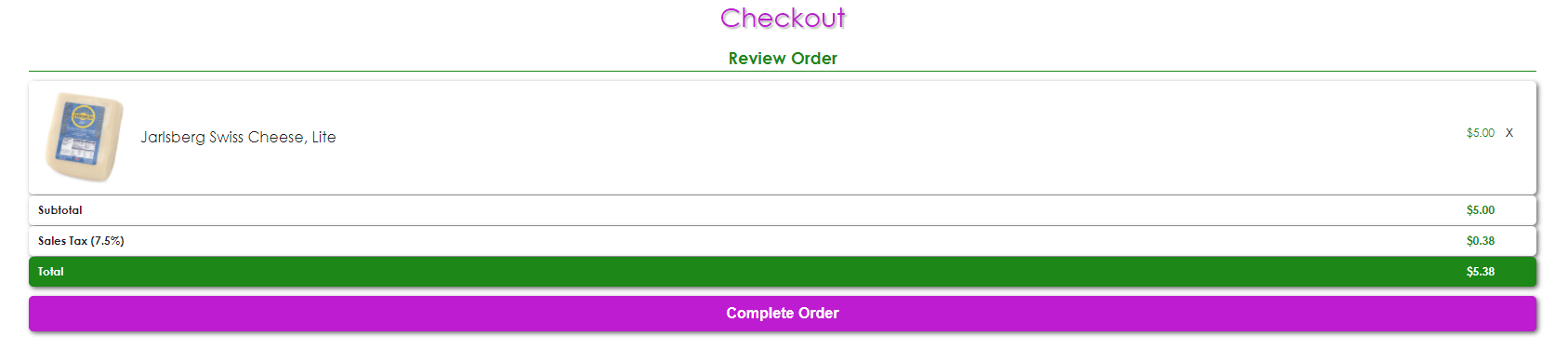}
\end{figure}

On either the product detail page or product listing page, a participant
can add a product to the shopping cart by clicking the product. After
adding products to the cart for purchase, the participant can self
direct to the \textit{checkout page} as shown in Figure \ref{fig:OSOG-Checkout-Page},
or continue shopping. Participants end the shopping session by clicking
on ``Complete Order''.

\noindent 
\noindent 
The OSOG platform includes 3,542 products across 4 categories: i)
produce; ii) meat, dairy, and eggs, iii) bakery, pasta and grains,
and iv) beverages. Each product is characterized by multiple attributes,
such as price, size, ingredients, nutrition tags (e.g., calories,
organic, health starpoints measuring the healthiness of the product,
etc.), and allergens. The final attribute space has 22-40 dimensions
which vary across categories, as detailed in Web Appendix \ref{subsec:Item-attributes}.

\subsection{Experimental Design and Task}

\subsubsection{Experimental Design\label{subsec:Design}}

To assess the relative performance of the CFB-A, we compare its performance
to three baseline models: the CFB (CF and bandit, no attribute), CFA
(CF and attribute, no bandit), and UCB (bandit, no data reduction).
This yields four experimental cells. Contrasts between these cells
help to highlight the relative importance of the CFB-A innovations:
i) the contrast between CFB-A and CFB highlights the role of attributes
in improving the cold-start recommendations, ii) the contrast between
CFB-A and CFA focuses upon the role of learning in improving recommendations,
and iii) the contrast between CFB-A and UCB informs the relative improvement
due to the factor space reduction. Notably, these contrasts enable
us to decompose the relative benefit of cold start, learning, and
dimension reduction to ascertain which is more important for making
recommendations in the experimental context.

The experiment consists of two phases, or waves: training and testing.
Each phase uses a different set of participants. The one-week training
stage, conducted in December of 2021 has two objectives: First, it
provides data to ``tune'' the population-level hyper-parameters
in the recommendation systems used in the testing phase, conducted
in January and February of 2022. For CFB-A these hyper-parameters
include the regularization terms $\Sigma=\left\{ \sigma^{2},\sigma_{d}^{2},\sigma_{a}^{2},\lambda_{u},\lambda_{v},\lambda_{w},\lambda_{\psi}\right\} $,
the dimension of latent user and product spaces, $K$, and the UCB
scale parameter $\alpha$ balancing exploration and exploitation.
The hyper-parameters for the other models and the tuning process are
detailed in Web Appendix \ref{sec:Hyper-parameter-Tuning}.\footnote{It is assumed that the population-level hyper-parameters in the training
and testing samples are the same across the two populations. While
the assumption is impossible to test, one can conduct an equality
of means test on the reported demographics; results presented in Web
Appendix \ref{subsec:Demographic-Comparison-Between} indicate that
the two samples are similar on these observables. Hence, it is reasonable
to assume the same population-level hyper-parameters in the training
and testing samples to the extent that hyper-parameters are functions
of demographics.} In addition, the training stage yields priors for the testing stage,
which alleviates the cold-start problem by transforming uninformative
priors on latent matrices $W$ and $\Psi$ (which project user or
product locations in the latent preference space onto user or product
attributes) to informative priors through Equations (\ref{W_posterior})
and (\ref{eq:Psi_posterior}).

\subsubsection{Task}

At the beginning of both phases of the experiment, participants complete
a survey to elicit their demographics and food preferences (e.g.,
age, gender, ethnicity, education, household income, household size,
state, religion, height, weight, and dietary restrictions). The complete
list of collected variables is detailed in Web Appendix \ref{subsec:Demographic-survey-questions}
and comprises 119 dimensions. Thus, the demographic data include 119
variables and the product attribute data include 22-40 variables,
which should be sufficient to illustrate the improvement of the CFB-A
over the CFB. This is likely a conservative test of our model, which
can scale readily to larger numbers of attributes and demographics.

After the survey, participants in the \emph{training} phase completed
a single shopping task with a budget of \$75 to ensure that they do
not select an inordinately large number of products.\footnote{This budget constraint is non-binding for 78\% of the shopping tasks
collected in the experiment.} The order of products presented to each participant in the training
phase was randomized within each category. Participants made product
selection decisions by adding products to cart, and were free to revisit
prior detail and listing pages before the checkout decision. To avoid
potential inventory effect suppressing choices of preferred products,
participants were requested to imagine they had no grocery at home
when completing the shopping task. The final participant decision
was to end the shopping visit, at which point the experimental session
ends.

Participants in the \emph{testing} phase were asked to revisit the
experiment one- , two- , and three-weeks after the demographic survey
in the first week, and complete a shopping task with a budget of \$75
in each visit. As with the training phase, participants' clicks on
products and purchases are recorded. Unlike the training phase where
product order is randomized, the order of products presented to participants
in the second week of the testing phase (i.e., one week after the
demographic survey was administered) was determined by the algorithm
for the given experimental cell based on reported demographics and
participant purchases in the training phase. The order of products
presented in the third and the fourth weeks of the testing phase further
considers participant purchases in the preceding weeks of the testing
phase.

\subsection{\label{subsec:Subjects}Subjects}

Two waves of participants were recruited via CloudResearch, corresponding
to the two experimental phases: training and testing. In total, 531
subjects participated in the training phase and 847 subjects participated
in the testing phase. Participants in the testing phase were randomly
assigned to one of the four experimental cells. The participants were
largely representative of the United States with respect to age, gender,
and ethnicity on Mturk, as directed by the CloudResearch recruitment
settings.\footnote{The average age of all participants is 39. As for gender distribution,
55\% are female, 44\% are male, and 1\% are third gender or non-binary.
As for ethnicity distribution, 81\% are White, 10\% are Black or African
American, 2\% are American Indian or Alaska Native, 10\% are Asian,
0.3\% are Native Hawaiian or Pacific Islander, and 2\% are others
including Hispanic, Latinx, and Middle Eastern. Note that participants
can choose more than one ethnicity option.} Participants were compensated \$0.12/minute and those in the testing
phase received a bonus of \$1 for completing all four sessions. On
average, participants spent 16.7 minutes completing a survey with
the shopping task. In addition, attention checks were used to ensure
data quality, and 1.1\% of participants failing to pass these checks
did not receive compensation (details on these checks are provided
in Web Appendix \ref{subsec:Attention-Check-Questions}).

\subsection{\label{subsec:metrics}Performance Metric\label{subsec:Performance-Metrics}}

Model performance is evaluated via homepage (the category home product
listing page) purchase rates (HPR) of each category, which captures
participants' tendency to purchase the 100 most highly recommended
products. HPR is an ideal performance metric for several reasons.
First, all participants browsed the home page, whereas many fewer
of them (from 45\% to 73\% depending on the category) browsed any
listing page beyond the home page. Thus the HPR metric is not affected
by products a participant does not typically see. Second, by focusing
on a specific set size of products, the metric is comparable in interpretation
across categories relative to one that uses different set sizes. Third,
the homepage performance is also of practical business interest because
the homepage sales rates are a KPI of interest and because including
less relevant products on the home page can induce customer churn.

The HPR for category $c$ in period $t$ is computed as follows:

\begin{eqnarray}
HPR_{t}^{c} & = & \frac{1}{I\times J}\sum_{i=1}^{I}\sum_{j=1}^{J}Purchases_{ijt}^{c},\label{eq:HPR}
\end{eqnarray}
where $i$ denotes participant and $j$ denotes product, $J=100$
is the total number of products on the home page. $Purchases_{ijt}^{c}$
is an indicator variable which takes the value of 1 if participant
$i$ purchased product $j$ in period $t$. The period-level purchase
rate can be extended to a cumulative (over all the shopping periods
thus far) purchase rate for category $c$ in period $t$ as follows:

\begin{equation}
cHPR_{t}^{c}=\frac{1}{t}\sum_{\tau=1}^{t}HPR_{\tau}^{c}
\end{equation}
The category-level HPR and cumulative HPR can be aggregated to a site-level
HPR measure by averaging or summing across the four categories.\footnote{A robustness check considers the NDCG metric (\citealp{jarvelin2002cumulated}),
which accounts for product rank and favors recommendation engines
that place chosen products earlier in the list. Findings are qualitatively
robust to this choice of alternative performance metric.}

\subsection{\label{subsec:Results-1}Results}

Using the HPR metric described in Section \ref{subsec:metrics}, this
section compares model performance across the experimental cells (corresponding
to the four methods outlined in Section \ref{subsec:Design}), and
decomposes the relative contributions of data reduction, cold-start,
and test and learn. We first report results aggregated across all
categories and then detail the category-specific results to assess
conditions under which any one of the three components (data reduction,
cold-start, and test and learn) contributes most in the CFB-A.

\subsubsection{\label{subsec:Aggregate-level-Results}Aggregate Site-Level Results}

\begin{figure}[h]
\caption{\label{fig:Homepage-Cumulative-Purchase}cHPR Combined Over Categories
By Period}

\begin{centering}
\includegraphics[scale=0.5]{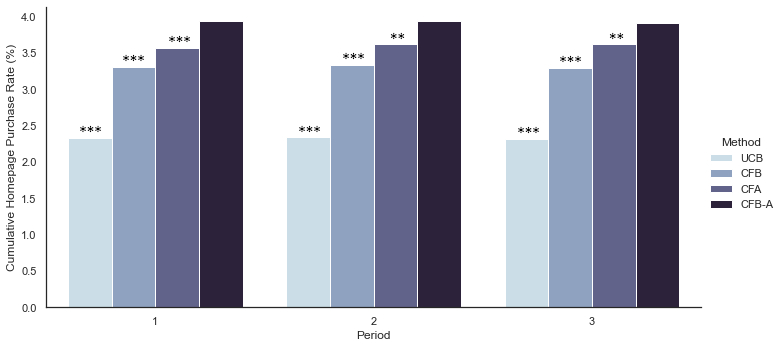}
\par\end{centering}
\raggedright{}\textit{\emph{\footnotesize{}Notes: Period represents
week, or choice occasion, in the experiment. Each bar represents the
cHPR for a given method and period. For example, the CFB-A yields
a cHPR of 3.91\% in period 3 as evidenced by the last bar. Independent
two-sample t-tests compare the cHPRs under CFB-A and each benchmark
method, and the t-values for the contrasts between methods are reported
in Web Appendix \ref{subsec:Full-results}}}\textit{\footnotesize{}.}\textit{\emph{\footnotesize{}
Significance levels are denoted as follows:}}\emph{\footnotesize{}
}{\footnotesize{}$**p<.05,***p<.01$}\textit{\footnotesize{}.}{\footnotesize\par}
\end{figure}

Figure \ref{fig:Homepage-Cumulative-Purchase} depicts the cumulative
HPR (cHPR), for each method and period across all categories in the
experiment. Asterisks indicate the statistical significance of the
difference between the CFB-A and the respective benchmark model as
determined by independent two-sample t-tests. The cHPRs of CFB-A,
CFA, CFB, and UCB at the conclusion of the experiment (period 3) are
respectively 3.91\%, 3.63\%, 3.29\%, and 2.32\%. The CFB-A significantly
outperforms the three benchmark models. In percentage terms, the CFB-A
outperforms the UCB by 69\%, (contribution due to data reduction),
the CFB by 19\% (contribution due to cold start), and the CFA by 8\%
(contribution due to test and learn). Compared to the worst performing
model, the CFB-A nearly doubles the cHPR. The larger improvement of
CFB-A over CFB relative to CFA implies that A (cold start) matters
more than B (test and learn), presumably due to informative priors
(that is, demographics and attributes can predict choices). However,
the comparative advantages of A and B can be context-dependent, which
motivates us to compare the performance of these methods for each
category.

The increased HPR translates to increased homepage retailer sales.
In particular, the CFB-A generates an average of \$52.39 in homepage
sales per subject, while the CFB, CFA, and UCB generate an average
homepage sales of \$50.14, \$50.79, and \$32.87 respectively. According
to the one-sided independent two-sample t-test, the increase in CFB-A
sales relative to the other approaches are all statistically significant
or marginally significant (with t-values as follows: CFB-A vs CFB
($t=1.47$), CFB-A vs CFA ($t=1.74$), and CFB-A vs UCB ($t=17.00$)).

To obtain a deeper insight into how the CFB-A recommendation algorithm
affects total consumer demand at the grocery site, we compare the
total number of products purchased and total revenue across all sessions
and pages within the session (i.e., not just the home pages). The
number of products sold under CFB-A is significantly or marginally
significantly higher than under CFB and UCB (the t-values for the
independent two-sample t-tests are as follows: CFB-A vs CFB (22.7
vs 19.7, $t=4.31$), CFB-A vs CFA (22.7 vs 22.0, $t=0.93$), and CFB-A
vs UCB (22.7 vs 21.6, $t=1.39$)). However, there is little difference
in total site revenue across the four benchmark methods (the t-values
for the independent two-sample t-tests as follows: CFB-A vs CFB (\$64.72
vs \$64.92, $t=\text{\textendash}0.13$), CFB-A vs CFA (\$64.72 vs
\$64.29, $t=0.26$), and CFB-A vs UCB (\$64.72 vs \$64.22, $t=0.31$)).
Given that more products are sold under CFB-A, but there is no difference
in revenue, it follows that lower priced products, on average are
sold under CFB-A. The average price of products on the home page under
CFB-A is significantly lower than under CFB and CFA (the t-values
for the independent two-sample t-tests are as follows: CFB-A vs CFB
(\$3.61 vs \$3.82, $t=-22.58$), CFB-A vs CFA (\$3.61 vs \$3.72, $t=-11.24$),
and CFB-A vs UCB (\$3.61 vs \$3.62, $t=-0.36$)). The finding that
the CFB-A serves the lowest prices suggests that it tends to capture
user preferences for lower prices better than the other algorithms.

In sum, the CFB-A yields two key implications for the experimental
site. First, as implied by the increase in the cHPR performance metric,
which focuses on purchases within the first 100 products, the CFB-A
recommends preferred products earlier in the ordered set of available
products relative to the other extant methods (i.e., the first 100
products contain more consumer matches). This result is also consistent
with the finding that the CFB-A outperforms other methods under the
NDCG metric (\citealp{jarvelin2002cumulated}), which accounts for
product rank and favors recommendation engines that place chosen products
earlier in the list. Second, because more products are sold in the
experimental grocery site while the grocer's revenue remains constant,
the CFB-A tends to recommend less expensive products, which could
limit the revenue implications of the recommender system in retail
settings.

We make several observations about these two implications. First,
the effect of price on revenue is a concern relatively unique to retail.
In settings without variable item prices, such as online news (which
is monetized by viewership) or movie recommendations and OTAs (where
prices or referral fees tend to be constant with clicks), sites are
unambiguously better off with higher cHPR because monetization increases
with items viewed or ordered. Second, the current experiment uses
cHPR as the performance measure in hyper-parameter tuning, and CFB-A
is likely to lead to higher revenue if renenue were used as the performance
measure. Third, even if actual site revenues were constant in practice
upon adopting the CFB-A, the finding that consumers obtained higher
utility from goods purchased indicates that the retailer is better
off because satisfied customers are less likely to attrite.

\subsubsection{\label{subsec:Category-level-Results}Category-level Results}

\begin{figure}[h]
\caption{\label{fig:Categorical-Results}Period 3 cHPR by Category}

\begin{centering}
\textcolor{white}{\textbackslash\textbackslash\textbackslash\textbackslash}
\par\end{centering}
\begin{centering}
\includegraphics[scale=0.5]{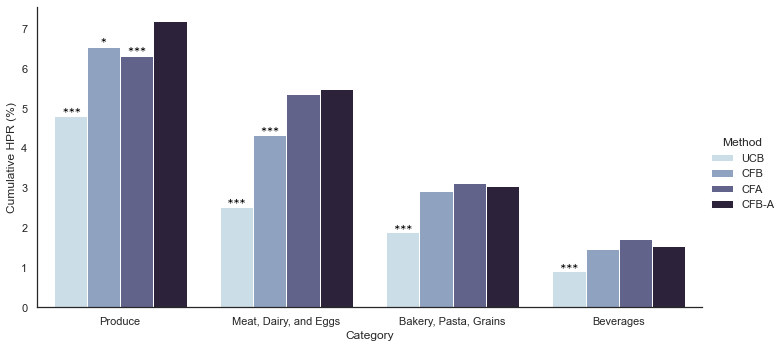}
\par\end{centering}
\raggedright{}\textit{\emph{\footnotesize{}Notes: Each bar represents
the cHPR for a given method and category over all periods in the experiment.
Specific cHPR values for each category and method are reported in
Web Appendix \ref{subsec:Full-results}. Independent two-sample t-tests
compare the cHPR under CFB-A and each benchmark method, and the t-values
for these method contrasts can be found in Web Appendix \ref{subsec:Full-results}.
Significance levels are denoted as follows: }}\emph{\footnotesize{}$*p<.10,***p<.01$}{\footnotesize{}.}{\footnotesize\par}
\end{figure}

Figure \ref{fig:Categorical-Results} portrays the cHPR of the CFB-A
and the three benchmark models at the conclusion of the experiment
(period 3) by product category. Consistent with the aggregate results,
a comparison between CFB-A and UCB indicates that data reduction yields
the largest gains in cHPR for each category. Notably, a comparison
of CFB-A and CFB indicates that cold start has the most salient role
in the category of meat, dairy, and eggs, while a comparison of CFB-A
and CFA indicates that test and learn only makes a difference in produce.

The differences of the relative importance of A (cold start) across
categories imply that the informativeness of attributes on preferences
differs across categories. The A part (cold start) is expected to
be more important than the B part (test and learn) when attributes
are perfectly informative about preferences. If attributes perfectly
predict preferences, there is nothing to learn. As attributes become
less predictive, there is, in general, increasing value in B.\footnote{When the attributes are completely uninformative about initial preference,
the value of B also degrades in the early periods. An efficient test-and-learn
experiment requires the exploration of alternatives with higher uncertainty.
When the values of all attributes are equally and highly uncertain,
test and learn has a difficult time ascertaining where to begin in
exploration.}

Assessing the role of the informativeness of attributes on preferences
requires a metric of informativeness. One such metric is available
from the one-period training data, which is based on the counterfactual
HPRs for the CFA and CF models (note there is no B component as the
training data consist of only one period). Specifically, the percentage
HPR lift in the training data for the counterfactual CFA relative
to the counterfactual CF measures how well attributes explain outcomes
(i.e., how much the attribute information matters). In other words,
when the CFA improves over the CF on the training sample, the quality
of priors embedding attributes is higher for predicting initial preferences
and making initial recommendations.

Table \ref{tab:HPR_train} reports, by category, both the informativeness
of attributes (by quality of prior) in the training data (column 2)
and the relative contribution of A to B in the testing data (columns
3 to 5). The correlation between the second column ($Gain_{A,Train}$)
and the last column ($Gain_{A}/Gain_{B}$) across the four categories
is 0.83 and there is a monotonic relation between the entries in the
two columns. Therefore, the relative importance of A to B in the testing
data increases as A becomes more informative in the training data,
as we have conjectured. In other words, as attributes become increasingly
informative of preferences, the cold-start component of the model
(A) becomes relatively more useful for recommendations than the test-and-learn
component (B) of the model.

\begin{table}[h]
{\footnotesize{}\caption{\label{tab:HPR_train}The Effect of Attribute Salience on the Relative
Contribution of A}
}{\footnotesize\par}

\vspace{0.2cm}

\begin{centering}
{\footnotesize{}}%
\begin{tabular}{l>{\centering}p{3cm}>{\centering}p{0.2cm}>{\centering}p{3cm}>{\centering}p{3cm}c}
\hline 
{\footnotesize{}Category} & {\footnotesize{}Training Data} &  & \multicolumn{3}{c}{{\footnotesize{}Testing Data}}\tabularnewline
\cline{2-2} \cline{4-6} 
 & {\footnotesize{}$Gain_{A,Train}=\frac{HPR_{CFA}-HPR_{CF}}{HPR_{CF}}$} &  & {\footnotesize{}$Gain_{A}=\frac{cHPR_{CFB-A}-cHPR_{CFB}}{cHPR_{CFB-A}}$} & {\footnotesize{}$Gain_{B}=\frac{cHPR_{CFB-A}-cHPR_{CFA}}{cHPR_{CFB-A}}$} & {\footnotesize{}$\frac{Gain_{A}}{Gain_{B}}$}\tabularnewline
\cline{1-2} \cline{4-6} 
{\footnotesize{}Meat, Dairy, and Eggs} & {\footnotesize{}469\%} &  & {\footnotesize{}21\%} & {\footnotesize{}2\%} & {\footnotesize{}10.5}\tabularnewline
{\footnotesize{}Produce} & {\footnotesize{}318\%} &  & {\footnotesize{}9\%} & {\footnotesize{}12\%} & {\footnotesize{}0.8}\tabularnewline
{\footnotesize{}Beverages} & {\footnotesize{}232\%} &  & {\footnotesize{}5\%} & {\footnotesize{}--11\%} & {\footnotesize{}--0.5}\tabularnewline
{\footnotesize{}Bakery, Pasta, and Grains} & {\footnotesize{}12\%} &  & {\footnotesize{}4\%} & {\footnotesize{}--3\%} & {\footnotesize{}--1.3}\tabularnewline
\hline 
\end{tabular}{\footnotesize\par}
\par\end{centering}
{\footnotesize{}Notes: The first column reports the category, the
second column reports the percentage increase in the HPR for the CFA
relative to CF in the training data, the third column reports the
increase in the cHPR for the CFB-A relative to the CFB (normalized
by the cHPR under CFB-A and representing the gain from A), the fourth
column reports the increase in the cHPR for the CFB-A relative to
the CFA (normalized by the cHPR under CFB-A and representing the gain
from B), and the final column reports the ratio of the gain from A
reported in the third column to the gain from B reported in the fourth
column (in other words, the last column is a measure of relative importance
of A to B in improving forecasts). Please see Table \ref{result_ctr}
in Web Appendix \ref{subsec:Full-results} for values used in the
testing data computations; for example in produce $Gain_{A}=(7.19-6.53)/7.19=9\%$.}{\footnotesize\par}
\end{table}
In summary, our results provide three main takeaways. First, CFB-A
has the best aggregate performance. Second, given that UCB has the
worst performance throughout, data reduction has the largest contribution.
Third, the relative importance of cold start and test and learn is
context-dependent: though the CFB-A tends to perform better when A
explains consumer choices, the test and learn component can perform
better when attributes are moderately informative of preferences,
because the bandit can better ascertain the dimensions along which
consumer preferences are uncertain (when A is entirely uninformative,
all error variances are high and B has a harder time assessing where
to explore).

\section{\label{sec:Discussions}Conclusion}

This paper considers recommendation systems common in many digital
contexts such as news, music, shopping, or streaming where new users
and new items are common, the space of users and items is large, and
the tags used to describe them are extensive. In such contexts, collaborative
filtering exhibits cold-start limitations as there is little history
upon which to base recommendations.

We develop a model specifically for these contexts that embeds two
key components. One component reduces the dimensionality of recommender
data, and thus enables the approach to scale to larger applications.
Specifically, this component factors the user-by-item preference matrix,
the user-by-demographic matrix, and the item-by-attribute matrix.
One key benefit of this factorization is the ability to address the
\emph{cold-start} problem by recommending items preferred by others
who are similar in their demographic composition (denoted as the CFA).
A second component involves a bandit approach to \emph{test and learn}
them (denoted as the CFB), and this becomes more relevant when demographics
are not sufficiently informative about user preferences. A key advance
is to combine the CFA and CFB into the CFB-A. One notable synergy
arising from this combination is that the bandit needs only search
for items in a much reduced space. Hence it scales well. A second
notable synergy is that the recommendation approach becomes more robust.
When observables such as demographics are predictive of preferences,
our model uses this information to make better predictions. When unobservables
are less informative, our model uses bandit exploration on past choices
to improve recommendations. In this sense, combining the two components
leads to a more robust recommendation approach.

As a result, the combination of the two components (CFA and CFB) into
a CFB-A yields material gains in the quality of consumer recommendations
in the context of collaborative filtering (CF). The model is benchmarked
against a wide array of extant alternatives that address scale, cold-start,
or bandit learning but not all three. We do so using standard benchmarking
data from MovieLens, a simulation, and a grocery experiment. In sum,
we benchmark over a large set of models (that contain components of
our innovations) and datasets (static, simulated, and live).

In each case, the CFB-A improves upon the alternative models, and
the degree of improvement varies across models and data sets (compared
to the next best approach for each data set, CFB-A improves by about
1\% in the MovieLens data, 19\% in the synthetic data, and 8\% in
the experimental data). In general, when attributes are informative
about preferences and feedback scales are informative, our model works
better than those that omit attributes. When attributes are less informative,
our model works better than those that omit the bandit. Moreover,
our model converges more quickly to the best option than other approaches,
because the bandit can learn user preferences more quickly in the
reduced user (item)-attribute spaces (in one case, user preferences
are learned in one to two recommendations whereas other approaches
require closer to ten). The empirical performance of the CFB-A suggests
future theoretical work on the regret bounds and convergence rate
of this method to formalize the mechanisms underpinning its advantages
in combining CF, B, and A.

The utility of our analysis is of straightforward benefit to practitioners.
Those facing issues of scale and cold start in recommender contexts
stand to gain directly from our algorithm. Our research may have policy
implications for an increasingly privacy-oriented environment. By
balancing learning from first-party data with demographics from first-
and third-party data, our approach is relatively robust to changes
in the data environment. It is even possible for third-party firms
to compute user factor ``scores'' to assure privacy on given covariates,
as the factors scores by themselves are not sufficient to impute demographics
that underpin them (as scores are a linear combination of demographics).
In other words, scores are more privacy compliant than demographics
and would still enable better cold-start recommendations.

Beyond these immediate applications, a set of broader questions of
relevance to our research exist. First, though the algorithm scales
well, it requires updating based on past choices. Given the engineering
challenges of updating item positions in a real-time environment,
assessing the optimal level of updating (such as real time, minute,
hour, or daily frequency) or how to batch recommendations is of interest.
Second, hybrid recommendation methods have recently combined state-of-the-art
representation learning techniques (e.g., autoencoders) with traditional
heuristics (e.g., matrix factorization, logistic regression) (\citealp{jannach2020deep,strub2016hybrid,dong2020hybrid,geng2022representation,zhou2023spoiled}),
suggesting that CFB-A could be improved by combining matrix factorization
with autoencoders. In addition, other inference techniques, such as
variational inference (\citealp{Ansari2018}), can be used as an alternative
to MCMC estimation for the Bayesian component of CFB-A. Related and
third, ours is a list recommender system. It is relatively straightforward
to extend the approach to menu recommendations by modeling user preference
for one item to be a function of all recommended items. This extension
would be an interesting new direction which captures complementarity
and substitutability between items (e.g., \citealp{kumar2020scalable}).
Fourth, while CFB-A improves the cumulative direct rewards such as
clicks, purchases, and ratings, it would be of interest to assess
how recommendation approaches affect indirect outcomes such as consumer
loyalty in the long run. Fifth, as the CFB-A combines both aggregate
preference information and user choices, a pertinent question is whether
it can mitigate tendencies to be polarizing or homogenizing (e.g.,
\citealp{hosanagar2014will,lee2019recommender,berman2020curation,su2016effect,zhang2022does})
and how it can be used to enhance diversity (e.g., \citealp{song2019and,zhou2023spoiled}).
Finally, in any recommendation system, algorithmic bias (e.g., \citealp{cowgill2019economics})
is a concern. Owing to its flexibility in combining information from
own and others' choices as well as item attributes and consumer characteristics,
the potential for the CFB-A to ameliorate algorithmic bias is a worthwhile
future direction. It is promising that the test-and-learn component
can enhance recommendation performance in the absence of third-party
data used in cold start (such as racial demographics), making the
CFB-A more robust in capturing preferences. We hope others will build
on the ideas in this paper to address these and other timely and relevant
issues in the context of recommendation systems.

\section*{Funding and Competing Interests}

All authors certify that they have no affiliations with or involvement
in any organization or entity with any financial interest or non-financial
interest in the subject matter or materials discussed in this manuscript.
The authors have no funding to report.

\pagebreak\bibliographystyle{mktsci}
\addcontentsline{toc}{section}{\refname}\bibliography{CFB}

\begin{thebibliography}{93}
\expandafter\ifx\csname natexlab\endcsname\relax\def\natexlab#1{#1}\fi
\expandafter\ifx\csname url\endcsname\relax
  \def\url#1{{\tt #1}}\fi
\expandafter\ifx\csname urlprefix\endcsname\relax\def\urlprefix{URL }\fi
\expandafter\ifx\csname urlstyle\endcsname\relax
  \expandafter\ifx\csname doi\endcsname\relax
  \def\doi#1{doi:\discretionary{}{}{}#1}\fi \else
  \expandafter\ifx\csname doi\endcsname\relax
  \def\doi{doi:\discretionary{}{}{}\begingroup \urlstyle{rm}\Url}\fi \fi

\bibitem[{Abernethy et~al.(2009)Abernethy, Bach, Evgeniou, and
  Vert}]{abernethy2009new}
Abernethy, Jacob, Francis Bach, Theodoros Evgeniou, Jean-Philippe Vert. 2009.
\newblock A new approach to collaborative filtering: Operator estimation with
  spectral regularization.
\newblock {\it Journal of Machine Learning Research\/} {\bf 10}(3).

\bibitem[{Agarwal et~al.(2009)Agarwal, Chen, Elango, Motgi, Park, Ramakrishnan,
  Roy, and Zachariah}]{Agarwal2009a}
Agarwal, Deepak, Bee~Chung Chen, Pradheep Elango, Nitin Motgi, Seung~Taek Park,
  Raghu Ramakrishnan, Scott Roy, Joe Zachariah. 2009.
\newblock {Online models for content optimization}.
\newblock {\it Advances in Neural Information Processing Systems 21 -
  Proceedings of the 2008 Conference\/}. 17--24.

\bibitem[{Ahn(2008)}]{Ahn2008}
Ahn, Hyung~Jun. 2008.
\newblock {A new similarity measure for collaborative filtering to alleviate
  the new user cold-starting problem}.
\newblock {\it Information Sciences\/} {\bf 178}(1) 37--51.

\bibitem[{Ansari et~al.(2018)Ansari, Li, and Zhang}]{Ansari2018}
Ansari, Asim, Yang Li, Jonathan~Z Zhang. 2018.
\newblock {Probabilistic topic model for hybrid recommender systems: A
  stochastic variational bayesian approach}.
\newblock {\it Marketing Science\/} {\bf 37}(6) 987--1008.

\bibitem[{Aramayo et~al.(Forthcoming)Aramayo, Schiappacasse, and
  Goic}]{aramayo2022multi}
Aramayo, Nicol{\'a}s, Mario Schiappacasse, Marcel Goic. Forthcoming.
\newblock A multi-armed bandit approach for house ads recommendations.
\newblock {\it Marketing Science\/} {\bf 0}(0) null.

\bibitem[{Auer(2002)}]{auer2002using}
Auer, Peter. 2002.
\newblock Using confidence bounds for exploitation-exploration trade-offs.
\newblock {\it Journal of Machine Learning Research\/} {\bf 3}(Nov) 397--422.

\bibitem[{Bayati et~al.(2022)Bayati, Cao, and Chen}]{bayati2022speed}
Bayati, Mohsen, Junyu Cao, Wanning Chen. 2022.
\newblock Speed up the cold-start learning in two-sided bandits with many arms.
\newblock arXiv preprint arXiv:2210.00340.

\bibitem[{Berman and Katona(2020)}]{berman2020curation}
Berman, Ron, Zsolt Katona. 2020.
\newblock Curation algorithms and filter bubbles in social networks.
\newblock {\it Marketing Science\/} {\bf 39}(2) 296--316.

\bibitem[{Bernstein et~al.(2019)Bernstein, Modaresi, and
  Saur{\'{e}}}]{Bernstein2019}
Bernstein, Fernando, Sajad Modaresi, Denis Saur{\'{e}}. 2019.
\newblock {A dynamic clustering approach to data-driven assortment
  personalization}.
\newblock {\it Management Science\/} {\bf 65}(5) 2095--2115.

\bibitem[{Bertsimas and Mersereau(2007)}]{Bertsimas2007}
Bertsimas, Dimitris, Adam~J. Mersereau. 2007.
\newblock {A learning approach for interactive marketing to a customer
  segment}.
\newblock {\it Operations Research\/} {\bf 55}(6) 1120--1135.

\bibitem[{Bobadilla et~al.(2012)Bobadilla, Ortega, Hernando, and
  Bernal}]{Bobadilla2012}
Bobadilla, Jes{\'{u}}s, Fernando Ortega, Antonio Hernando, Jes{\'{u}}s Bernal.
  2012.
\newblock {A collaborative filtering approach to mitigate the new user cold
  start problem}.
\newblock {\it Knowledge-Based Systems\/} {\bf 26} 225--238.

\bibitem[{Cakanlar et~al.(2022)Cakanlar, Trudel, and
  White}]{cakanlar2022political}
Cakanlar, Aylin, Remi Trudel, Katherine White. 2022.
\newblock Political ideology and the perceived impact of coronavirus prevention
  behaviors for the self and others.
\newblock {\it Journal of the Association for Consumer Research\/} {\bf 7}(1)
  36--44.

\bibitem[{Chen et~al.(2019)Chen, Li, Li, Jiang, Qi, and
  Song}]{chen2019generative}
Chen, Xinshi, Shuang Li, Hui Li, Shaohua Jiang, Yuan Qi, Le~Song. 2019.
\newblock Generative adversarial user model for reinforcement learning based
  recommendation system.
\newblock {\it International Conference on Machine Learning\/}. PMLR,
  1052--1061.

\bibitem[{Christakopoulou and Banerjee(2018)}]{Christakopoulou2018}
Christakopoulou, Konstantina, Arindam Banerjee. 2018.
\newblock {Learning to interact with users: A collaborative-bandit approach}.
\newblock {\it SIAM International Conference on Data Mining, SDM 2018\/}
  (Section 3) 612--620.

\bibitem[{Christakopoulou et~al.(2016)Christakopoulou, Radlinski, and
  Hofmann}]{Christakopoulou2016TowardsCR}
Christakopoulou, Konstantina, Filip Radlinski, Katja Hofmann. 2016.
\newblock Towards conversational recommender systems.
\newblock {\it Proceedings of the 22nd ACM SIGKDD International Conference on
  Knowledge Discovery and Data Mining\/}.

\bibitem[{Chu et~al.(2011)Chu, Li, Reyzin, and Schapire}]{chu2011contextual}
Chu, Wei, Lihong Li, Lev Reyzin, Robert Schapire. 2011.
\newblock Contextual bandits with linear payoff functions.
\newblock {\it Proceedings of the 14th International Conference on Artificial
  Intelligence and Statistics\/}. JMLR Workshop and Conference Proceedings,
  208--214.

\bibitem[{Cortes(2018)}]{cortes2018cold}
Cortes, David. 2018.
\newblock Cold-start recommendations in collective matrix factorization.
\newblock arXiv preprint arXiv:1809.00366.

\bibitem[{Cowgill and Tucker(2019)}]{cowgill2019economics}
Cowgill, Bo, Catherine~E Tucker. 2019.
\newblock Economics, fairness and algorithmic bias.
\newblock preparation for: Journal of Economic Perspectives.

\bibitem[{Dong et~al.(2020)Dong, Zhu, Li, and Wu}]{dong2020hybrid}
Dong, Bingbing, Yi~Zhu, Lei Li, Xindong Wu. 2020.
\newblock Hybrid collaborative recommendation via dual-autoencoder.
\newblock {\it IEEE Access\/} {\bf 8} 46030--46040.

\bibitem[{Farias and Li(2019)}]{farias2019learning}
Farias, Vivek~F, Andrew~A Li. 2019.
\newblock Learning preferences with side information.
\newblock {\it Management Science\/} {\bf 65}(7) 3131--3149.

\bibitem[{Filippi et~al.(2010)Filippi, Cappe, Garivier, and
  Szepesv{\'a}ri}]{filippi2010parametric}
Filippi, Sarah, Olivier Cappe, Aur{\'e}lien Garivier, Csaba Szepesv{\'a}ri.
  2010.
\newblock Parametric bandits: The generalized linear case.
\newblock {\it Advances in Neural Information Processing Systems\/} {\bf 23}.

\bibitem[{Gangan et~al.(2021)Gangan, Kudus, and Ilyushin}]{elena2021survey}
Gangan, Elena, Milos Kudus, Eugene Ilyushin. 2021.
\newblock Survey of multi-armed bandit algorithms applied to recommendation
  systems.
\newblock {\it International Journal of Open Information Technologies\/} {\bf
  9}(4) 12--27.

\bibitem[{Gardete and Santos(2020)}]{gardete2020no}
Gardete, Pedro~M, Carlos~D Santos. 2020.
\newblock {No data? No problem! A search-based recommendation system with cold
  starts}.
\newblock arXiv preprint arXiv:2010.03455.

\bibitem[{Geng et~al.(2022)Geng, Xiao, Sun, and Zhu}]{geng2022representation}
Geng, Yishuai, Xiao Xiao, Xiaobing Sun, Yi~Zhu. 2022.
\newblock Representation learning: Recommendation with knowledge graph via
  triple-autoencoder.
\newblock {\it Frontiers in Genetics\/} {\bf 13}.

\bibitem[{Gomez-Uribe and Hunt(2015)}]{Gomez-Uribe2015}
Gomez-Uribe, Carlos~A., Neil Hunt. 2015.
\newblock {The netflix recommender system: Algorithms, business value, and
  innovation}.
\newblock {\it ACM Transactions on Management Information Systems\/} {\bf 6}(4)
  1--19.

\bibitem[{{Gonzalez Camacho} and Alves-Souza(2018)}]{GonzalezCamacho2018}
{Gonzalez Camacho}, Lesly~Alejandra, Solange~Nice Alves-Souza. 2018.
\newblock {Social network data to alleviate cold-start in recommender system: A
  systematic review}.
\newblock {\it Information Processing {\&} Management\/} {\bf 54}(4) 529--544.

\bibitem[{Gordon et~al.(2021)Gordon, Jerath, Katona, Narayanan, Shin, and
  Wilbur}]{Gordon_et_al_2021}
Gordon, Brett~R., Kinshuk Jerath, Zsolt Katona, Sridhar Narayanan, Jiwoong
  Shin, Kenneth~C. Wilbur. 2021.
\newblock Inefficiencies in digital advertising markets.
\newblock {\it Journal of Marketing\/} {\bf 85}(1) 7--25.

\bibitem[{Guo et~al.(2020)Guo, Ktena, Myana, Huszar, Shi, Tejani, Kneier, and
  Das}]{guo2020deep}
Guo, Dalin, Sofia~Ira Ktena, Pranay~Kumar Myana, Ferenc Huszar, Wenzhe Shi,
  Alykhan Tejani, Michael Kneier, Sourav Das. 2020.
\newblock {Deep bayesian bandits: Exploring in online personalized
  recommendations}.
\newblock {\it Proceedings of the 14th ACM Conference on Recommender
  Systems\/}. 456--461.

\bibitem[{Harpale and Yang(2008)}]{harpale2008personalized}
Harpale, Abhay~S, Yiming Yang. 2008.
\newblock Personalized active learning for collaborative filtering.
\newblock {\it Proceedings of the 31st annual international ACM SIGIR
  conference on Research and development in information retrieval\/}. 91--98.

\bibitem[{Harper and Konstan(2015)}]{Harper2015TheMD}
Harper, F.~Maxwell, Joseph~A. Konstan. 2015.
\newblock {The MovieLens datasets: History and context}.
\newblock {\it ACM Transactions on Interactive Intelligent Systems\/} {\bf 5}
  19:1--19:19.

\bibitem[{He et~al.(2020)He, Deng, Wang, Li, Zhang, and Wang}]{He2020}
He, Xiangnan, Kuan Deng, Xiang Wang, Yan Li, Yongdong Zhang, Meng Wang. 2020.
\newblock Lightgcn: Simplifying and powering graph convolution network for
  recommendation.
\newblock {\it Proceedings of the 43rd International ACM SIGIR conference on
  research and development in Information Retrieval\/}. 639--648.

\bibitem[{Hosanagar et~al.(2014)Hosanagar, Fleder, Lee, and
  Buja}]{hosanagar2014will}
Hosanagar, Kartik, Daniel Fleder, Dokyun Lee, Andreas Buja. 2014.
\newblock {Will the global village fracture into tribes? Recommender systems
  and their effects on consumer fragmentation}.
\newblock {\it Management Science\/} {\bf 60}(4) 805--823.

\bibitem[{Howe et~al.(2022)Howe, Fitzsimons, and Ubel}]{howeopen}
Howe, Holly~Samantha, Gavan~J Fitzsimons, Peter Ubel. 2022.
\newblock Open science online grocery: A tool for studying choice context and
  food choice.
\newblock {\it Journal of the Association of Consumer Research\/}.

\bibitem[{Hu et~al.(2019)Hu, Du, Hu, and Li}]{hu2019hybrid}
Hu, Peng, Rong Du, Yao Hu, Nan Li. 2019.
\newblock Hybrid item-item recommendation via semi-parametric embedding.
\newblock {\it IJCAI\/}. 2521--2527.

\bibitem[{Hu et~al.(2022)Hu, Zhang, and Zhu}]{hu2022zero}
Hu, Saiquan, Juanjuan Zhang, Yuting Zhu. 2022.
\newblock Zero to one: Sales prospecting with augmented recommendation.
\newblock Available at SSRN 4006841.

\bibitem[{Jain and Pal(2022)}]{jain2022online}
Jain, Prateek, Soumyabrata Pal. 2022.
\newblock Online low rank matrix completion.
\newblock arXiv preprint arXiv:2209.03997.

\bibitem[{Jannach et~al.(2020)Jannach, de~Souza P.~Moreira, and
  Oldridge}]{jannach2020deep}
Jannach, Dietmar, Gabriel de~Souza P.~Moreira, Even Oldridge. 2020.
\newblock {Why are deep learning models not consistently winning recommender
  systems competitions yet? A position paper}.
\newblock {\it Proceedings of the Recommender Systems Challenge 2020\/}.
  44--49.

\bibitem[{J{\"a}rvelin and Kek{\"a}l{\"a}inen(2002)}]{jarvelin2002cumulated}
J{\"a}rvelin, Kalervo, Jaana Kek{\"a}l{\"a}inen. 2002.
\newblock {Cumulated gain-based evaluation of IR techniques}.
\newblock {\it ACM Transactions on Information Systems (TOIS)\/} {\bf 20}(4)
  422--446.

\bibitem[{Johari et~al.(2021)Johari, Kamble, and Kanoria}]{johari2021matching}
Johari, Ramesh, Vijay Kamble, Yash Kanoria. 2021.
\newblock Matching while learning.
\newblock {\it Operations Research\/} {\bf 69}(2) 655--681.

\bibitem[{Katehakis and Veinott(1987)}]{Katehakis1987}
Katehakis, Michael~N, Arthur~F Veinott. 1987.
\newblock {The multi-armed bandit problem: Decomposition and computation}.
\newblock {\it Mathematics of Operations Research\/} {\bf 12}(2) 262--268.

\bibitem[{Kawale et~al.(2015)Kawale, Bui, Kveton, Tran-Thanh, and
  Chawla}]{kawale2015efficient}
Kawale, Jaya, Hung~H Bui, Branislav Kveton, Long Tran-Thanh, Sanjay Chawla.
  2015.
\newblock {Efficient Thompson sampling for online? Matrix factorization
  recommendation}.
\newblock {\it Advances in Neural Information Processing Systems\/} {\bf 28}.

\bibitem[{Keskin et~al.(2022)Keskin, Li, and Sunar}]{keskin2022data}
Keskin, N~Bora, Yuexing Li, Nur Sunar. 2022.
\newblock Data-driven clustering and feature-based retail electricity pricing
  with smart meters.
\newblock Available at SSRN 3686518.

\bibitem[{Kille et~al.(2015)Kille, Lommatzsch, and Brodt}]{kille2015news}
Kille, Benjamin, Andreas Lommatzsch, Torben Brodt. 2015.
\newblock News recommendation in real-time.
\newblock {\it Smart Information Systems\/}. Springer, 149--180.

\bibitem[{Kokkodis and Ipeirotis(Forthcoming)}]{kokkodis2020good}
Kokkodis, Marios, Panagiotis~G Ipeirotis. Forthcoming.
\newblock The good, the bad, and the unhirable: Recommending job applicants in
  online labor markets.
\newblock {\it Management Science\/} {\bf 0}(0) null.

\bibitem[{Koren et~al.(2009)Koren, Bell, and Volinsky}]{koren2009matrix}
Koren, Yehuda, Robert Bell, Chris Volinsky. 2009.
\newblock Matrix factorization techniques for recommender systems.
\newblock {\it Computer\/} {\bf 42}(8) 30--37.

\bibitem[{Korkut and Li(2021)}]{korkut2021disposable}
Korkut, Melda, Andrew Li. 2021.
\newblock Disposable linear bandits for online recommendations.
\newblock {\it Proceedings of the AAAI Conference on Artificial
  Intelligence\/}, vol.~35. 4172--4180.

\bibitem[{Kotkov et~al.(2016)Kotkov, Wang, and Veijalainen}]{kotkov2016survey}
Kotkov, Denis, Shuaiqiang Wang, Jari Veijalainen. 2016.
\newblock A survey of serendipity in recommender systems.
\newblock {\it Knowledge-Based Systems\/} {\bf 111} 180--192.

\bibitem[{Kumar et~al.(2020{\natexlab{a}})Kumar, Eckles, and
  Aral}]{kumar2020scalable}
Kumar, Madhav, Dean Eckles, Sinan Aral. 2020{\natexlab{a}}.
\newblock Scalable bundling via dense product embeddings.
\newblock arXiv preprint arXiv:2002.00100.

\bibitem[{Kumar et~al.(2020{\natexlab{b}})Kumar, Bala, and
  Mukherjee}]{Kumar2020}
Kumar, Rahul, Pradip~Kumar Bala, Shubhadeep Mukherjee. 2020{\natexlab{b}}.
\newblock {A new neighborhood formation approach for solving cold-start user
  problem in collaborative filtering}.
\newblock {\it International Journal of Applied Management Science\/} {\bf
  12}(2) 118--141.

\bibitem[{Kveton et~al.(2017)Kveton, Szepesv{\'a}ri, Rao, Wen, Abbasi-Yadkori,
  and Muthukrishnan}]{kveton2017stochastic}
Kveton, Branislav, Csaba Szepesv{\'a}ri, Anup Rao, Zheng Wen, Yasin
  Abbasi-Yadkori, S~Muthukrishnan. 2017.
\newblock Stochastic low-rank bandits.
\newblock arXiv preprint arXiv:1712.04644.

\bibitem[{Lee and Hosanagar(2019)}]{lee2019recommender}
Lee, Dokyun, Kartik Hosanagar. 2019.
\newblock {How do recommender systems affect sales diversity? A cross-category
  investigation via randomized field experiment}.
\newblock {\it Information Systems Research\/} {\bf 30}(1) 239--259.

\bibitem[{Li(2013)}]{li2013generalized}
Li, Lihong. 2013.
\newblock Generalized thompson sampling for contextual bandits.
\newblock arXiv preprint arXiv:1310.7163.

\bibitem[{Li et~al.(2012)Li, Chu, Langford, Moon, and Wang}]{Li2012}
Li, Lihong, Wei Chu, John Langford, Taesup Moon, Xuanhui Wang. 2012.
\newblock {An unbiased offline evaluation of contextual bandit algorithms with
  generalized linear models}.
\newblock {\it JMLR: Workshop and Conference Proceedings\/} {\bf 26} 19--36.

\bibitem[{Li et~al.(2010)Li, Chu, Langford, and Schapire}]{Li2010a}
Li, Lihong, Wei Chu, John Langford, Robert~E. Schapire. 2010.
\newblock {A contextual-bandit approach to personalized news article
  recommendation}.
\newblock {\it Proceedings of the 19th International Conference on World Wide
  Web, WWW '10\/}  661--670.

\bibitem[{Li et~al.(2011)Li, Chu, Langford, and Wang}]{Li2011}
Li, Lihong, Wei Chu, John Langford, Xuanhui Wang. 2011.
\newblock {Unbiased offline evaluation of contextual-bandit-based news article
  recommendation algorithms}.
\newblock {\it Proceedings of the 4th ACM International Conference on Web
  Search and Data Mining, WSDM 2011\/}. 297--306.

\bibitem[{Li et~al.(2016)Li, Karatzoglou, and Gentile}]{li2016collaborative}
Li, Shuai, Alexandros Karatzoglou, Claudio Gentile. 2016.
\newblock Collaborative filtering bandits.
\newblock {\it Proceedings of the 39th International ACM SIGIR conference on
  Research and Development in Information Retrieval\/}. 539--548.

\bibitem[{Liberali and Ferecatu(2022)}]{liberali2022morphing}
Liberali, Gui, Alina Ferecatu. 2022.
\newblock Morphing for consumer dynamics: Bandits meet hidden markov models.
\newblock {\it Marketing Science\/} {\bf 41}(4) 769--794.

\bibitem[{Lu et~al.(2021)Lu, Meisami, and Tewari}]{lu2021low}
Lu, Yangyi, Amirhossein Meisami, Ambuj Tewari. 2021.
\newblock Low-rank generalized linear bandit problems.
\newblock {\it International Conference on Artificial Intelligence and
  Statistics\/}. PMLR, 460--468.

\bibitem[{McInerney et~al.(2018)McInerney, Lacker, Hansen, Higley, Bouchard,
  Gruson, and Mehrotra}]{McInerney2018}
McInerney, James, Benjamin Lacker, Samantha Hansen, Karl Higley, Hugues
  Bouchard, Alois Gruson, Rishabh Mehrotra. 2018.
\newblock {Explore, exploit, and explain: Personalizing explainable
  recommendations with bandits}.
\newblock {\it RecSys 2018 - 12th ACM Conference on Recommender Systems\/}
  31--39.

\bibitem[{Misra et~al.(2019)Misra, Schwartz, and Abernethy}]{Misra2019}
Misra, Kanishka, Eric~M. Schwartz, Jacob Abernethy. 2019.
\newblock {Dynamic online pricing with incomplete information using multi-armed
  bandit experiments}.
\newblock {\it Marketing Science\/} {\bf 38}(2) 226--252.

\bibitem[{Nabi et~al.(2022)Nabi, Nassif, Hong, Mamani, and
  Imbens}]{nabi2022bayesian}
Nabi, Sareh, Houssam Nassif, Joseph Hong, Hamed Mamani, Guido Imbens. 2022.
\newblock Bayesian meta-prior learning using empirical bayes.
\newblock {\it Management Science\/} {\bf 68}(3) 1737--1755.

\bibitem[{Padilla and Ascarza(2021)}]{padilla2021overcoming}
Padilla, Nicolas, Eva Ascarza. 2021.
\newblock Overcoming the cold start problem of customer relationship management
  using a probabilistic machine learning approach.
\newblock {\it Journal of Marketing Research\/} {\bf 58}(5) 981--1006.

\bibitem[{Resnick et~al.(1994)Resnick, Iacovou, Suchak, Bergstrom, and
  Riedl}]{Resnick1994}
Resnick, Paul, Neophytos Iacovou, Mitesh Suchak, Peter Bergstrom, John Riedl.
  1994.
\newblock {GroupLens: An open architecture for collaborative filtering of
  netnews}.
\newblock {\it Proceedings of the 1994 ACM Conference on Computer Supported
  Cooperative Work, CSCW 1994\/}. 175--186.

\bibitem[{Rubens et~al.(2015)Rubens, Elahi, Sugiyama, and
  Kaplan}]{rubens2015active}
Rubens, Neil, Mehdi Elahi, Masashi Sugiyama, Dain Kaplan. 2015.
\newblock Active learning in recommender systems.
\newblock {\it Recommender systems handbook\/}  809--846.

\bibitem[{Sardianos et~al.(2017)Sardianos, Varlamis, and
  Eirinaki}]{sardianos2017scaling}
Sardianos, Christos, Iraklis Varlamis, Magdalini Eirinaki. 2017.
\newblock Scaling collaborative filtering to large--scale bipartite rating
  graphs using lenskit and spark.
\newblock {\it 2017 IEEE Third International Conference on Big Data Computing
  Service and Applications (BigDataService)\/}. IEEE, 70--79.

\bibitem[{Sarwar et~al.(2001)Sarwar, Karypis, Konstan, and Riedl}]{Sarwar2001}
Sarwar, Badrul, George Karypis, Joseph Konstan, John Riedl. 2001.
\newblock {Item-based collaborative filtering recommendation algorithms}.
\newblock {\it Proceedings of the 10th International Conference on World Wide
  Web, WWW 2001\/}. 285--295.

\bibitem[{Schwartz et~al.(2017)Schwartz, Bradlow, and Fader}]{Schwartz2017}
Schwartz, Eric~M., Eric~T. Bradlow, Peter~S. Fader. 2017.
\newblock {Customer acquisition via display advertising using multi-armed
  bandit experiments}.
\newblock {\it Marketing Science\/} {\bf 36}(4) 500--522.

\bibitem[{Shi et~al.(2014)Shi, Larson, and Hanjalic}]{shi2014collaborative}
Shi, Yue, Martha Larson, Alan Hanjalic. 2014.
\newblock Collaborative filtering beyond the user-item matrix: A survey of the
  state of the art and future challenges.
\newblock {\it ACM Computing Surveys (CSUR)\/} {\bf 47}(1) 1--45.

\bibitem[{Silva et~al.(2022)Silva, Werneck, Silva, Pereira, and
  Rocha}]{silva2022multi}
Silva, N{\'\i}collas, Heitor Werneck, Thiago Silva, Adriano~CM Pereira,
  Leonardo Rocha. 2022.
\newblock Multi-armed bandits in recommendation systems: A survey of the
  state-of-the-art and future directions.
\newblock {\it Expert Systems with Applications\/} {\bf 197} 116669.

\bibitem[{Song et~al.(2019)Song, Sahoo, and Ofek}]{song2019and}
Song, Yicheng, Nachiketa Sahoo, Elie Ofek. 2019.
\newblock {When and how to diversify? A multicategory utility model for
  personalized content recommendation}.
\newblock {\it Management Science\/} {\bf 65}(8) 3737--3757.

\bibitem[{Strub et~al.(2016)Strub, Gaudel, and Mary}]{strub2016hybrid}
Strub, Florian, Romaric Gaudel, J{\'e}r{\'e}mie Mary. 2016.
\newblock Hybrid recommender system based on autoencoders.
\newblock {\it Proceedings of the 1st workshop on deep learning for recommender
  systems\/}. 11--16.

\bibitem[{Su et~al.(2016)Su, Sharma, and Goel}]{su2016effect}
Su, Jessica, Aneesh Sharma, Sharad Goel. 2016.
\newblock The effect of recommendations on network structure.
\newblock {\it Proceedings of the 25th international conference on World Wide
  Web\/}. 1157--1167.

\bibitem[{Tang et~al.(2014)Tang, Jiang, Li, and Li}]{tang2014ensemble}
Tang, Liang, Yexi Jiang, Lei Li, Tao Li. 2014.
\newblock Ensemble contextual bandits for personalized recommendation.
\newblock {\it Proceedings of the 8th ACM Conference on Recommender Systems\/}.
  73--80.

\bibitem[{Tang and Zhou(2012)}]{tang2012dynamic}
Tang, Xiangyu, Jie Zhou. 2012.
\newblock Dynamic personalized recommendation on sparse data.
\newblock {\it IEEE transactions on knowledge and data engineering\/} {\bf
  25}(12) 2895--2899.

\bibitem[{Thompson(1933)}]{Samples1933}
Thompson, William~R. 1933.
\newblock {On the likelihood that one unknown probability exceeds another in
  view of the evidence of two samples}.
\newblock {\it Biometrika\/} {\bf 25}(3) 285--294.

\bibitem[{Trinh et~al.(2020)Trinh, Kaufmann, Vernade, and
  Combes}]{trinh2020solving}
Trinh, Cindy, Emilie Kaufmann, Claire Vernade, Richard Combes. 2020.
\newblock Solving bernoulli rank-one bandits with unimodal thompson sampling.
\newblock {\it Algorithmic Learning Theory\/}. PMLR, 862--889.

\bibitem[{Waisman et~al.(2019)Waisman, Nair, Carrion, and
  Xu}]{waisman2019online}
Waisman, Caio, Harikesh~S Nair, Carlos Carrion, Nan Xu. 2019.
\newblock Online causal inference for advertising in real-time bidding
  auctions.
\newblock arXiv preprint arXiv:1908.08600.

\bibitem[{Wang and Blei(2011)}]{Wang2011}
Wang, Chong, David~M Blei. 2011.
\newblock {Collaborative topic modeling for recommending scientific articles}.
\newblock {\it Proceedings of the ACM SIGKDD International Conference on
  Knowledge Discovery and Data Mining\/}. 448--456.

\bibitem[{Wang et~al.(2016)Wang, Wu, and Wang}]{wang2016learning}
Wang, Huazheng, Qingyun Wu, Hongning Wang. 2016.
\newblock Learning hidden features for contextual bandits.
\newblock {\it Proceedings of the 25th ACM international on conference on
  information and knowledge management\/}. 1633--1642.

\bibitem[{Wang et~al.(2017)Wang, Wu, and Wang}]{Wang_Wu_Wang_2017}
Wang, Huazheng, Qingyun Wu, Hongning Wang. 2017.
\newblock Factorization bandits for interactive recommendation.
\newblock {\it Proceedings of the AAAI Conference on Artificial Intelligence\/}
  {\bf 31}(1).

\bibitem[{Wang et~al.(2018)Wang, Zeng, Zhou, Li, Iyengar, Shwartz, and
  Grabarnik}]{wang2018online}
Wang, Qing, Chunqiu Zeng, Wubai Zhou, Tao Li, S~Sitharama Iyengar, Larisa
  Shwartz, Genady~Ya Grabarnik. 2018.
\newblock Online interactive collaborative filtering using multi-armed bandit
  with dependent arms.
\newblock {\it IEEE Transactions on Knowledge and Data Engineering\/} {\bf
  31}(8) 1569--1580.

\bibitem[{Wang et~al.(2022)Wang, Tao, and Zhang}]{wang2022}
Wang, Y., L.~Tao, X.~Zhang. 2022.
\newblock Recommending for a three-sided food delivery marketplace: A
  multi-objective hierarchical approach.
\newblock Working paper.

\bibitem[{Wei et~al.(2017)Wei, He, Chen, Zhou, and Tang}]{Wei2017}
Wei, Jian, Jianhua He, Kai Chen, Yi~Zhou, Zuoyin Tang. 2017.
\newblock {Collaborative filtering and deep learning based recommendation
  system for cold start items}.
\newblock {\it Expert Systems with Applications\/} {\bf 69} 29--39.

\bibitem[{Wu(2018)}]{Wu2018InteractiveOL}
Wu, Qingyun. 2018.
\newblock Interactive online learning with incomplete knowledge.
\newblock Working paper.

\bibitem[{Xing et~al.(2014)Xing, Wang, and Wang}]{Xing2014}
Xing, Zhe, Xinxi Wang, Ye~Wang. 2014.
\newblock {Enhancing collaborative filtering music recommendation by balancing
  exploration and exploitation}.
\newblock {\it Proceedings of the 15th International Society for Music
  Information Retrieval Conference, ISMIR 2014\/} (Ismir) 445--450.

\bibitem[{Ye et~al.(2020)Ye, Zhang, Zhang, Zhang, Chen, and Xu}]{ye2020cold}
Ye, Zikun, Dennis Zhang, Heng Zhang, Renyu~Philip Zhang, Xin Chen, Zhiwei Xu.
  2020.
\newblock Cold start to improve market thickness on online advertising
  platforms: Data-driven algorithms and field experiments.
\newblock Available at SSRN 3702786.

\bibitem[{Ying et~al.(2006)Ying, Feinberg, and Wedel}]{Ying2006}
Ying, Yuanping, Fred Feinberg, Michel Wedel. 2006.
\newblock Leveraging missing ratings to improve online recommendation systems.
\newblock {\it Journal of Marketing Research\/} {\bf 43}(3) 355--365.

\bibitem[{Zhang et~al.(2022)Zhang, Moon, and Veeraraghavan}]{zhang2022does}
Zhang, Jiding, Ken Moon, Senthil~K Veeraraghavan. 2022.
\newblock Does fake news create echo chambers?
\newblock Available at SSRN 4144897.

\bibitem[{Zhang et~al.(2020)Zhang, Tsang, and Duan}]{zhang2020collaborative}
Zhang, Yan, Ivor~W Tsang, Lixin Duan. 2020.
\newblock Collaborative generative hashing for marketing and fast cold-start
  recommendation.
\newblock {\it IEEE Intelligent Systems\/} {\bf 35}(5) 84--95.

\bibitem[{Zhao et~al.(2013)Zhao, Zhang, and Wang}]{Zhao2013}
Zhao, Xiaoxue, Weinan Zhang, Jun Wang. 2013.
\newblock Interactive collaborative filtering.
\newblock {\it Proceedings of the 22nd ACM international conference on
  Information \& Knowledge Management\/}. 1411--1420.

\bibitem[{Zheng et~al.(2018)Zheng, Lu, Jiang, Zhang, and
  Yu}]{zheng2018spectral}
Zheng, Lei, Chun-Ta Lu, Fei Jiang, Jiawei Zhang, Philip~S Yu. 2018.
\newblock Spectral collaborative filtering.
\newblock {\it Proceedings of the 12th ACM conference on recommender
  systems\/}. 311--319.

\bibitem[{Zhou(2015)}]{Zhou2015}
Zhou, Li. 2015.
\newblock {A survey on contextual multi-armed bandits}.
\newblock arXiv preprint arXiv:1508.03326.

\bibitem[{Zhou et~al.(2023)Zhou, Wang, Yan, and Tan}]{zhou2023spoiled}
Zhou, Tongxin, Yingfei Wang, Lu~Yan, Yong Tan. 2023.
\newblock Spoiled for choice? personalized recommendation for healthcare
  decisions: A multiarmed bandit approach.
\newblock {\it Information Systems Research\/} .

\end{thebibliography}

\pagebreak{}

\appendix

\part*{Appendix}

\setcounter{figure}{0} \renewcommand{\thefigure}{A.\arabic{figure}}
\setcounter{table}{0} \renewcommand{\thetable}{A.\arabic{table}}
\setcounter{equation}{0}

\section{\label{sec:Notations}Model Notation}

Table \ref{tab:Notation-1} details the notational definitions used
throughout the paper.

\begin{table}[h]
\caption{\label{tab:Notation-1}Notation}

\vspace{0.2cm}

\noindent \centering{}%
\begin{tabular}{>{\raggedright}p{4cm}>{\raggedright}p{12cm}}
\hline 
\centering{}{\footnotesize{}Notation} & \centering{}{\footnotesize{}Definition}\tabularnewline
\hline 
\centering{}{\footnotesize{}For CFA (\S\ref{subsec:CF})} & \centering{}\tabularnewline
\hline 
\centering{}{\footnotesize{}$K$} & \centering{}{\footnotesize{}The dimension of latent representation
space determined by CF}\tabularnewline
\centering{}{\footnotesize{}$I$} & \centering{}{\footnotesize{}The number of users}\tabularnewline
\centering{}{\footnotesize{}$J$} & \centering{}{\footnotesize{}The number of items}\tabularnewline
\centering{}{\footnotesize{}$P$} & \centering{}{\footnotesize{}The dimension of user feature vector}\tabularnewline
\centering{}{\footnotesize{}$Q$} & \centering{}{\footnotesize{}The dimension of item feature vector}\tabularnewline
\centering{}{\footnotesize{}$D$, $A$} & \centering{}{\footnotesize{}Matrices of user demographics and item
attributes, respectively}\tabularnewline
\centering{}{\footnotesize{}$\mu_{ij}$} & \centering{}{\footnotesize{}The mean feedback of user $i$ to item
$j$ (e.g. rankings, choices, etc.)}\tabularnewline
\centering{}{\footnotesize{}$y_{ij}$} & \centering{}{\footnotesize{}An indicator which equals to 1 if feedback
of user $i$ to item $j$ is observed, and 0 otherwise}\tabularnewline
\centering{}{\footnotesize{}$U=\left[\boldsymbol{u}_{1},\boldsymbol{u}_{2},\cdots,\boldsymbol{u}_{I}\right]^{T}$} & \centering{}{\footnotesize{}Latent representations for user location
in preference space}\tabularnewline
\centering{}{\footnotesize{}$V=\left[\boldsymbol{v}_{1},\boldsymbol{v}_{2},\cdots,\boldsymbol{v}_{J}\right]$} & \centering{}{\footnotesize{}Latent representation for item location
in preference space}\tabularnewline
\centering{}{\footnotesize{}$W=\left[\boldsymbol{w}_{1},\boldsymbol{w}_{2,}\cdots,\boldsymbol{w}_{P}\right]$} & \centering{}{\footnotesize{}Matrix projecting user locations in the
latent preference space onto user demographics}\tabularnewline
\centering{}{\footnotesize{}$\Psi=\left[\boldsymbol{\psi}_{1},\boldsymbol{\psi}_{2},\cdots,\boldsymbol{\psi}_{Q}\right]$} & \centering{}{\footnotesize{}Matrix projecting item locations in the
latent preference space onto item attributes}\tabularnewline
\centering{}{\footnotesize{}$\Sigma=\{\sigma^{2},\sigma_{d}^{2},\sigma_{a}^{2},\lambda_{u},\lambda_{v},\lambda_{w},\lambda_{\psi}\}$} & \centering{}{\footnotesize{}CFA hyper-parameters}\tabularnewline
\centering{}{\footnotesize{}$\overline{\boldsymbol{v}},\overline{\boldsymbol{u}},\overline{\boldsymbol{w}},\overline{\boldsymbol{\psi}},\Sigma_{v},\Sigma_{u},\Sigma_{w},\Sigma_{\psi}$} & \centering{}{\footnotesize{}Posterior means and covariances of $U,V,W,\Psi$,
respectively}\tabularnewline
\hline 
\centering{}{\footnotesize{}For Bandit Learning (\S\ref{subsec:Bandit-Learning-Stage})} & \centering{}\tabularnewline
\hline 
\centering{}{\footnotesize{}$T$} & \centering{}{\footnotesize{}The total number of feedback occasions}\tabularnewline
\centering{}{\footnotesize{}$CAR$} & \centering{}{\footnotesize{}The expected cumulative rewards at time
$t$}\tabularnewline
\hline 
\centering{}{\footnotesize{}For Experiments (\S\ref{sec:Simulations}
and \S\ref{sec:Live-Experiment})} & \centering{}\tabularnewline
\hline 
\centering{}{\footnotesize{}$\Gamma$} & \centering{}{\footnotesize{}Matrix projecting preferences to demographics
in the linear setting}\tabularnewline
\centering{}{\footnotesize{}$\boldsymbol{\epsilon}$}, {\footnotesize{}$\boldsymbol{e}$},{\footnotesize{}
$\boldsymbol{\boldsymbol{\varepsilon}}$} & \centering{}{\footnotesize{}Random }\textit{\footnotesize{}iid}{\footnotesize{}
normal errors}\tabularnewline
\centering{}{\footnotesize{}$HPR_{t}^{c}$} & \centering{}{\footnotesize{}Purchase rate on the home page of category
$c$ in period $t$}\tabularnewline
\hline 
\centering{}{\footnotesize{}Throughout the Paper} & \centering{}\tabularnewline
\hline 
\centering{}{\footnotesize{}{*}} & \centering{}{\footnotesize{}A superscribe to remark the optimal solutions
of the corresponding variable}\tabularnewline
\centering{}{\footnotesize{}$\widehat{}$} & \centering{}{\footnotesize{}A symbol to remark the estimate of the
corresponding variable}\tabularnewline
\hline 
\end{tabular}
\end{table}
\newpage{}

\begin{landscape}

\section{\label{sec:CFB-A-algorithm-flow}The CFB-A Algorithm}

Algorithms \ref{alg:CFB-A_ucb} (for UCB) and \ref{alg:CFB-A_ts}
(for TS) outline the CFB-A algorithms for learning either new users'
preferences on existing items or for learning existing users' preferences
on new items. We exemplify the algorithm with the case of new users.\footnote{In the case of new items, user demographics $\boldsymbol{d}_{i,t}$
in Algorithms \ref{alg:CFB-A_ucb} and \ref{alg:CFB-A_ts} is replaced
by item attributes $\boldsymbol{a}_{j,t}$ for item $j$ at time $t$.
Correspondingly, the pre-computed variables from training set are
the CFA-estimated user latent factors $\boldsymbol{u}$ (distribution
$\mathcal{F}_{U}$) and item attribute latent factors $\boldsymbol{\psi}_{0}$
(prior distribution $\mathcal{F}_{\Psi}^{0}$).}

\begin{algorithm}[H]
\caption{\label{alg:CFB-A_ucb}CFB-A (UCB)}

\textbf{Pre-determined hyper-parameters:} $K,\sigma^{2},\sigma_{d}^{2},\lambda_{u},\lambda_{w},\alpha$

\textbf{Observed variables:} $\mu_{ij,t}$ for user $i$, item $j$,
and user demographics $\boldsymbol{d}_{i,t}$ at time $t$

\textbf{Variables pre-computed with existing users and items (training
set):} CFA-estimated item latent factors $\boldsymbol{v}$ and user
demographic latent factors $\boldsymbol{w}_{0}$

\textbf{Initialization of regularization parameters and priors:} $S\leftarrow\lambda_{u}\sigma^{2}I_{K};\boldsymbol{b}\leftarrow S(\boldsymbol{d}_{i,0}^{T}\boldsymbol{w}_{0}^{g})^{T};S_{u}\leftarrow\lambda_{w}\sigma_{d}^{2}I_{K};\boldsymbol{b}_{u}\leftarrow S_{u}\boldsymbol{w}_{0}$

\medskip{}

\textbf{for} $t=1,2,...,T$

\textbf{\quad{}}Calculate $\boldsymbol{w}_{t}=S_{u}^{-1}\boldsymbol{b}_{u};\boldsymbol{u}_{i,t}=S^{-1}\boldsymbol{b};\Sigma_{i,t}=S^{-1}\sigma^{2}$
(\textbf{CFA estimation})

\textbf{\quad{}}Select the item $j^{*}=\arg\max_{j}\left(\boldsymbol{u}_{i,t}^{T}\boldsymbol{v}_{j}+\alpha_{t}\cdot\sqrt{\boldsymbol{v}_{j}^{T}\Sigma_{i,t}\boldsymbol{v}_{j}}\right)$
for each user $i$ (\textbf{UCB Bandit allocation})

\textbf{\quad{}}Observe user feedback to recommended items $\overrightarrow{\mu}_{ij^{*}t}$,
and incorporate newly observed user demographics $\boldsymbol{d}_{i,t}$

\textbf{\quad{}}Update $S_{u}\leftarrow S_{u}+\mathbb{I}\left[i\thinspace makes\thinspace any\thinspace choices\right]\cdot\boldsymbol{u}_{i,t}\boldsymbol{u}_{i,t}^{T}$
(\textbf{CFA estimation})

\textbf{\hspace{4em}}$\boldsymbol{b}_{u}\leftarrow\boldsymbol{b}_{u}+\mathbb{I}\left[i\thinspace chooses\thinspace j^{*}\right]\cdot\boldsymbol{u}_{i,t}\boldsymbol{d}_{i,t}^{T}$(\textbf{CFA
estimation})

\textbf{\hspace{4em}}$S\leftarrow S+\mathbb{I}\left[i\thinspace chooses\thinspace j^{*}\right]\cdot\boldsymbol{v}_{j^{*}}\boldsymbol{v}_{j^{*}}^{T}+\frac{\sigma^{2}}{\sigma_{d}^{2}}\boldsymbol{w}_{t}\boldsymbol{w}_{t}^{T}$
(\textbf{CFA estimation})

\textbf{\hspace{4em}}$\boldsymbol{b}\leftarrow\boldsymbol{b}+\mathbb{I}\left[i\thinspace chooses\thinspace j^{*}\right]\cdot\boldsymbol{v}_{j^{*}}\cdot\mu_{ij^{*}}+\frac{\sigma^{2}}{\sigma_{d}^{2}}\boldsymbol{w}_{t}\boldsymbol{d}_{i,t}^{T}$
(\textbf{CFA estimation})

\textbf{\hspace{4em}}$t\leftarrow t+1$

\textbf{end for}

\bigskip{}

\textit{\emph{\footnotesize{}Notes: }}{\footnotesize{}rate $\alpha_{t}$
is a slowly increasing function of $t$ to maintain exploration as
uncertainties $\sqrt{\boldsymbol{v}_{j}^{T}\Sigma_{i,t}\boldsymbol{v}_{j}}$
decrease over time (\citealp{Zhao2013,filippi2010parametric}). $\alpha_{t}$
is specified as $\alpha\sqrt{\log t}$ and $\alpha\sqrt{t}$ in practice.}{\footnotesize\par}
\end{algorithm}
\noindent \begin{flushleft}
\par\end{flushleft}

\end{landscape}

\pagebreak{}

\begin{landscape}

\begin{algorithm}[H]
\caption{\label{alg:CFB-A_ts}CFB-A (TS)}

\textbf{Pre-determined hyper-parameters:} $K,\sigma^{2},\sigma_{d}^{2},\lambda_{u},\lambda_{w}$

\textbf{Observed variables:} $\mu_{ij,t}$ for user $i$, item $j$,
and user demographics $\boldsymbol{d}_{i,t}$ at time $t$

\textbf{Variables pre-computed with existing users and items (training
set):} CFA-estimated distributions of item latent space $\mathcal{F}_{V}$
(item $j$ has $\boldsymbol{v}_{j}$ randomly drawn from $\mathcal{N}\left(\overline{\boldsymbol{v}}_{j},\Sigma_{\boldsymbol{v}_{j}}\right)$)
and user demographic latent space $\mathcal{F}_{W}^{0}$ (priors,
demographic features $\boldsymbol{w}_{0}$ are randomly drawn from
$\mathcal{N}\left(\overline{\boldsymbol{w}}_{0},\Sigma_{0}^{w}\right)$).

\medskip{}

\textbf{Initialization of regularization parameters and priors:} $S\leftarrow\lambda_{u}\sigma^{2}I_{K};\boldsymbol{b}\leftarrow S(\boldsymbol{d}_{i,0}^{T}\boldsymbol{w}_{0}^{g})^{T};S_{u}\leftarrow\lambda_{w}\sigma_{d}^{2}I_{K};\boldsymbol{b}_{u}\leftarrow S_{u}\boldsymbol{w}_{0}$

\medskip{}

\textbf{for} $t=1,2,...,T$

\textbf{\quad{}}Calculate $\overline{\boldsymbol{w}}_{t}=S_{u}^{-1}\boldsymbol{b}_{u};\text{\ensuremath{\Sigma_{t}^{w}=}\ensuremath{S_{u}^{-1}}\ensuremath{\ensuremath{\sigma_{d}^{2}}}};\overline{\boldsymbol{u}}_{i,t}=S^{-1}\boldsymbol{b};\Sigma_{i,t}=S^{-1}\sigma^{2}$
(\textbf{CFA estimation})

\textbf{\quad{}}Select the item $j^{*}=\arg\max_{j}\boldsymbol{u}_{i,t}^{T}\boldsymbol{v}_{j}$
for each user $i$, where $\boldsymbol{u}_{i,t}$ is randomly drawn
from $\mathcal{N}\left(\overline{\boldsymbol{u}}_{i,t},\Sigma_{i,t}\right)$
(\textbf{TS Bandit allocation})

\textbf{\quad{}}Observe user feedback to recommended items $\overrightarrow{\mu}_{ij^{*}t}$,
and incorporate newly observed user demographics $\boldsymbol{d}_{i,t}$

\textbf{\quad{}}Update $S_{u}\leftarrow S_{u}+\mathbb{I}\left[i\thinspace makes\thinspace any\thinspace choices\right]\cdot\boldsymbol{u}_{i,t}\boldsymbol{u}_{i,t}^{T}$
(\textbf{CFA estimation})

\textbf{\hspace{4em}}$\boldsymbol{b}_{u}\leftarrow\boldsymbol{b}_{u}+\mathbb{I}\left[i\thinspace chooses\thinspace j^{*}\right]\cdot\boldsymbol{u}_{i,t}\boldsymbol{d}_{i,t}^{T}$(\textbf{CFA
estimation})

\textbf{\hspace{4em}}$S\leftarrow S+\mathbb{I}\left[i\thinspace chooses\thinspace j^{*}\right]\cdot\boldsymbol{v}_{j^{*}}\boldsymbol{v}_{j^{*}}^{T}+\frac{\sigma^{2}}{\sigma_{d}^{2}}\boldsymbol{w}_{t}\boldsymbol{w}_{t}^{T}$,
where $\boldsymbol{w}_{t}$ is randomly drawn from $\mathcal{N}\left(\overline{\boldsymbol{w}}_{t},\Sigma_{t}^{w}\right)$
(\textbf{CFA estimation})

\textbf{\hspace{4em}}$\boldsymbol{b}\leftarrow\boldsymbol{b}+\mathbb{I}\left[i\thinspace chooses\thinspace j^{*}\right]\cdot\boldsymbol{v}_{j^{*}}\cdot\mu_{ij^{*}}+\frac{\sigma^{2}}{\sigma_{d}^{2}}\boldsymbol{w}_{t}\boldsymbol{d}_{i,t}^{T}$
(\textbf{CFA estimation})

\textbf{\hspace{4em}}$t\leftarrow t+1$

\textbf{end for}
\end{algorithm}
\end{landscape}

\part*{\protect\pagebreak Web Appendix}

\setcounter{figure}{0} \renewcommand{\thefigure}{A.\arabic{figure}}
\setcounter{table}{0} \renewcommand{\thetable}{A.\arabic{table}}
\setcounter{section}{0}
\setcounter{equation}{0}
\setcounter{page}{1}

\section{\label{sec:Derivations-of-Posteriors_CF}Posterior Distributions
for $U,V$ in the CF Model}

The likelihood of preference $\boldsymbol{\mu}$ over users and items
given $U$, $V$ and hyper-parameters is:

\begin{equation}
P\left(\boldsymbol{\mu}\mid U,V,\lambda_{u},\lambda_{v},\sigma^{2}\right)=\Pi_{i=1}^{I}\Pi_{j=1}^{J}\left[\mathcal{N}\left(\mu_{ij}\mid\boldsymbol{u}_{i}^{T}\boldsymbol{v}_{j},\lambda_{u},\lambda_{v},\sigma^{2}\right)\right]^{y_{ij}}
\end{equation}

\subsection{Posterior of $U$}

The posterior of $U$ is as follows:

\begin{eqnarray}
p\left(U\mid\boldsymbol{\mu},V,\lambda_{u},\lambda_{v},\sigma^{2}\right) & \propto & p\left(\boldsymbol{\mu}\mid U,V,\lambda_{u},\lambda_{v},\sigma^{2}\right)\cdot p\left(U\mid\lambda_{u},\lambda_{v},\sigma^{2}\right)\\
 & \propto & \Pi_{i=1}^{I}\mathcal{N}\left(\boldsymbol{u}_{i}\mid\textbf{0},\sigma_{u}^{2}\mathbf{I}_{K}\right)\cdot\nonumber \\
 &  & \Pi_{i=1}^{I}\Pi_{j=1}^{J}\left[\mathcal{N}\left(\mu_{ij}\mid U,V,\lambda_{u},\lambda_{v},\sigma^{2}\right)\right]^{y_{ij}=1}\cdot\nonumber \\
 & \propto & \Pi_{i=1}^{I}\exp\left\{ -\frac{1}{2}\lambda_{u}\boldsymbol{u}_{i}^{T}\boldsymbol{u}_{i}-\frac{1}{2}\left[\frac{1}{\sigma^{2}}\sum_{y_{ij}=1}\left(\mu_{ij}-\boldsymbol{u}_{i}^{T}\boldsymbol{v}_{j}\right)^{2}\right]\right\} \nonumber \\
 & \propto & \Pi_{i=1}^{I}\exp\left\{ -\frac{1}{2}\left[\boldsymbol{u}_{i}^{T}\left(\lambda_{u}\mathbf{I}_{K}+\frac{1}{\sigma^{2}}\sum_{y_{ij}=1}\boldsymbol{v}_{j}\boldsymbol{v}_{j}^{T}\right)\boldsymbol{u}_{i}-2\frac{1}{\sigma^{2}}\sum_{y_{ij}=1}\mu_{ij}\boldsymbol{u}_{i}^{T}\boldsymbol{v}_{j}\right]\right\} \nonumber \\
 & \propto & \Pi_{i=1}^{I}\mathcal{N}\left(\boldsymbol{u}_{i}\mid\overline{\boldsymbol{u}}_{i},\Sigma_{\boldsymbol{u}_{i}}\right)\nonumber 
\end{eqnarray}
Thus, $\boldsymbol{u}_{i}\sim\mathcal{N}\left(\overline{\boldsymbol{u}}_{i},\Sigma_{\boldsymbol{u}_{i}}\right)$,
where

\begin{equation}
\overline{\boldsymbol{u}}_{i}=\left(\frac{1}{\sigma^{2}}\sum_{j}y_{ij}\cdot\mu_{ij}\boldsymbol{v}_{j}\right)\Sigma_{\boldsymbol{u}_{i}},\label{u_posterior_mean_cf}
\end{equation}

\begin{equation}
\Sigma_{\boldsymbol{u}_{i}}=\left(\lambda_{u}\mathbf{I}_{K}+\sum_{j}y_{ij}\cdot\frac{1}{\sigma^{2}}\boldsymbol{v}_{j}\boldsymbol{v}_{j}^{T}\right)^{-1}\label{u_posterior_covariance_cf}
\end{equation}

\subsection{Posterior of $V$}

The posterior of $V$ is as follows:

\begin{eqnarray}
p\left(V\mid\boldsymbol{\mu},U,\lambda_{u},\lambda_{v},\sigma^{2}\right) & \propto & p\left(\boldsymbol{\mu}\mid U,V,\lambda_{u},\lambda_{v},\sigma^{2}\right)\cdot p\left(V\mid\lambda_{u},\lambda_{v},\sigma^{2}\right)\label{eq:posterior_v_cf}\\
 & \propto & \Pi_{j=1}^{J}\mathcal{N}\left(\boldsymbol{v}_{j}\mid\textbf{0},\lambda_{v}^{-1}\mathbf{I}_{K}\right)\cdot\nonumber \\
 &  & \Pi_{i=1}^{I}\Pi_{j=1}^{J}\left[\mathcal{N}(\mu_{ij}\mid U,V,\lambda_{u},\lambda_{v},\sigma^{2})\right]^{y_{ij}=1}\cdot\nonumber \\
 & \propto & \Pi_{j=1}^{J}\exp\left\{ -\frac{1}{2}\lambda_{v}\boldsymbol{v}_{j}^{T}\boldsymbol{v}_{j}-\frac{1}{2}\left[\frac{1}{\sigma^{2}}\sum_{y_{ij}=1}\left(\mu_{ij}-\boldsymbol{u}_{i}^{T}\boldsymbol{v}_{j}\right)^{2}\right]\right\} \nonumber \\
 & \propto & \Pi_{j=1}^{J}\exp\left\{ -\frac{1}{2}\left[\boldsymbol{v}_{j}^{T}\left(\lambda_{v}\mathbf{I}_{K}+\frac{1}{\sigma^{2}}\sum_{y_{ij}=1}\boldsymbol{u}_{i}\boldsymbol{u}_{i}^{T}\right)\boldsymbol{v}_{j}-2\frac{1}{\sigma^{2}}\sum_{y_{ij}=1}\mu_{ij}\boldsymbol{v}_{j}^{T}\boldsymbol{u}_{i}\right]\right\} \nonumber \\
 & \propto & \Pi_{j=1}^{J}\mathcal{N}\left(\boldsymbol{v}_{j}\mid\overline{\boldsymbol{v}}_{j},\Sigma_{\boldsymbol{v}_{j}}\right)\nonumber 
\end{eqnarray}

\noindent Thus, $\boldsymbol{v}_{j}\sim\mathcal{N}\left(\overline{\boldsymbol{v}}_{j},\Sigma_{\boldsymbol{v}_{j}}\right)$,
where

\noindent 
\begin{equation}
\overline{\boldsymbol{v}}_{j}=\left(\frac{1}{\sigma^{2}}\sum_{i}y_{ij}\cdot\mu_{ij}\boldsymbol{u}_{i}\right)\Sigma_{\boldsymbol{v}_{j}}\label{eq:v_post_mean_cf}
\end{equation}
\begin{equation}
\Sigma_{\boldsymbol{v}_{j}}=\left(\lambda_{v}\mathbf{I}_{K}+\sum_{i}y_{ij}\cdot\frac{1}{\sigma^{2}}\boldsymbol{u}_{i}\boldsymbol{u}_{i}^{T}\right)^{-1}\label{eq:v_post_cov_cf}
\end{equation}

\section{\label{sec:Derivations-of-Posteriors}Posterior Distributions for
$U,V,W,\Psi$ in the CFA Model}

The joint likelihood of $\left\{ \boldsymbol{\mu},D,A\right\} $ over
users and items given latent spaces $\left\{ U,V,W,\Psi\right\} $
and hyper-parameters is:

\begin{eqnarray}
P\left(\boldsymbol{\mu},D,A\mid U,V,W,\Psi,\Sigma\right) & = & P\left(\boldsymbol{\mu}\mid D,A,U,V,W,\Psi,\Sigma\right)P\left(D,A\mid U,V,W,\Psi,\Sigma\right)\\
 & = & \Pi_{i=1}^{I}\Pi_{j=1}^{J}\left[\mathcal{N}(\mu_{ij}\mid\boldsymbol{a}_{j},\boldsymbol{d}_{i},U,V,W,\Psi,\Sigma)\right]^{y_{ij}=1}\cdot\nonumber \\
 &  & \Pi_{p=1}^{P}\mathcal{N}(d_{ip}\mid U,W,\Sigma)\Pi_{q=1}^{Q}\mathcal{N}(a_{jq}\mid V,\Psi,\Sigma)\nonumber 
\end{eqnarray}

\noindent where $\Sigma=\left\{ \sigma^{2},\sigma_{d}^{2},\sigma_{a}^{2},\lambda_{u},\lambda_{v},\lambda_{w},\lambda_{\psi}\right\} $.

\subsection{Posterior of $U$}

The posterior of $U$ is calculated as follows:

\begin{eqnarray}
p\left(U\mid\boldsymbol{\mu},D,A,U,W,\Psi,\Sigma\right) & \propto & p\left(\boldsymbol{\mu},D,A\mid U,V,W,\Psi,\Sigma\right)\cdot p\left(U\mid\Sigma\right)\\
 & \propto & \Pi_{i=1}^{I}\mathcal{N}\left(\boldsymbol{u}_{i}\mid\textbf{0},\lambda_{u}^{-1}\mathbf{I}_{K}\right)\cdot\nonumber \\
 &  & \Pi_{i=1}^{I}\Pi_{j=1}^{J}\left[\mathcal{N}\left(\mu_{ij}\mid\boldsymbol{a}_{j},\boldsymbol{d}_{i},U,V,W,\Psi,\Sigma\right)\right]^{y_{ij}=1}\cdot\nonumber \\
 &  & \Pi_{p=1}^{P}\mathcal{N}\left(d_{ip}\mid U,W,\Sigma\right)\Pi_{q=1}^{Q}\mathcal{N}\left(a_{jq}\mid V,\Psi,\Sigma\right)\nonumber \\
 & \propto & \Pi_{i=1}^{I}\exp\Biggl\{-\frac{1}{2}\lambda_{u}\boldsymbol{u}_{i}^{T}\boldsymbol{u}_{i}-\frac{1}{2}\Biggl[\frac{1}{\sigma^{2}}\sum_{y_{ij}=1}\left(\mu_{ij}-\boldsymbol{u}_{i}^{T}\boldsymbol{v}_{j}\right)^{2}\nonumber \\
 &  & +\frac{1}{\sigma_{d}^{2}}\sum_{p=1}^{P}\left(d_{ip}-\boldsymbol{w}_{p}^{T}\boldsymbol{u}_{i}\right)^{2}\Biggr]\Biggr\}\nonumber \\
 & \propto & \Pi_{i=1}^{I}\exp\Biggl\{-\frac{1}{2}\Biggl[\boldsymbol{u}_{i}^{T}\left(\lambda_{u}\mathbf{I}_{K}+\frac{1}{\sigma^{2}}\sum_{y_{ij}=1}\boldsymbol{v}_{j}\boldsymbol{v}_{j}^{T}+\frac{1}{\sigma_{d}^{2}}\sum_{p=1}^{P}\boldsymbol{w}_{p}\boldsymbol{w}_{p}^{T}\right)\boldsymbol{u}_{i}\nonumber \\
 &  & -2\left(\frac{1}{\sigma^{2}}\sum_{y_{ij}=1}\mu_{ij}\boldsymbol{u}_{i}^{T}\boldsymbol{v}_{j}+\frac{1}{\sigma_{d}^{2}}\sum_{p=1}^{P}d_{ip}\boldsymbol{u}_{i}^{T}\boldsymbol{w}_{p}\right)\Biggr]\Biggr\}\nonumber \\
 & \propto & \Pi_{i=1}^{I}\mathcal{N}\left(\boldsymbol{u}_{i}\mid\overline{\boldsymbol{u}}_{i},\Sigma_{\boldsymbol{u}_{i}}\right)\nonumber 
\end{eqnarray}

Thus, $\boldsymbol{u}_{i}\sim\mathcal{N}\left(\overline{\boldsymbol{u}}_{i},\Sigma_{\boldsymbol{u}_{i}}\right)$,
where

\begin{equation}
\overline{\boldsymbol{u}}_{i}=\left(\frac{1}{\sigma^{2}}\sum_{j}y_{ij}\cdot\mu_{ij}\boldsymbol{v}_{j}+\frac{1}{\sigma_{d}^{2}}\sum_{p=1}^{P}d_{ip}\boldsymbol{w}_{p}\right)\Sigma_{\boldsymbol{u}_{i}},\label{u_posterior_mean-1}
\end{equation}

\begin{equation}
\Sigma_{\boldsymbol{u}_{i}}=\left(\lambda_{u}\mathbf{I}_{K}+\sum_{j}y_{ij}\cdot\frac{1}{\sigma^{2}}\boldsymbol{v}_{j}\boldsymbol{v}_{j}^{T}+\frac{1}{\sigma_{d}^{2}}\sum_{p=1}^{P}\boldsymbol{w}_{p}\boldsymbol{w}_{p}^{T}\right)^{-1}\label{u_posterior_covariance-1}
\end{equation}

\subsection{Posterior of $V$}

The posterior of $V$ is calculated as follows:

\begin{eqnarray}
p\left(V\mid\boldsymbol{\mu},D,A,U,W,\Psi,\Sigma\right) & \propto & p\left(\boldsymbol{\mu},D,A\mid U,V,W,\Psi,\Sigma\right)\cdot p\left(V\mid\Sigma\right)\label{eq:cfba_posterior_v}\\
 & \propto & \Pi_{j=1}^{J}\mathcal{N}\left(\boldsymbol{v}_{j}\mid\textbf{0},\lambda_{v}^{-1}\mathbf{I}_{K}\right)\cdot\nonumber \\
 &  & \Pi_{i=1}^{I}\Pi_{j=1}^{J}\left[\mathcal{N}(\mu_{ij}\mid\boldsymbol{a}_{j},\boldsymbol{d}_{i},U,V,W,\Psi,\Sigma)\right]^{y_{ij}=1}\cdot\nonumber \\
 &  & \Pi_{p=1}^{P}\mathcal{N}(d_{ip}\mid U,W,\Sigma)\Pi_{q=1}^{Q}\mathcal{N}(a_{jq}\mid V,\Psi,\Sigma)\nonumber \\
 & \propto & \Pi_{j=1}^{J}\exp\Biggl\{-\frac{1}{2}\lambda_{v}\boldsymbol{v}_{j}^{T}\boldsymbol{v}_{j}-\frac{1}{2}\Biggl[\frac{1}{\sigma^{2}}\sum_{y_{ij}=1}\left(\mu_{ij}-\boldsymbol{u}_{i}^{T}\boldsymbol{v}_{j}\right)^{2}\nonumber \\
 &  & +\frac{1}{\sigma_{a}^{2}}\sum_{q=1}^{Q}\left(a_{jq}-\boldsymbol{\psi}_{q}^{T}\boldsymbol{v}_{j}\right)^{2}\Biggr]\Biggr\}\nonumber \\
 & \propto & \Pi_{j=1}^{J}\exp\Biggl\{-\frac{1}{2}\left[\boldsymbol{v}_{j}^{T}\left(\lambda_{v}\mathbf{I}_{K}+\frac{1}{\sigma^{2}}\sum_{y_{ij}=1}\boldsymbol{u}_{i}\boldsymbol{u}_{i}^{T}+\frac{1}{\sigma_{a}^{2}}\sum_{q=1}^{Q}\boldsymbol{\psi}_{q}\boldsymbol{\psi}_{q}^{T}\right)\boldsymbol{v}_{j}\right.\nonumber \\
 &  & \left.\left.-2\left(\frac{1}{\sigma^{2}}\sum_{y_{ij}=1}\mu_{ij}\boldsymbol{v}_{j}^{T}\boldsymbol{u}_{i}+\frac{1}{\sigma_{a}^{2}}\sum_{q=1}^{Q}a_{jq}\boldsymbol{v}_{j}^{T}\boldsymbol{\psi}_{q}\right)\right]\right\} \nonumber \\
 & \propto & \Pi_{j=1}^{J}\mathcal{N}\left(\boldsymbol{v}_{j}\mid\overline{\boldsymbol{v}}_{j},\Sigma_{\boldsymbol{v}_{j}}\right)\nonumber 
\end{eqnarray}

\noindent Thus, $\boldsymbol{v}_{j}\sim\mathcal{N}\left(\overline{\boldsymbol{v}}_{j},\Sigma_{\boldsymbol{v}_{j}}\right)$,
where

\noindent 
\begin{equation}
\overline{\boldsymbol{v}}_{j}=\left(\frac{1}{\sigma^{2}}\sum_{i}y_{ij}\cdot\mu_{ij}\boldsymbol{u}_{i}+\frac{1}{\sigma_{a}^{2}}\sum_{q=1}^{Q}a_{jq}\boldsymbol{\psi}_{q}\right)\Sigma_{\boldsymbol{v}_{j}}\label{eq:v_post_mean}
\end{equation}
\begin{equation}
\Sigma_{\boldsymbol{v}_{j}}=\left(\lambda_{v}\mathbf{I}_{K}+\sum_{i}y_{ij}\cdot\frac{1}{\sigma^{2}}\boldsymbol{u}_{i}\boldsymbol{u}_{i}^{T}+\frac{1}{\sigma_{a}^{2}}\sum_{q=1}^{Q}\boldsymbol{\psi}_{q}\boldsymbol{\psi}_{q}^{T}\right)^{-1}\label{eq:v_post_cov}
\end{equation}

\subsection{Posterior of $W$}

The posterior of $W$ is calculated as follows:

\noindent {\small{}
\begin{eqnarray}
p\left(W\mid\boldsymbol{\mu},D,A,U,W,\Psi,\Sigma\right) & \propto & p\left(\boldsymbol{\mu},D,A\mid U,V,W,\Psi,\Sigma\right)\cdot p\left(W\mid\Sigma\right)\\
 & \propto & \Pi_{p=1}^{P}\mathcal{N}\left(\boldsymbol{w}_{p}\mid\textbf{0},\lambda_{w}^{-1}\mathbf{I}_{K}\right)\cdot\nonumber \\
 &  & \Pi_{i=1}^{I}\Pi_{j=1}^{J}\left[\mathcal{N}\left(\mu_{ij}\mid\boldsymbol{a}_{j},\boldsymbol{d}_{i},U,V,W,\Psi,\Sigma\right)\right]^{y_{ij}=1}\cdot\nonumber \\
 &  & \Pi_{p=1}^{P}\mathcal{N}\left(d_{ip}\mid U,W,\Sigma\right)\Pi_{q=1}^{Q}\mathcal{N}\left(a_{jq}\mid V,\Psi,\Sigma\right)\nonumber \\
 & \propto & \Pi_{p=1}^{P}\exp\Biggl\{-\frac{1}{2}\lambda_{w}\boldsymbol{w}_{p}^{T}\boldsymbol{w}_{p}-\frac{1}{2}\Biggl[\frac{1}{\sigma^{2}}\sum_{y_{ij}=1}\left(\mu_{ij}-\boldsymbol{u}_{i}^{T}\boldsymbol{v}_{j}\right)^{2}\nonumber \\
 &  & +\frac{1}{\sigma_{d}^{2}}\sum_{p=1}^{P}\left(d_{ip}-\boldsymbol{u}_{i}^{T}\boldsymbol{w}_{p}\right)^{2}\Biggr]\Biggr\}\nonumber \\
 & \propto & \Pi_{p=1}^{P}\exp\left\{ -\frac{1}{2}\left[\boldsymbol{w}_{p}^{T}\left(\lambda_{w}\mathbf{I}_{K}+\sum_{i=1}^{I}\boldsymbol{u}_{i}\boldsymbol{u}_{i}^{T}\right)\boldsymbol{w}_{p}-2\frac{1}{\sigma_{d}^{2}}\sum_{i=1}^{I}d_{ip}\boldsymbol{u}_{i}^{T}\boldsymbol{w}_{p}\right]\right\} \nonumber \\
 & \propto & \Pi_{p=1}^{P}\mathcal{N}\left(\boldsymbol{w}_{p}\mid\overline{\boldsymbol{w}}_{p},\Sigma_{\boldsymbol{w}_{p}}\right)\nonumber 
\end{eqnarray}
}{\small\par}

Thus, $\textbf{w}_{p}\sim\mathcal{N}\left(\overline{\boldsymbol{w}}_{p},\Sigma_{\boldsymbol{w}_{p}}\right)$,
where

\noindent 
\begin{equation}
\overline{\boldsymbol{w}}_{p}=\frac{1}{\sigma_{d}^{2}}\left(\sum_{i=1}^{I}d_{ip}\boldsymbol{u}_{i}^{T}\right)\Sigma_{\boldsymbol{w}_{p}}
\end{equation}
\begin{equation}
\Sigma_{\boldsymbol{w}_{p}}=\left(\lambda_{w}\mathbf{I}_{K}+\frac{1}{\sigma_{d}^{2}}\sum_{i=1}^{I}\boldsymbol{u}_{i}\boldsymbol{u}_{i}^{T}\right)^{-1}
\end{equation}

\subsection{Posterior of $\Psi$}

The posterior of $\Psi$ is calculated as follows:

\noindent 
\begin{eqnarray}
p\left(\Psi\mid\boldsymbol{\mu},D,A,U,W,\Psi,\Sigma\right) & \propto & p\left(\boldsymbol{\mu},D,A\mid U,V,W,\Psi,\Sigma\right)\cdot p\left(\Psi\mid\Sigma\right)\\
 & \propto & \Pi_{q=1}^{Q}\mathcal{N}\left(\boldsymbol{\textbf{\ensuremath{\psi}}}_{p}\mid\textbf{0},\lambda_{\psi}^{-1}\mathbf{I}_{K}\right)\cdot\nonumber \\
 &  & \Pi_{i=1}^{I}\Pi_{j=1}^{J}\left[\mathcal{N}\left(\mu_{ij}\mid\boldsymbol{a}_{j},\boldsymbol{d}_{i},U,V,W,\Psi,\Sigma\right)\right]^{y_{ij}=1}\cdot\nonumber \\
 &  & \Pi_{p=1}^{P}\mathcal{N}\left(d_{ip}\mid U,W,\Sigma\right)\Pi_{q=1}^{Q}\mathcal{N}\left(a_{jq}\mid V,\Psi,\Sigma\right)\nonumber \\
 & \propto & \Pi_{q=1}^{Q}\exp\Biggl\{-\frac{1}{2}\lambda_{\psi}\boldsymbol{\psi}_{q}^{T}\boldsymbol{\psi}_{q}-\frac{1}{2}\Biggl[\frac{1}{\sigma^{2}}\sum_{y_{ij}=1}\left(\mu_{ij}-\boldsymbol{u}_{i}^{T}\boldsymbol{v}_{j}\right)^{2}\nonumber \\
 &  & +\frac{1}{\sigma_{a}^{2}}\sum_{q=1}^{Q}\left(a_{jq}-\boldsymbol{v}_{j}^{T}\boldsymbol{\psi}_{q}\right)^{2}\Biggr]\Biggr\}\nonumber \\
 & \propto & \Pi_{q=1}^{Q}\exp\left\{ -\frac{1}{2}\left[\boldsymbol{\psi}_{q}^{T}\left(\lambda_{\psi}\mathbf{I}_{K}+\frac{1}{\sigma_{a}^{2}}\sum_{j=1}^{J}\boldsymbol{v}_{j}\boldsymbol{v}_{j}^{T}\right)\boldsymbol{\psi}_{q}-2\frac{1}{\sigma_{a}^{2}}\sum_{j=1}^{J}a_{jq}\boldsymbol{v}_{j}^{T}\boldsymbol{\psi}_{q}\right]\right\} \nonumber \\
 & \propto & \Pi_{q=1}^{Q}\mathcal{N}\left(\boldsymbol{\textbf{\ensuremath{\psi}}}_{d}\mid\overline{\boldsymbol{\psi}}_{q},\Sigma_{\boldsymbol{\psi}_{q}}\right)\nonumber 
\end{eqnarray}

Thus, $\boldsymbol{\textbf{\ensuremath{\psi}}}_{d}\sim\mathcal{N}\left(\overline{\boldsymbol{\psi}}_{d},\Sigma_{\boldsymbol{\psi}_{d}}\right)$,
where

\noindent 
\begin{equation}
\overline{\boldsymbol{\psi}}_{q}=\frac{1}{\sigma_{a}^{2}}\left(\sum_{j=1}^{J}a_{jd}\boldsymbol{v}_{j}^{T}\right)\Sigma_{\boldsymbol{\psi}_{q}}
\end{equation}
\begin{equation}
\Sigma_{\boldsymbol{\psi}_{d}}=\left(\lambda_{\psi}\mathbf{I}_{K}+\frac{1}{\sigma_{a}^{2}}\sum_{j=1}^{J}\boldsymbol{v}_{j}\boldsymbol{v}_{j}^{T}\right)^{-1}\label{eq:psi_post-1}
\end{equation}

\section{\label{sec:Hyper-parameter-Tuning}Hyper-parameter Tuning}

We detail hyper-parameter tuning for all competing methods. There
are 6 types of hyper-parameters: i) the prior of item popularity score
in POP; ii) the dimension of latent spaces $K$ in AL, CFB, and CFB-A;
iii) the exploration rate $\alpha$ in all methods that involve test
and learn; iv) the variance of user-item feedback ($\boldsymbol{\mu}$),
$\sigma^{2}$, in all methods except Random and POP; v) the variances
of latent vectors, $1/\lambda_{m}$, $m\in\left\{ U,V\right\} $ in
CFB and $m\in\left\{ U,V,W,\Psi\right\} $ in CFB-A; and vi) the variances
of user demographics and item attributes $\left\{ \sigma_{d}^{2},\sigma_{a}^{2}\right\} $.
Three methods are implemented to determine the values of different
hyper-parameters.

First, training data (existing users) are used to approximate the
item popularity score and $1/\lambda_{m}$. Specifically, the mean
feedback of each item is computed across existing users to initialize
the popularity score in POP for new users. An item with a higher mean
feedback among existing users is considered more popular. $1/\lambda_{U}$
and $1/\lambda_{V}$ are set as the mean variances of feedback across
users and items, respectively. $1/\lambda_{W}$ and $1/\lambda_{\Psi}$
are set as the mean variances across user demographics and item attributes,
respectively.

Second, for each hyper-parameter in $\left\{ K,\alpha,\sigma^{2}\right\} $,
a grid search is conducted within a value range\footnote{Specifically, $K$ ranges from 2 to the half of the minimum dimension
of user demographics and item attributes, $\alpha$ ranges from 0
to 10, and $\sigma^{2}$ ranges from 0.1 to 100.}, and select the best-performing value (e.g., cHPR in the live experiment)
for each method based on the training set (split into a calibration
and a validation set).\footnote{We apply the replay method (\citealp{Li2012}) described in Section
\ref{sec:Simulations} to the random-ordered training data to tune
hyper-parameters in the live experiment. That is, we compute the hyper-parameters
as if assignments were not actually random but were instead ordered
using the four alternative algorithms in the experiment and choices
were observed via the replay method. Consider the number of latent
factors, $K$, as an example. This exercise in hyper-prior setting
would yield, for example, user locations in the latent factor space
under the assumption of a specific number of factors, $K$, and a
specific method.

To determine the best fitting set of hyper-priors, we break the training
set into a calibration and a validation set and then select the best
fitting hyper-parameters for the validation set. Continuing with our
example for $K,$ we would compute under each level of $K$, user
locations in the matrix factorization space via the calibration set,
and then use the locations to predict purchases in the validation
set. We then retain the level of $K$ that best predicts observed
choices in the validation set.

An alternative to the replay method for setting hyper-parameters would
be to create testing data sets (the testing phase in the live experiment,
as described in Section \ref{subsec:Design}) for all combinations
of models and hyper-parameters and then find the best fitting hyper-parameters
for each model. However, this approach is economically infeasible
given the large number of cells and costs of subjects for each cell.}

Third, the $\left\{ \sigma_{d}^{2},\sigma_{a}^{2}\right\} $ in the
CFB-A are normalized to 1 for identification.

\section{\label{sec:Synthetic-Data-Distribution}Synthetic Data Distribution}

\begin{figure}[H]
\begin{centering}
\caption{\label{fig:Utility-Distribution}Utility Distribution in Linear and
Non-linear Settings}
\begin{tabular}{cc}
\includegraphics[scale=0.5]{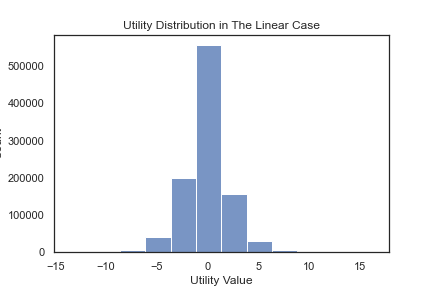} & \includegraphics[scale=0.5]{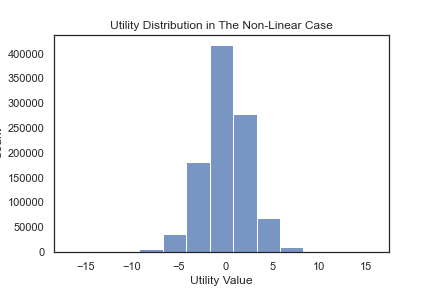}\tabularnewline
(a) & (b)\tabularnewline
\end{tabular}
\par\end{centering}
{\footnotesize{}Notes: This figure depicts the distribution of users'
simulated utility in the linear case (Panel a) and the non-linear
case (Panel b).}{\footnotesize\par}
\end{figure}

Figure \ref{fig:Utility-Distribution} shows the distribution of simulated
user utility in Section \ref{subsec:Synthetic-Data-Simulation}. Panel
(a) shows the linear case and Panel (b) shows the non-linear case.

\section{\label{sec:Additional-Details-in-experiment}Supplemental Experimental
Details}

We provide additional details in the live experiment, including attention
check questions (Section \ref{subsec:Subjects}), detailed item attributes
in the four product categories (Section \ref{subsec:Item-attributes}),
demographic survey and comparison of demographics between the training
and testing sets (Section \ref{subsec:Demographic-survey-questions}),
and full results (Section \ref{subsec:Full-results}).

\subsection{\label{subsec:Item-attributes}Item Attributes}

The following attributes are collected for each item for each category.
\begin{itemize}
\item Produce (22 attributes)
\begin{itemize}
\item Dummy variables: organic; four allergens (milk, soy, egg, and fish)
\item Numerical variables: 2 size metrics (servingSizeGrams and servings);
13 ingredients (calories, caloricDensity, totalFat, saturatedFat,
polyFat, monoFat, cholesterol, sodium, potassium, carbs, fiber, sugar,
and protein); price; and health starpoints ranging from 0 to 4 with
a higher value indicating a healthier option (see details in \citealt{howeopen}).
\end{itemize}
\item Meat, Dairy, and Eggs (40 attributes)
\begin{itemize}
\item Dummy variables: organic; 14 subcategories; 9 allergens (nuts, coconut,
almond, soy, sulfites, milk, wheat, fish, and shellfish)
\item Numerical variables: 2 size metrics (servingSizeGrams and servings);
14 ingredients (calories, caloricDensity, transFat, totalFat, saturatedFat,
polyFat, monoFat, cholesterol, sodium, potassium, carbs, fiber, sugar,
and protein); price; health starpoints
\end{itemize}
\item Bakery, Pasta, and Grains (31 attributes)
\begin{itemize}
\item Dummy variables: organic; 12 allergens (milk, egg, nuts, wheat, soy,
sesame, almond, coconut, gluten, peanuts, pecans, and sulfites)
\item Numerical variables: 2 size metrics (servingSizeGrams and servings);
14 ingredients (calories, caloricDensity, transFat, totalFat, saturatedFat,
polyFat, monoFat, cholesterol, sodium, potassium, carbs, fiber, sugar,
and protein); price; health starpoints
\end{itemize}
\item Beverage (38 attributes)
\begin{itemize}
\item Dummy variables: organic; 8 subcategories; 12 allergens (milk, egg,
nuts, wheat, soy, seasame, sulphur dioxide, coconut, gluten, sulfites,
phenylalanine, and caffeine)
\item Numerical variables: 2 size metrics (servingSizeGrams and servings);
13 ingredients (calories, caloricDensity, totalFat, saturatedFat,
polyFat, monoFat, cholesterol, sodium, potassium, carbs, fiber, sugar,
and protein); price; health starpoints
\end{itemize}
\end{itemize}

\subsection{\label{subsec:Attention-Check-Questions}Attention Check Questions}

After reading the grocery task introduction, a participant is asked
to answer the following questions
\begin{itemize}
\item In the task, you will imagine to be? Four options include evaluating
pictures, watching videos, grocery shopping (the correct answer),
and talking to a friend.
\item In the grocery store task, you will imagine shopping with a budget
of? Four options include \$30, \$50, \$75 (the correct answer), and
\$100.
\end{itemize}
If a participant cannot correctly answer two questions above, he or
she will be asked to read the instruction and answer the comprehension
questions again. Upon answering the two questions correctly, the participant
will proceed to the rest of the study.

\subsection{\label{subsec:Demographic-survey-questions}Demographic Survey Questions}

The demographic survey questions are as follows.

\subsubsection*{Single answer questions}
\begin{itemize}
\item How would you describe your gender? Three options include male, female,
and third gender (non-binary)
\item What is the highest level of school you have completed or the highest
degree you have received? 8 options include less than high school
degree, high school graduate (high school diploma or equivalent including
GED), some college but no degree, associate degree in college (2-year),
bachelor's degree in college (4-year), master's degree, doctoral degree,
and professional degree (JD, MD).
\item Please indicate the answer that includes your entire household income
in (previous year) before taxes. 13 options include less than \$10,000;
\$10,000 to \$19,999; \$20,000 to \$29,999; \$30,000 to \$39,999;
\$40,000 to \$49,999; \$50,000 to \$59,999; \$60,000 to \$69,999;
\$70,000 to \$79,999; \$80,000 to \$89,999; \$90,000 to \$99,999;
\$100,000 to \$149,999; \$150,000 or more; Prefer not to say.
\item In which state do you currently reside? 52 options including fifty
states, D.C. and Puerto Rico
\item What religious family do you belong to or identify yourself most close
to? 6 options include Buddhism, Jewish, Muslim, Christian (Catholic
protestant or any other Christian denominations) , non-religious,
and others (need specification).
\item What is your height in feet and inches? Please choose your best estimate
from 8 options include shorter than 5'1'', 5'1'' to 5'2'', 5'3'' to
5'4'', 5'5'' to 5'6'', 5'7'' to 5'8'', 5'9'' to 5'10'', 5'11'' to
6'0'', and taller than 6'0''.
\item Indicate the extent (seven levels) to which you agree or disagree
with each statement: i) I am on a diet; ii) I prefer organic food;
iii) I like sweets; iv) I limit my calorie intake; v) I avoid eating
too many carbs; vi) I avoid eating too much fat; vii) I avoid eating
too much cholesterol; viii) I avoid eating too much salt.
\end{itemize}

\subsubsection*{Multiple answer questions}
\begin{itemize}
\item Choose one or more races that you consider yourself to be. 6 options
include White, Black or African American, American Indian or Alaska
Native, Asian, Native Hawaiian or Pacific Islander , and others (need
specification).
\item Do you (or any immediate family member) have any dietary restrictions?
9 options include gluten-free, sugar-free, vegan, vegetarian, kosher,
allergic to nuts, lactose intolerant, no restrictions, and others
(need specification).
\end{itemize}

\subsubsection*{Open-ended questions}
\begin{itemize}
\item What is your age?
\item How many people live or stay in this household at least half the time?
\item What is your current weight in pounds?
\end{itemize}

\subsection{\label{subsec:Demographic-Comparison-Between}Demographic Comparison
Between the Training and Testing Data}

The experimental analyses presumes that the population-level hyper-parameters
in the training and testing samples are the same. While the assumption
is impossible to test, it is possible to assess the equality of means
on the reported demographics.

\begin{figure}[H]
\caption{\label{fig:Distribution-of-P-value}Distribution of p-value of t-Test
on Demographic Differences}

\begin{centering}
\includegraphics[scale=0.6]{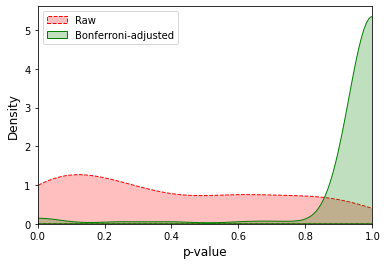}
\par\end{centering}
{\footnotesize{}Note: This figure portrays the distributions of the
original p-values and Bonferroni-adjusted p-values for the t-test
on the values of the demographic variables in the training set and
in the testing set. The distributions are over the 119 tests corresponding
to the 119 demographic variables.}{\footnotesize\par}
\end{figure}

Based on participants' survey responses, we construct 119 demographic
variables. For each variable, we conduct a t-test to compare the values
for the training set and the testing set. Figure \ref{fig:Distribution-of-P-value}
shows the distributions of the original p-values and Bonferroni-adjusted
p-values (\citealp{cakanlar2022political}) across 119 tests. The
Bonferroni-adjusted p-values indicate that participants in the training
and testing sets are not statistically different.

\subsection{\label{subsec:Full-results}Full Results}

Table \ref{result_ctr} reports the cumulative HPR for each product
category and aggregated across them.

\newpage{}

\begin{landscape}

\begin{table}[H]
{\scriptsize{}\caption{\label{result_ctr}Cumulative HPR (\%) Results}
}{\scriptsize\par}
\begin{centering}
{\scriptsize{}}%
\begin{tabular}{>{\raggedright}m{1.2cm}c>{\centering}p{1.5cm}>{\centering}p{1.5cm}>{\centering}p{2.5cm}c>{\centering}p{1.5cm}>{\centering}p{1.5cm}>{\centering}p{2.5cm}c>{\centering}p{1.5cm}>{\centering}p{1.5cm}>{\centering}p{2.5cm}}
\toprule 
\centering{}{\scriptsize{}Category} & {\scriptsize{}Method} & \multicolumn{4}{c}{{\scriptsize{}$T=1$}} & \multicolumn{3}{c}{{\scriptsize{}$T=2$}} &  & \multicolumn{3}{c}{{\scriptsize{}$T=3$}}\tabularnewline
\midrule
\centering{} &  & {\scriptsize{}Num.Obs} & {\scriptsize{}cHPR} & {\scriptsize{}Improvement (\%) by CFB-A} &  & {\scriptsize{}Num.Obs} & {\scriptsize{}cHPR} & {\scriptsize{}Improvement (\%) by CFB-A} &  & {\scriptsize{}Num.Obs} & {\scriptsize{}cHPR} & {\scriptsize{}Improvement (\%) by CFB-A}\tabularnewline
\midrule
\multirow{4}{1.2cm}{\centering{}{\scriptsize{}All}} & {\scriptsize{}UCB} & {\scriptsize{}155} & {\scriptsize{}2.33} & {\scriptsize{}69.10 (12.21){*}{*}{*}} &  & {\scriptsize{}127} & {\scriptsize{}2.34} & {\scriptsize{}68.38 (12.14){*}{*}{*}} &  & {\scriptsize{}110} & {\scriptsize{}2.32} & {\scriptsize{}68.53 (11.31){*}{*}{*}}\tabularnewline
 & {\scriptsize{}CFB} & {\scriptsize{}156} & {\scriptsize{}3.31} & {\scriptsize{}19.03 (4.37){*}{*}{*}} &  & {\scriptsize{}126} & {\scriptsize{}3.34} & {\scriptsize{}17.96 (4.61){*}{*}{*}} &  & {\scriptsize{}110} & {\scriptsize{}3.29} & {\scriptsize{}18.84 (4.51){*}{*}{*}}\tabularnewline
 & {\scriptsize{}CFA} & {\scriptsize{}142} & {\scriptsize{}3.57} & {\scriptsize{}9.39 (2.58){*}{*}{*}} &  & {\scriptsize{}121} & {\scriptsize{}3.62} & {\scriptsize{}8.84 (2.22){*}{*}} &  & {\scriptsize{}109} & {\scriptsize{}3.63} & {\scriptsize{}7.71 (1.83){*}{*}}\tabularnewline
 & {\scriptsize{}CFB-A} & {\scriptsize{}149} & \textbf{\scriptsize{}3.94} & {\scriptsize{}--} &  & {\scriptsize{}124} & \textbf{\scriptsize{}3.94} & {\scriptsize{}--} &  & {\scriptsize{}104} & \textbf{\scriptsize{}3.91} & {\scriptsize{}--}\tabularnewline
\midrule
\multirow{4}{1.2cm}{\centering{}{\scriptsize{}Produce}} & {\scriptsize{}UCB} & {\scriptsize{}153} & {\scriptsize{}5.34} & {\scriptsize{}35.96 (4.85){*}{*}{*}} &  & {\scriptsize{}126} & {\scriptsize{}4.93} & {\scriptsize{}47.67 (6.32){*}{*}{*}} &  & {\scriptsize{}109} & {\scriptsize{}4.80} & {\scriptsize{}49.79 (6.59){*}{*}{*}}\tabularnewline
 & {\scriptsize{}CFB} & {\scriptsize{}153} & {\scriptsize{}6.90} & {\scriptsize{}5.22 (0.80)} &  & {\scriptsize{}123} & {\scriptsize{}6.75} & {\scriptsize{}7.85 (1.24)} &  & {\scriptsize{}104} & {\scriptsize{}6.53} & {\scriptsize{}10.11 (1.56){*}}\tabularnewline
 & {\scriptsize{}CFA} & {\scriptsize{}139} & {\scriptsize{}6.52} & {\scriptsize{}11.35 (1.93){*}{*}} &  & {\scriptsize{}123} & {\scriptsize{}6.54} & {\scriptsize{}11.31 (1.94){*}{*}} &  & {\scriptsize{}107} & {\scriptsize{}6.31} & {\scriptsize{}13.95 (2.38){*}{*}{*}}\tabularnewline
 & {\scriptsize{}CFB-A} & {\scriptsize{}147} & \textbf{\scriptsize{}7.26} & {\scriptsize{}--} &  & {\scriptsize{}126} & \textbf{\scriptsize{}7.28} & {\scriptsize{}--} &  & {\scriptsize{}102} & \textbf{\scriptsize{}7.19} & {\scriptsize{}--}\tabularnewline
\midrule
\multirow{4}{1.2cm}{\centering{}{\scriptsize{}Meat, Dairy, and Eggs}} & {\scriptsize{}UCB} & {\scriptsize{}134} & {\scriptsize{}2.31} & {\scriptsize{}59.40 (13.26){*}{*}{*}} &  & {\scriptsize{}117} & {\scriptsize{}2.48} & {\scriptsize{}120.16 (12.25){*}{*}{*}} &  & {\scriptsize{}103} & {\scriptsize{}2.52} & {\scriptsize{}112.70 (11.50){*}{*}{*}}\tabularnewline
 & {\scriptsize{}CFB} & {\scriptsize{}131} & {\scriptsize{}4.16} & {\scriptsize{}36.78 (5.31){*}{*}{*}} &  & {\scriptsize{}113} & {\scriptsize{}4.31} & {\scriptsize{}26.68 (4.21){*}{*}{*}} &  & {\scriptsize{}101} & {\scriptsize{}4.32} & {\scriptsize{}26.62 (4.03){*}{*}{*}}\tabularnewline
 & {\scriptsize{}CFA} & {\scriptsize{}123} & {\scriptsize{}5.29} & {\scriptsize{}7.56 (1.24)} &  & {\scriptsize{}109} & {\scriptsize{}5.34} & {\scriptsize{}2.25 (0.45)} &  & {\scriptsize{}101} & {\scriptsize{}5.36} & {\scriptsize{}2.05 (0.34)}\tabularnewline
 & {\scriptsize{}CFB-A} & {\scriptsize{}134} & \textbf{\scriptsize{}5.69} & {\scriptsize{}--} &  & {\scriptsize{}113} & \textbf{\scriptsize{}5.46} & {\scriptsize{}--} &  & {\scriptsize{}97} & \textbf{\scriptsize{}5.47} & {\scriptsize{}--}\tabularnewline
\midrule
\multirow{4}{1.2cm}{\centering{}{\scriptsize{}Bakery, Pasta, Grains}} & {\scriptsize{}UCB} & {\scriptsize{}120} & {\scriptsize{}1.95} & {\scriptsize{}57.95 (6.34){*}{*}{*}} &  & {\scriptsize{}107} & {\scriptsize{}2.09} & {\scriptsize{}44.50 (5.17){*}{*}{*}} &  & {\scriptsize{}98} & {\scriptsize{}1.90} & {\scriptsize{}60.00 (6.53){*}{*}{*}}\tabularnewline
 & {\scriptsize{}CFB} & {\scriptsize{}121} & {\scriptsize{}2.83} & {\scriptsize{}8.83 (1.21)} &  & {\scriptsize{}97} & {\scriptsize{}2.97} & {\scriptsize{}1.68 (0.29)} &  & {\scriptsize{}92} & {\scriptsize{}2.92} & {\scriptsize{}4.11 (0.60)}\tabularnewline
 & {\scriptsize{}CFA} & {\scriptsize{}117} & {\scriptsize{}2.91} & {\scriptsize{}5.84 (0.89)} &  & {\scriptsize{}99} & {\scriptsize{}2.98} & {\scriptsize{}1.34 (0.20)} &  & {\scriptsize{}94} & \textbf{\scriptsize{}3.13} & {\scriptsize{}-2.88 (-0.49)}\tabularnewline
 & {\scriptsize{}CFB-A} & {\scriptsize{}126} & \textbf{\scriptsize{}3.08} & {\scriptsize{}--} &  & {\scriptsize{}111} & \textbf{\scriptsize{}3.02} & {\scriptsize{}--} &  & {\scriptsize{}90} & {\scriptsize{}3.04} & {\scriptsize{}--}\tabularnewline
\midrule
\multirow{4}{1.2cm}{\centering{}{\scriptsize{}Beverages}} & {\scriptsize{}UCB} & {\scriptsize{}81} & {\scriptsize{}1.04} & {\scriptsize{}37.5 (2.42){*}{*}{*}} &  & {\scriptsize{}73} & {\scriptsize{}0.97} & {\scriptsize{}55.67 (3.52){*}{*}{*}} &  & {\scriptsize{}73} & {\scriptsize{}0.92} & {\scriptsize{}68.48 (3.93){*}{*}{*}}\tabularnewline
 & {\scriptsize{}CFB} & {\scriptsize{}74} & \textbf{\scriptsize{}1.68} & {\scriptsize{}-14.88 (-1.33){*}} &  & {\scriptsize{}65} & \textbf{\scriptsize{}1.68} & {\scriptsize{}-10.12 (-1.11)} &  & {\scriptsize{}64} & {\scriptsize{}1.47} & {\scriptsize{}5.44 (0.56)}\tabularnewline
 & {\scriptsize{}CFA} & {\scriptsize{}87} & {\scriptsize{}1.48} & {\scriptsize{}-3.38 (-0.30)} &  & {\scriptsize{}71} & {\scriptsize{}1.65} & {\scriptsize{}-8.48 (-1.01)} &  & {\scriptsize{}69} & \textbf{\scriptsize{}1.72} & {\scriptsize{}-9.88 (-1.07)}\tabularnewline
 & {\scriptsize{}CFB-A} & {\scriptsize{}92} & {\scriptsize{}1.43} & {\scriptsize{}--} &  & {\scriptsize{}83} & {\scriptsize{}1.51} & {\scriptsize{}--} &  & {\scriptsize{}66} & {\scriptsize{}1.55} & {\scriptsize{}--}\tabularnewline
\bottomrule
\end{tabular}{\scriptsize{}}{\scriptsize\par}
\par\end{centering}
{\footnotesize{}Notes: For a baseline model $m\in\left\{ UCB,CFB,CFA\right\} $,
the improvement (\%) by CFB-A over $m$ is defined as $\frac{cHPR_{CFB-A}-cHPR_{m}}{cHPR_{m}}$;
the best-performing model of each case is highlighted in bold. }\textit{\emph{\footnotesize{}Significance
levels of independent two-sample t-tests are denoted as follows: }}\emph{\footnotesize{}$*p<.10,**p<0.05,***p<.01$}{\footnotesize{},
and t-values are shown in parentheses.}{\footnotesize\par}
\end{table}
\end{landscape}
\end{document}